\documentclass{lmcs} 
\pdfoutput=1

\usepackage{lastpage}
\lmcsdoi{22}{2}{28}
\lmcsheading{}{\pageref{LastPage}}{}{}%
{Aug.~04,~2025}{Jun.~15,~2026}{}


\usepackage{hyperref}
\usepackage{graphicx}
\usepackage{tikz}
\usetikzlibrary{math}
\usetikzlibrary{patterns}
\usetikzlibrary{patterns.meta}
\usepackage{tikz-cd}

\usepackage{pgfplots}
\usepackage{tikz-3dplot}

\tdplotsetmaincoords{60}{115}
\pgfplotsset{compat=newest}

\usepackage{graphicx}%
\usepackage{multirow}%
\usepackage{amsmath,amssymb,amsfonts}%
\usepackage{mathrsfs}%
\usepackage{stmaryrd}
\usepackage[title]{appendix}%
\usepackage{xcolor}%
\usepackage{textcomp}%
\usepackage{manyfoot}%
\usepackage{mathtools}%
\usepackage{booktabs}%
\usepackage{algorithm}%
\usepackage{algorithmicx}%
\usepackage[noend]{algpseudocode}%
\usepackage{listings}%
\usepackage{mathpartir}%
\usepackage{xparse}
\usepackage{xspace}

\usepackage{bussproofs}
\usepackage{stackengine}

\algrenewcommand\algorithmicrequire{\textbf{Input:}}
\algrenewcommand\algorithmicensure{\textbf{Output:}}

\usepackage[labelformat=simple]{subcaption}

\usepackage{agda/latex/agda}
\usepackage[utf8]{inputenc}
\usepackage{newunicodechar}

\tdplotsetmaincoords{60}{115}
\pgfplotsset{compat=newest}

\usepackage{todonotes} 

\NewDocumentCommand{\newtodoer}{m m m}{%
  \NewDocumentCommand{#1}{m}{\todo[inline,color=#3]{\textbf{#2}: ##1}}
  \NewEnviron{#2}{%
    \todo[inline,color=#3,caption={}]{%
      \begin{minipage}{\linewidth}
        \textbf{#2}: \BODY
      \end{minipage}
    }
  }
}

\newcommand{\systemname}[1]{\texttt{\color{darkgray}#1}\xspace}
\newcommand{\Agda}{\systemname{Agda}}
\newcommand{\agdaCubical}{\systemname{agda/cubical}}
\newcommand{\onelab}{\systemname{1lab}}
\newcommand{\CubicalAgda}{\systemname{Cubical} \systemname{Agda}}
\newcommand{\Haskell}{\systemname{Haskell}}

\newcommand{\redtt}{\systemname{redtt}}

\newcommand{\Cartesian}{\mproblem{Cartesian}}
\newcommand{\Disjunctive}{\mproblem{Disjunctive}}
\newcommand{\Dedekind}{\mproblem{Dedekind}}
\newcommand{\DeMorgan}{\mproblem{DeMorgan}}
\newcommand{\KanCubicalCell}{\mproblem{Kan}}
\newcommand{\Conts}[2]{\ensuremath{\mathsf{Conts}(#1, #2)}}
\NewDocumentCommand{\mproblem}{m o}{\textsc{{#1}}\IfValueT{#2}{(\ensuremath{{#2}})}\xspace}

\makeatletter
\DeclareRobustCommand{\sqcdot}{\mathbin{\mathpalette\morphic@sqcdot\relax}}
\newcommand{\morphic@sqcdot}[2]{%
  \sbox\z@{$\m@th#1\centerdot$}%
  \ht\z@=.33333\ht\z@
  \vcenter{\box\z@}%
}
\makeatother

\newcommand{\ctxtjudg}{\ensuremath{\ \mathsf{ctxt}}}
\newcommand{\ictxtjudg}{\ensuremath{\ \mathsf{dctxt}}}
\newcommand{\dimjudg}{\ensuremath{\ \mathsf{dim}}}
\newcommand{\typejudg}{\ensuremath{\ \mathsf{bdy}}}
\newcommand{\termjudg}{\ensuremath{\ \mathsf{cell}}}

\newcommand{\assign}[3]{\ensuremath{#1 = #2 \mapsto #3}}
\newcommand{\assignbdy}[2]{\ensuremath{\partial #1 \mapsto #2}}
\newcommand{\se}{\mid} 
\NewDocumentCommand{\boundary}{s m}{\IfBooleanTF{#1}{\left[\,#2\,\right]}{[\,#2\,]}}

\newcommand{\contortlabel}{\mathrm{c}}
\newcommand{\fillinglabel}{}

\NewDocumentCommand{\isictxt}{m o}{{#1}\IfValueT{#2}{ = {#2}} \ictxtjudg}
\NewDocumentCommand{\isisubst}{m o m m}{%
  {#1}\IfValueT{#2}{ = {#2}} \colon {#3} \rightsquigarrow {#4}}
\NewDocumentCommand{\isincl}{m o m m}{%
  {#1}\IfValueT{#2}{ = {#2}} \colon {#3} \to {#4}}
\NewDocumentCommand{\isdim}{o m o}{\IfValueT{#1}{{#1} \vdash} {#2}\IfValueT{#3}{ = {#3}}\ \mathsf{atom}}

\NewDocumentCommand{\iscdim}{o m o}{\IfValueT{#1}{{#1} \vdash} {#2}\IfValueT{#3}{ = {#3}} \dimjudg}
\NewDocumentCommand{\iscterm}{s o o m o d<>}{%
  \IfValueT{#2}{{#2} \mid {#3} \vdash_{\contortlabel}}{#4}\IfValueT{#5}{ = {#5}}\IfValueTF{#6}{ : \IfBooleanTF{#1}{\boundary*{#6}}{\boundary{#6}}}{\termjudg}}
\NewDocumentCommand{\iscbdy}{o o o m o}{\IfValueT{#1}{{#1} \mid {#2}\IfValueT{#3}{ \mid\!\mid {#3}} \vdash_{\contortlabel}}{#4}\IfValueT{#5}{ = {#5}} \typejudg}
\NewDocumentCommand{\isctxt}{m o}{{#1}\IfValueT{#2}{ = {#2}} \ctxtjudg}

\NewDocumentCommand{\isfbdy}{o o o m o}{\IfValueT{#1}{{#1} \mid {#2}\IfValueT{#3}{ \mid\!\mid {#3}} \vdash_{\fillinglabel}}{#4}\IfValueT{#5}{ = {#5}} \typejudg}
\NewDocumentCommand{\isfterm}{s o o m o d<>}{%
  \IfValueT{#2}{{#2} \mid {#3} \vdash_{\fillinglabel}}{#4}\IfValueT{#5}{ = {#5}}\IfValueTF{#6}{ : \IfBooleanTF{#1}{\boundary*{#6}}{\boundary{#6}}}{\termjudg}}
\NewDocumentCommand{\isof}{s m g o}{%
  {#2}\IfValueT{#3}{({#3})} : \IfBooleanTF{#1}{\boundary*{\IfValueT{#4}{#4}}}{\boundary{\IfValueT{#4}{#4}}}}

\newcommand{\ictxtcon}[3]{{#1}[#2 = #3]}

\newcommand{\inclface}[2]{{#1} \mapsto {#2}}
\newcommand{\inclcon}[2]{{#1} = {#2}}

\newcommand{\isubstface}[2]{{#1} \mapsto {#2}}

\newcommand{\ffillop}{\mathsf{fill}}
\NewDocumentCommand{\ffill}{s m m m m m}{\ffillop^{{#2} \to {#3}}~#4.\IfBooleanTF{#1}{\boundary*{#5}}{\boundary{#5}}~#6}

\newcommand{\face}[2]{\ensuremath{{#1}[{#2}]}}
\newcommand{\cont}[2]{\ensuremath{{#1}\langle{#2}\rangle}}

\newcommand{\csvar}[2]{\ensuremath{ ({#1}={#2}) }}

\newcommand{\hole}{\textbf{?}}

\NewDocumentCommand{\pcomp}{m m}{{#1} \sqcdot {#2}}

\NewDocumentCommand{\pthreecomp}{o m m m}{({#2} \sqcdot {#3} \sqcdot {#4})^\IfValueTF{#1}{#1}{\izero}}
\NewDocumentCommand{\pthreecompfill}{o m m m m}{\mathsf{fill}^{\IfValueTF{#1}{#1}{\izero} \to {#2}}\ ({#3} \sqcdot {#4} \sqcdot {#5})}

\newcommand{\izero}{{\mathchoice{\scriptstyle\mathbf{0}}{\scriptstyle\mathbf{0}}{\scriptscriptstyle\mathbf{0}}{\scriptscriptstyle\mathbf{0}}}}
\newcommand{\ione}{{\mathchoice{\scriptstyle\mathbf{1}}{\scriptstyle\mathbf{1}}{\scriptscriptstyle\mathbf{1}}{\scriptscriptstyle\mathbf{1}}}}
\newcommand{\negI}[1]{\overline{#1}}
\newcommand{\inv}{{\sim}}
\newcommand{\join}{\vee}
\newcommand{\meet}{\wedge}

\newcommand{\pint}[1]{\mathbf{I}^{#1}}

\newcommand{\pintrestr}[3]{\pint{#1}_{{#2}={#3}}}

\newcommand{\pmtoformula}[1]{\ensuremath{{#1}_{\join \meet}}}
\newcommand{\formulatopm}[1]{\ensuremath{{#1}_\pint{}}}

\newcommand{\restrict}[2]{{#1}|_{#2}}

\newcommand{\card}[1]{\ensuremath{| {#1} |}}
\newcommand{\pow}[1]{\mathcal{P}({#1})}

\DeclarePairedDelimiter\brackss\llbracket\rrbracket

\DeclarePairedDelimiter\braces\lbrace\rbrace
\DeclarePairedDelimiterX\set[2]\lbrace\rbrace{#1 \mathrel{\delimsize\vert} #2}

\newcommand{\dimsquare}[2]{
  \begin{tikzpicture}[scale=0.5]
    \draw[->,>=stealth] (0,0) -- (0,1) node [near end,left, fill=none] {$#1$};
    \draw[->,>=stealth] (0,0) -- (1,0) node [near end,below, fill=none] {$#2$};
  \end{tikzpicture}
}

\newcommand{\dimcube}[3]{
  \begin{tikzpicture}[scale=0.6]
    \draw[->,>=stealth] (0,0) -- (0,1) node [near end,left, fill=none] {$#1$};
    \draw[->,>=stealth] (0,0) -- (1,0) node [near end,below, fill=none] {$#2$};
    \draw[->,>=stealth] (0,0) -- (-.71,-.71) node [near end,below=.02cm, fill=none] {$#3$};
  \end{tikzpicture}
}

\newunicodechar{λ}{\ensuremath{\mathnormal\lambda}}
\newunicodechar{←}{\ensuremath{\mathnormal\from}}
\newunicodechar{→}{\ensuremath{\mathnormal\to}}
\newunicodechar{∀}{\ensuremath{\mathnormal\forall}}
\newunicodechar{Ω}{\ensuremath{\mathnormal\Omega}}
\newunicodechar{ℓ}{\ensuremath{l}}
\newunicodechar{≡}{\ensuremath{\mathnormal\equiv}}
\newunicodechar{∨}{\ensuremath{\mathnormal\vee}}
\newunicodechar{∧}{\ensuremath{\mathnormal\wedge}}
\newunicodechar{⊔}{\ensuremath{\mathnormal\sqcup}}
\newunicodechar{↦}{\ensuremath{\mathnormal\mapsto}}
\newunicodechar{φ}{\ensuremath{\mathnormal\phi}}
\newunicodechar{∙}{\ensuremath{\mathnormal\sqcdot}}
\newunicodechar{Σ}{\ensuremath{\mathnormal\Sigma}}
\newunicodechar{α}{\ensuremath{\mathnormal\alpha}}
\newunicodechar{₁}{\ensuremath{_1}}
\newunicodechar{₂}{\ensuremath{_2}}
\newunicodechar{₃}{\ensuremath{_3}}
\newunicodechar{₄}{\ensuremath{_4}}
\newunicodechar{⊢}{\ensuremath{\vdash}}

\newcommand{\present}[2]{\langle #1 {\mid} #2 \rangle}
\newcommand{\presentctxt}[2]{\ulcorner #1 {\mid} #2 \urcorner}
\newcommand{\tmword}[2]{\ulcorner #1 \urcorner(#2)}
\newcommand{\tmcancel}[5]{\mathsf{cancel}^{#2}_{#3,#5}(#1,#4)}
\newcommand{\tmrewrite}[4]{\mathsf{rew}^{#2}_{#3,#4}(#1)}
\newcommand{\tmcnxor}[3]{\mathsf{cnx}_{\lor}(#1)(#2,#3)}

\NewDocumentCommand{\gsem}{s m m}{\llbracket {#2} \rrbracket\IfBooleanT{#1}{^\ast}_{#3}}

\begin{document}

\newtodoer{\todomax}{Max}{cyan!20}
\newtodoer{\todoevan}{Evan}{magenta!20}
\newtodoer{\todoanders}{Anders}{lime!20}

\begin{code}[hide]%
\>[0]\AgdaSymbol{\{-\#}\AgdaSpace{}%
\AgdaKeyword{OPTIONS}\AgdaSpace{}%
\AgdaPragma{--cubical}\AgdaSpace{}%
\AgdaSymbol{\#-\}}\<%
\\
\>[0]\AgdaKeyword{module}\AgdaSpace{}%
\AgdaModule{Examples}\AgdaSpace{}%
\AgdaKeyword{where}\<%
\\
\\[\AgdaEmptyExtraSkip]%
\>[0]\AgdaKeyword{open}\AgdaSpace{}%
\AgdaKeyword{import}\AgdaSpace{}%
\AgdaModule{Agda.Primitive}\AgdaSpace{}%
\AgdaKeyword{public}\<%
\\
\>[0][@{}l@{\AgdaIndent{0}}]%
\>[2]\AgdaKeyword{using}%
\>[11]\AgdaSymbol{(}\AgdaSpace{}%
\AgdaPostulate{Level}\<%
\\
\>[2]\AgdaSymbol{;}\AgdaSpace{}%
\AgdaPrimitive{SSet}\AgdaSpace{}%
\AgdaSymbol{)}\<%
\\
\>[2]\AgdaKeyword{renaming}\AgdaSpace{}%
\AgdaSymbol{(}\AgdaSpace{}%
\AgdaPrimitive{lzero}\AgdaSpace{}%
\AgdaSymbol{to}\AgdaSpace{}%
\AgdaPrimitive{ℓ-zero}\<%
\\
\>[2]\AgdaSymbol{;}\AgdaSpace{}%
\AgdaPrimitive{lsuc}%
\>[10]\AgdaSymbol{to}\AgdaSpace{}%
\AgdaPrimitive{ℓ-suc}\<%
\\
\>[2]\AgdaSymbol{;}\AgdaSpace{}%
\AgdaOperator{\AgdaPrimitive{\AgdaUnderscore{}⊔\AgdaUnderscore{}}}%
\>[10]\AgdaSymbol{to}\AgdaSpace{}%
\AgdaOperator{\AgdaPrimitive{ℓ-max}}\<%
\\
\>[2]\AgdaSymbol{;}\AgdaSpace{}%
\AgdaPrimitive{Set}%
\>[10]\AgdaSymbol{to}\AgdaSpace{}%
\AgdaPrimitive{Type}\<%
\\
\>[2]\AgdaSymbol{;}\AgdaSpace{}%
\AgdaPrimitive{Setω}%
\>[10]\AgdaSymbol{to}\AgdaSpace{}%
\AgdaPrimitive{Typeω}\AgdaSpace{}%
\AgdaSymbol{)}\<%
\\
\>[0]\AgdaKeyword{open}\AgdaSpace{}%
\AgdaKeyword{import}\AgdaSpace{}%
\AgdaModule{Agda.Builtin.Cubical.Sub}\AgdaSpace{}%
\AgdaKeyword{public}\<%
\\
\>[0][@{}l@{\AgdaIndent{0}}]%
\>[2]\AgdaKeyword{renaming}\AgdaSpace{}%
\AgdaSymbol{(}\AgdaPrimitive{primSubOut}\AgdaSpace{}%
\AgdaSymbol{to}\AgdaSpace{}%
\AgdaPrimitive{outS}\AgdaSymbol{)}\<%
\\
\\[\AgdaEmptyExtraSkip]%
\>[0]\AgdaKeyword{open}\AgdaSpace{}%
\AgdaKeyword{import}\AgdaSpace{}%
\AgdaModule{Agda.Builtin.Cubical.Path}\AgdaSpace{}%
\AgdaKeyword{public}\<%
\\
\>[0]\AgdaKeyword{open}\AgdaSpace{}%
\AgdaKeyword{import}\AgdaSpace{}%
\AgdaModule{Agda.Builtin.Cubical.Sub}\AgdaSpace{}%
\AgdaKeyword{public}\<%
\\
\>[0]\AgdaKeyword{open}\AgdaSpace{}%
\AgdaKeyword{import}\AgdaSpace{}%
\AgdaModule{Agda.Primitive.Cubical}\AgdaSpace{}%
\AgdaKeyword{renaming}\AgdaSpace{}%
\AgdaSymbol{(}\AgdaPrimitive{primINeg}\AgdaSpace{}%
\AgdaSymbol{to}\AgdaSpace{}%
\AgdaPrimitive{\textasciitilde{}\AgdaUnderscore{}}\AgdaSymbol{;}\AgdaSpace{}%
\AgdaPrimitive{primIMax}\AgdaSpace{}%
\AgdaSymbol{to}\AgdaSpace{}%
\AgdaPrimitive{\AgdaUnderscore{}∨\AgdaUnderscore{}}\AgdaSymbol{;}\AgdaSpace{}%
\AgdaPrimitive{primIMin}\AgdaSpace{}%
\AgdaSymbol{to}\AgdaSpace{}%
\AgdaPrimitive{\AgdaUnderscore{}∧\AgdaUnderscore{}}\AgdaSymbol{;}\<%
\\
\>[0][@{}l@{\AgdaIndent{0}}]%
\>[2]\AgdaPrimitive{primHComp}\AgdaSpace{}%
\AgdaSymbol{to}\AgdaSpace{}%
\AgdaPrimitive{hcomp}\AgdaSymbol{;}\AgdaSpace{}%
\AgdaPrimitive{primTransp}\AgdaSpace{}%
\AgdaSymbol{to}\AgdaSpace{}%
\AgdaPrimitive{transp}\AgdaSymbol{;}\AgdaSpace{}%
\AgdaPrimitive{primComp}\AgdaSpace{}%
\AgdaSymbol{to}\AgdaSpace{}%
\AgdaPrimitive{comp}\AgdaSymbol{;}\<%
\\
\>[2]\AgdaPostulate{itIsOne}\AgdaSpace{}%
\AgdaSymbol{to}\AgdaSpace{}%
\AgdaPostulate{1=1}\AgdaSymbol{)}\AgdaSpace{}%
\AgdaKeyword{public}\<%
\\
\\[\AgdaEmptyExtraSkip]%
\>[0]\AgdaKeyword{open}\AgdaSpace{}%
\AgdaKeyword{import}\AgdaSpace{}%
\AgdaModule{Agda.Builtin.Sigma}\AgdaSpace{}%
\AgdaKeyword{public}\<%
\\
\\[\AgdaEmptyExtraSkip]%
\\[\AgdaEmptyExtraSkip]%
\>[0]\AgdaFunction{refl}\AgdaSpace{}%
\AgdaSymbol{:}\AgdaSpace{}%
\AgdaSymbol{∀\{}\AgdaBound{ℓ}\AgdaSymbol{\}}\AgdaSpace{}%
\AgdaSymbol{\{}\AgdaBound{A}\AgdaSpace{}%
\AgdaSymbol{:}\AgdaSpace{}%
\AgdaPrimitive{Set}\AgdaSpace{}%
\AgdaBound{ℓ}\AgdaSymbol{\}}\AgdaSpace{}%
\AgdaSymbol{\{}\AgdaBound{x}\AgdaSpace{}%
\AgdaSymbol{:}\AgdaSpace{}%
\AgdaBound{A}\AgdaSymbol{\}}\AgdaSpace{}%
\AgdaSymbol{→}\AgdaSpace{}%
\AgdaBound{x}\AgdaSpace{}%
\AgdaOperator{\AgdaFunction{≡}}\AgdaSpace{}%
\AgdaBound{x}\<%
\\
\>[0]\AgdaFunction{refl}\AgdaSpace{}%
\AgdaSymbol{\{}\AgdaArgument{x}\AgdaSpace{}%
\AgdaSymbol{=}\AgdaSpace{}%
\AgdaBound{x}\AgdaSymbol{\}}\AgdaSpace{}%
\AgdaSymbol{\AgdaUnderscore{}}\AgdaSpace{}%
\AgdaSymbol{=}\AgdaSpace{}%
\AgdaBound{x}\<%
\\
\>[0]\AgdaSymbol{\{-\#}\AgdaSpace{}%
\AgdaKeyword{INLINE}\AgdaSpace{}%
\AgdaFunction{refl}\AgdaSpace{}%
\AgdaSymbol{\#-\}}\<%
\\
\\[\AgdaEmptyExtraSkip]%
\>[0]\AgdaFunction{Path}\AgdaSpace{}%
\AgdaSymbol{:}\AgdaSpace{}%
\AgdaSymbol{∀}\AgdaSpace{}%
\AgdaSymbol{\{}\AgdaBound{ℓ}\AgdaSymbol{\}}\AgdaSpace{}%
\AgdaSymbol{(}\AgdaBound{A}\AgdaSpace{}%
\AgdaSymbol{:}\AgdaSpace{}%
\AgdaPrimitive{Set}\AgdaSpace{}%
\AgdaBound{ℓ}\AgdaSymbol{)}\AgdaSpace{}%
\AgdaSymbol{→}\AgdaSpace{}%
\AgdaBound{A}\AgdaSpace{}%
\AgdaSymbol{→}\AgdaSpace{}%
\AgdaBound{A}\AgdaSpace{}%
\AgdaSymbol{→}\AgdaSpace{}%
\AgdaPrimitive{Set}\AgdaSpace{}%
\AgdaBound{ℓ}\<%
\\
\>[0]\AgdaFunction{Path}\AgdaSpace{}%
\AgdaBound{A}\AgdaSpace{}%
\AgdaBound{a}\AgdaSpace{}%
\AgdaBound{b}\AgdaSpace{}%
\AgdaSymbol{=}\AgdaSpace{}%
\AgdaPostulate{PathP}\AgdaSpace{}%
\AgdaSymbol{(λ}\AgdaSpace{}%
\AgdaBound{\AgdaUnderscore{}}\AgdaSpace{}%
\AgdaSymbol{→}\AgdaSpace{}%
\AgdaBound{A}\AgdaSymbol{)}\AgdaSpace{}%
\AgdaBound{a}\AgdaSpace{}%
\AgdaBound{b}\<%
\\
\\[\AgdaEmptyExtraSkip]%
\>[0]\AgdaOperator{\AgdaFunction{\AgdaUnderscore{}[\AgdaUnderscore{}↦\AgdaUnderscore{}]}}\AgdaSpace{}%
\AgdaSymbol{:}\AgdaSpace{}%
\AgdaSymbol{∀}\AgdaSpace{}%
\AgdaSymbol{\{}\AgdaBound{ℓ}\AgdaSymbol{\}}\AgdaSpace{}%
\AgdaSymbol{(}\AgdaBound{A}\AgdaSpace{}%
\AgdaSymbol{:}\AgdaSpace{}%
\AgdaPrimitive{Set}\AgdaSpace{}%
\AgdaBound{ℓ}\AgdaSymbol{)}\AgdaSpace{}%
\AgdaSymbol{(}\AgdaBound{φ}\AgdaSpace{}%
\AgdaSymbol{:}\AgdaSpace{}%
\AgdaDatatype{I}\AgdaSymbol{)}\AgdaSpace{}%
\AgdaSymbol{(}\AgdaBound{u}\AgdaSpace{}%
\AgdaSymbol{:}\AgdaSpace{}%
\AgdaPrimitive{Partial}\AgdaSpace{}%
\AgdaBound{φ}\AgdaSpace{}%
\AgdaBound{A}\AgdaSymbol{)}\AgdaSpace{}%
\AgdaSymbol{→}\AgdaSpace{}%
\AgdaSymbol{\AgdaUnderscore{}}\<%
\\
\>[0]\AgdaBound{A}\AgdaSpace{}%
\AgdaOperator{\AgdaFunction{[}}\AgdaSpace{}%
\AgdaBound{φ}\AgdaSpace{}%
\AgdaOperator{\AgdaFunction{↦}}\AgdaSpace{}%
\AgdaBound{u}\AgdaSpace{}%
\AgdaOperator{\AgdaFunction{]}}\AgdaSpace{}%
\AgdaSymbol{=}\AgdaSpace{}%
\AgdaPostulate{Sub}\AgdaSpace{}%
\AgdaBound{A}\AgdaSpace{}%
\AgdaBound{φ}\AgdaSpace{}%
\AgdaBound{u}\<%
\\
\\[\AgdaEmptyExtraSkip]%
\>[0]\AgdaKeyword{private}\<%
\\
\>[0][@{}l@{\AgdaIndent{0}}]%
\>[2]\AgdaKeyword{variable}\<%
\\
\>[2][@{}l@{\AgdaIndent{0}}]%
\>[4]\AgdaGeneralizable{ℓ}\AgdaSpace{}%
\AgdaSymbol{:}\AgdaSpace{}%
\AgdaPostulate{Level}\<%
\\
\>[4]\AgdaGeneralizable{A}\AgdaSpace{}%
\AgdaSymbol{:}\AgdaSpace{}%
\AgdaPrimitive{Type}\AgdaSpace{}%
\AgdaGeneralizable{ℓ}\<%
\\
\>[0]\<%
\end{code}
\newcommand{\agdapathcomp}{%
\begin{code}%
\>[0]\AgdaOperator{\AgdaFunction{\AgdaUnderscore{}∙\AgdaUnderscore{}}}\AgdaSpace{}%
\AgdaSymbol{:}\AgdaSpace{}%
\AgdaSymbol{\{}\AgdaBound{x}\AgdaSpace{}%
\AgdaBound{y}\AgdaSpace{}%
\AgdaBound{z}\AgdaSpace{}%
\AgdaSymbol{:}\AgdaSpace{}%
\AgdaGeneralizable{A}\AgdaSymbol{\}}\AgdaSpace{}%
\AgdaSymbol{→}\AgdaSpace{}%
\AgdaBound{x}\AgdaSpace{}%
\AgdaOperator{\AgdaFunction{≡}}\AgdaSpace{}%
\AgdaBound{y}\AgdaSpace{}%
\AgdaSymbol{→}\AgdaSpace{}%
\AgdaBound{y}\AgdaSpace{}%
\AgdaOperator{\AgdaFunction{≡}}\AgdaSpace{}%
\AgdaBound{z}\AgdaSpace{}%
\AgdaSymbol{→}\AgdaSpace{}%
\AgdaBound{x}\AgdaSpace{}%
\AgdaOperator{\AgdaFunction{≡}}\AgdaSpace{}%
\AgdaBound{z}\<%
\\
\>[0]\AgdaSymbol{(}\AgdaBound{p}\AgdaSpace{}%
\AgdaOperator{\AgdaFunction{∙}}\AgdaSpace{}%
\AgdaBound{q}\AgdaSymbol{)}\AgdaSpace{}%
\AgdaBound{i}\AgdaSpace{}%
\AgdaSymbol{=}\AgdaSpace{}%
\AgdaPrimitive{hcomp}\AgdaSpace{}%
\AgdaSymbol{(λ}\AgdaSpace{}%
\AgdaBound{j}\AgdaSpace{}%
\AgdaSymbol{→}\AgdaSpace{}%
\AgdaSymbol{λ}\AgdaSpace{}%
\AgdaSymbol{\{}\AgdaSpace{}%
\AgdaSymbol{(}\AgdaBound{i}\AgdaSpace{}%
\AgdaSymbol{=}\AgdaSpace{}%
\AgdaInductiveConstructor{i0}\AgdaSymbol{)}\AgdaSpace{}%
\AgdaSymbol{→}\AgdaSpace{}%
\AgdaBound{p}\AgdaSpace{}%
\AgdaInductiveConstructor{i0}\AgdaSpace{}%
\AgdaSymbol{;}\AgdaSpace{}%
\AgdaSymbol{(}\AgdaBound{i}\AgdaSpace{}%
\AgdaSymbol{=}\AgdaSpace{}%
\AgdaInductiveConstructor{i1}\AgdaSymbol{)}\AgdaSpace{}%
\AgdaSymbol{→}\AgdaSpace{}%
\AgdaBound{q}\AgdaSpace{}%
\AgdaBound{j}\AgdaSpace{}%
\AgdaSymbol{\})}\AgdaSpace{}%
\AgdaSymbol{(}\AgdaBound{p}\AgdaSpace{}%
\AgdaBound{i}\AgdaSymbol{)}\<%
\end{code}}
\begin{code}[hide]%
\>[0]\<%
\\
\>[0]\AgdaFunction{sym}\AgdaSpace{}%
\AgdaSymbol{:}\AgdaSpace{}%
\AgdaSymbol{\{}\AgdaBound{x}\AgdaSpace{}%
\AgdaBound{y}\AgdaSpace{}%
\AgdaSymbol{:}\AgdaSpace{}%
\AgdaGeneralizable{A}\AgdaSymbol{\}}\AgdaSpace{}%
\AgdaSymbol{→}\AgdaSpace{}%
\AgdaBound{x}\AgdaSpace{}%
\AgdaOperator{\AgdaFunction{≡}}\AgdaSpace{}%
\AgdaBound{y}\AgdaSpace{}%
\AgdaSymbol{→}\AgdaSpace{}%
\AgdaBound{y}\AgdaSpace{}%
\AgdaOperator{\AgdaFunction{≡}}\AgdaSpace{}%
\AgdaBound{x}\<%
\\
\>[0]\AgdaFunction{sym}\AgdaSpace{}%
\AgdaBound{p}\AgdaSpace{}%
\AgdaBound{i}\AgdaSpace{}%
\AgdaSymbol{=}\AgdaSpace{}%
\AgdaPrimitive{hcomp}\AgdaSpace{}%
\AgdaSymbol{(λ}\AgdaSpace{}%
\AgdaBound{j}\AgdaSpace{}%
\AgdaSymbol{→}\AgdaSpace{}%
\AgdaSymbol{λ}\AgdaSpace{}%
\AgdaSymbol{\{}\AgdaSpace{}%
\AgdaSymbol{(}\AgdaBound{i}\AgdaSpace{}%
\AgdaSymbol{=}\AgdaSpace{}%
\AgdaInductiveConstructor{i0}\AgdaSymbol{)}\AgdaSpace{}%
\AgdaSymbol{→}\AgdaSpace{}%
\AgdaBound{p}\AgdaSpace{}%
\AgdaBound{j}\AgdaSpace{}%
\AgdaSymbol{;}\AgdaSpace{}%
\AgdaSymbol{(}\AgdaBound{i}\AgdaSpace{}%
\AgdaSymbol{=}\AgdaSpace{}%
\AgdaInductiveConstructor{i1}\AgdaSymbol{)}\AgdaSpace{}%
\AgdaSymbol{→}\AgdaSpace{}%
\AgdaBound{p}\AgdaSpace{}%
\AgdaInductiveConstructor{i0}\AgdaSpace{}%
\AgdaSymbol{\})}\AgdaSpace{}%
\AgdaSymbol{(}\AgdaBound{p}\AgdaSpace{}%
\AgdaInductiveConstructor{i0}\AgdaSymbol{)}\<%
\\
\\[\AgdaEmptyExtraSkip]%
\>[0]\<%
\end{code}
\newcommand{\agdahfill}{%
\begin{code}%
\>[0]\AgdaFunction{hfill}\AgdaSpace{}%
\AgdaSymbol{:}\AgdaSpace{}%
\AgdaSymbol{\{}\AgdaBound{A}\AgdaSpace{}%
\AgdaSymbol{:}\AgdaSpace{}%
\AgdaPrimitive{Type}\AgdaSpace{}%
\AgdaGeneralizable{ℓ}\AgdaSymbol{\}}\AgdaSpace{}%
\AgdaSymbol{\{}\AgdaBound{φ}\AgdaSpace{}%
\AgdaSymbol{:}\AgdaSpace{}%
\AgdaDatatype{I}\AgdaSymbol{\}}\AgdaSpace{}%
\AgdaSymbol{(}\AgdaBound{u}\AgdaSpace{}%
\AgdaSymbol{:}\AgdaSpace{}%
\AgdaSymbol{∀}\AgdaSpace{}%
\AgdaBound{i}\AgdaSpace{}%
\AgdaSymbol{→}\AgdaSpace{}%
\AgdaPrimitive{Partial}\AgdaSpace{}%
\AgdaBound{φ}\AgdaSpace{}%
\AgdaBound{A}\AgdaSymbol{)}\AgdaSpace{}%
\AgdaSymbol{(}\AgdaBound{u0}\AgdaSpace{}%
\AgdaSymbol{:}\AgdaSpace{}%
\AgdaBound{A}\AgdaSpace{}%
\AgdaOperator{\AgdaFunction{[}}\AgdaSpace{}%
\AgdaBound{φ}\AgdaSpace{}%
\AgdaOperator{\AgdaFunction{↦}}\AgdaSpace{}%
\AgdaBound{u}\AgdaSpace{}%
\AgdaInductiveConstructor{i0}\AgdaSpace{}%
\AgdaOperator{\AgdaFunction{]}}\AgdaSymbol{)}\<%
\\
\>[0][@{}l@{\AgdaIndent{0}}]%
\>[2]\AgdaSymbol{(}\AgdaBound{i}\AgdaSpace{}%
\AgdaSymbol{:}\AgdaSpace{}%
\AgdaDatatype{I}\AgdaSymbol{)}\AgdaSpace{}%
\AgdaSymbol{→}\AgdaSpace{}%
\AgdaBound{A}\<%
\\
\>[0]\AgdaFunction{hfill}%
\>[252I]\AgdaSymbol{\{}\AgdaArgument{φ}\AgdaSpace{}%
\AgdaSymbol{=}\AgdaSpace{}%
\AgdaBound{φ}\AgdaSymbol{\}}\AgdaSpace{}%
\AgdaBound{u}\AgdaSpace{}%
\AgdaBound{u0}\AgdaSpace{}%
\AgdaBound{i}\AgdaSpace{}%
\AgdaSymbol{=}\AgdaSpace{}%
\AgdaPrimitive{hcomp}\AgdaSpace{}%
\AgdaSymbol{(λ}\AgdaSpace{}%
\AgdaBound{j}\AgdaSpace{}%
\AgdaSymbol{→}\AgdaSpace{}%
\AgdaSymbol{λ}\AgdaSpace{}%
\AgdaSymbol{\{}\<%
\\
\>[.][@{}l@{}]\<[252I]%
\>[6]\AgdaSymbol{(}\AgdaBound{φ}\AgdaSpace{}%
\AgdaSymbol{=}\AgdaSpace{}%
\AgdaInductiveConstructor{i1}\AgdaSymbol{)}\AgdaSpace{}%
\AgdaSymbol{→}\AgdaSpace{}%
\AgdaBound{u}\AgdaSpace{}%
\AgdaSymbol{(}\AgdaBound{i}\AgdaSpace{}%
\AgdaOperator{\AgdaPrimitive{∧}}\AgdaSpace{}%
\AgdaBound{j}\AgdaSymbol{)}\AgdaSpace{}%
\AgdaPostulate{1=1}\<%
\\
\>[0][@{}l@{\AgdaIndent{0}}]%
\>[4]\AgdaSymbol{;}\AgdaSpace{}%
\AgdaSymbol{(}\AgdaBound{i}\AgdaSpace{}%
\AgdaSymbol{=}\AgdaSpace{}%
\AgdaInductiveConstructor{i0}\AgdaSymbol{)}\AgdaSpace{}%
\AgdaSymbol{→}\AgdaSpace{}%
\AgdaPrimitive{outS}\AgdaSpace{}%
\AgdaBound{u0}\AgdaSpace{}%
\AgdaSymbol{\})}\<%
\\
\>[4]\AgdaSymbol{(}\AgdaPrimitive{outS}\AgdaSpace{}%
\AgdaBound{u0}\AgdaSymbol{)}\<%
\end{code}}
\begin{code}[hide]%
\>[0]\<%
\\
\\[\AgdaEmptyExtraSkip]%
\\[\AgdaEmptyExtraSkip]%
\>[0]\AgdaKeyword{module}\AgdaSpace{}%
\AgdaModule{\AgdaUnderscore{}}\AgdaSpace{}%
\AgdaSymbol{(}\AgdaBound{A}\AgdaSpace{}%
\AgdaSymbol{:}\AgdaSpace{}%
\AgdaPrimitive{Type}\AgdaSpace{}%
\AgdaGeneralizable{ℓ}\AgdaSymbol{)}\AgdaSpace{}%
\AgdaSymbol{(}\AgdaBound{w}\AgdaSpace{}%
\AgdaBound{x}\AgdaSpace{}%
\AgdaBound{y}\AgdaSpace{}%
\AgdaBound{z}\AgdaSpace{}%
\AgdaSymbol{:}\AgdaSpace{}%
\AgdaBound{A}\AgdaSymbol{)}\AgdaSpace{}%
\AgdaSymbol{(}\AgdaBound{p}\AgdaSpace{}%
\AgdaSymbol{:}\AgdaSpace{}%
\AgdaBound{w}\AgdaSpace{}%
\AgdaOperator{\AgdaFunction{≡}}\AgdaSpace{}%
\AgdaBound{x}\AgdaSymbol{)}\AgdaSpace{}%
\AgdaSymbol{(}\AgdaBound{q}\AgdaSpace{}%
\AgdaSymbol{:}\AgdaSpace{}%
\AgdaBound{x}\AgdaSpace{}%
\AgdaOperator{\AgdaFunction{≡}}\AgdaSpace{}%
\AgdaBound{y}\AgdaSymbol{)}\AgdaSpace{}%
\AgdaSymbol{(}\AgdaBound{r}\AgdaSpace{}%
\AgdaSymbol{:}\AgdaSpace{}%
\AgdaBound{y}\AgdaSpace{}%
\AgdaOperator{\AgdaFunction{≡}}\AgdaSpace{}%
\AgdaBound{z}\AgdaSymbol{)}\AgdaSpace{}%
\AgdaKeyword{where}\<%
\end{code}
\newcommand{\pcompassoc}{%
\begin{code}%
\>[0][@{}l@{\AgdaIndent{1}}]%
\>[2]\AgdaFunction{assoc}\AgdaSpace{}%
\AgdaSymbol{:}\AgdaSpace{}%
\AgdaSymbol{(}\AgdaBound{p}\AgdaSpace{}%
\AgdaOperator{\AgdaFunction{∙}}\AgdaSpace{}%
\AgdaBound{q}\AgdaSymbol{)}\AgdaSpace{}%
\AgdaOperator{\AgdaFunction{∙}}\AgdaSpace{}%
\AgdaBound{r}\AgdaSpace{}%
\AgdaOperator{\AgdaFunction{≡}}\AgdaSpace{}%
\AgdaBound{p}\AgdaSpace{}%
\AgdaOperator{\AgdaFunction{∙}}\AgdaSpace{}%
\AgdaSymbol{(}\AgdaBound{q}\AgdaSpace{}%
\AgdaOperator{\AgdaFunction{∙}}\AgdaSpace{}%
\AgdaBound{r}\AgdaSymbol{)}\<%
\\
\>[2]\AgdaFunction{assoc}\AgdaSpace{}%
\AgdaBound{i}\AgdaSpace{}%
\AgdaBound{j}\AgdaSpace{}%
\AgdaSymbol{=}\AgdaSpace{}%
\AgdaPrimitive{hcomp}\AgdaSpace{}%
\AgdaSymbol{(λ}\AgdaSpace{}%
\AgdaBound{k}\AgdaSpace{}%
\AgdaSymbol{→}\<%
\\
\>[2][@{}l@{\AgdaIndent{0}}]%
\>[6]\AgdaSymbol{λ}%
\>[327I]\AgdaSymbol{\{}\AgdaSpace{}%
\AgdaSymbol{(}\AgdaBound{j}\AgdaSpace{}%
\AgdaSymbol{=}\AgdaSpace{}%
\AgdaInductiveConstructor{i0}\AgdaSymbol{)}\AgdaSpace{}%
\AgdaSymbol{→}\AgdaSpace{}%
\AgdaBound{w}\<%
\\
\>[.][@{}l@{}]\<[327I]%
\>[8]\AgdaSymbol{;}\AgdaSpace{}%
\AgdaSymbol{(}\AgdaBound{j}\AgdaSpace{}%
\AgdaSymbol{=}\AgdaSpace{}%
\AgdaInductiveConstructor{i1}\AgdaSymbol{)}\AgdaSpace{}%
\AgdaSymbol{→}%
\>[337I]\AgdaPrimitive{hcomp}\AgdaSpace{}%
\AgdaSymbol{(λ}\AgdaSpace{}%
\AgdaBound{l}\AgdaSpace{}%
\AgdaSymbol{→}\<%
\\
\>[337I][@{}l@{\AgdaIndent{0}}]%
\>[23]\AgdaSymbol{λ}%
\>[341I]\AgdaSymbol{\{}\AgdaSpace{}%
\AgdaSymbol{(}\AgdaBound{i}\AgdaSpace{}%
\AgdaSymbol{=}\AgdaSpace{}%
\AgdaInductiveConstructor{i0}\AgdaSymbol{)}\AgdaSpace{}%
\AgdaSymbol{→}%
\>[346I]\AgdaPrimitive{hcomp}\AgdaSpace{}%
\AgdaSymbol{(λ}\AgdaSpace{}%
\AgdaBound{m}\AgdaSpace{}%
\AgdaSymbol{→}\<%
\\
\>[346I][@{}l@{\AgdaIndent{0}}]%
\>[40]\AgdaSymbol{λ}%
\>[350I]\AgdaSymbol{\{}\AgdaSpace{}%
\AgdaSymbol{(}\AgdaBound{l}\AgdaSpace{}%
\AgdaSymbol{=}\AgdaSpace{}%
\AgdaInductiveConstructor{i0}\AgdaSymbol{)}\AgdaSpace{}%
\AgdaSymbol{→}\AgdaSpace{}%
\AgdaBound{q}\AgdaSpace{}%
\AgdaBound{k}\<%
\\
\>[.][@{}l@{}]\<[350I]%
\>[42]\AgdaSymbol{;}\AgdaSpace{}%
\AgdaSymbol{(}\AgdaBound{l}\AgdaSpace{}%
\AgdaSymbol{=}\AgdaSpace{}%
\AgdaInductiveConstructor{i1}\AgdaSymbol{)}\AgdaSpace{}%
\AgdaSymbol{→}\AgdaSpace{}%
\AgdaBound{r}\AgdaSpace{}%
\AgdaSymbol{(}\AgdaBound{k}\AgdaSpace{}%
\AgdaOperator{\AgdaPrimitive{∧}}\AgdaSpace{}%
\AgdaBound{m}\AgdaSymbol{)}\<%
\\
\>[42]\AgdaSymbol{;}\AgdaSpace{}%
\AgdaSymbol{(}\AgdaBound{k}\AgdaSpace{}%
\AgdaSymbol{=}\AgdaSpace{}%
\AgdaInductiveConstructor{i0}\AgdaSymbol{)}\AgdaSpace{}%
\AgdaSymbol{→}\AgdaSpace{}%
\AgdaBound{q}\AgdaSpace{}%
\AgdaBound{l}\<%
\\
\>[42]\AgdaSymbol{;}\AgdaSpace{}%
\AgdaSymbol{(}\AgdaBound{k}\AgdaSpace{}%
\AgdaSymbol{=}\AgdaSpace{}%
\AgdaInductiveConstructor{i1}\AgdaSymbol{)}\AgdaSpace{}%
\AgdaSymbol{→}\AgdaSpace{}%
\AgdaBound{r}\AgdaSpace{}%
\AgdaSymbol{(}\AgdaBound{l}\AgdaSpace{}%
\AgdaOperator{\AgdaPrimitive{∧}}\AgdaSpace{}%
\AgdaBound{m}\AgdaSymbol{)}\AgdaSpace{}%
\AgdaSymbol{\})}\AgdaSpace{}%
\AgdaSymbol{(}\AgdaBound{q}\AgdaSpace{}%
\AgdaSymbol{(}\AgdaBound{l}\AgdaSpace{}%
\AgdaOperator{\AgdaPrimitive{∨}}\AgdaSpace{}%
\AgdaBound{k}\AgdaSymbol{))}\<%
\\
\>[.][@{}l@{}]\<[341I]%
\>[25]\AgdaSymbol{;}\AgdaSpace{}%
\AgdaSymbol{(}\AgdaBound{k}\AgdaSpace{}%
\AgdaSymbol{=}\AgdaSpace{}%
\AgdaInductiveConstructor{i0}\AgdaSymbol{)}\AgdaSpace{}%
\AgdaSymbol{→}%
\>[388I]\AgdaPrimitive{hcomp}\AgdaSpace{}%
\AgdaSymbol{(λ}\AgdaSpace{}%
\AgdaBound{m}\AgdaSpace{}%
\AgdaSymbol{→}\<%
\\
\>[388I][@{}l@{\AgdaIndent{0}}]%
\>[40]\AgdaSymbol{λ}%
\>[392I]\AgdaSymbol{\{}\AgdaSpace{}%
\AgdaSymbol{(}\AgdaBound{i}\AgdaSpace{}%
\AgdaSymbol{=}\AgdaSpace{}%
\AgdaInductiveConstructor{i0}\AgdaSymbol{)}\AgdaSpace{}%
\AgdaSymbol{→}\AgdaSpace{}%
\AgdaBound{q}\AgdaSpace{}%
\AgdaSymbol{(}\AgdaBound{l}\AgdaSpace{}%
\AgdaOperator{\AgdaPrimitive{∧}}\AgdaSpace{}%
\AgdaBound{m}\AgdaSymbol{)}\<%
\\
\>[.][@{}l@{}]\<[392I]%
\>[42]\AgdaSymbol{;}\AgdaSpace{}%
\AgdaSymbol{(}\AgdaBound{i}\AgdaSpace{}%
\AgdaSymbol{=}\AgdaSpace{}%
\AgdaInductiveConstructor{i1}\AgdaSymbol{)}\AgdaSpace{}%
\AgdaSymbol{→}\AgdaSpace{}%
\AgdaBound{x}\<%
\\
\>[42]\AgdaSymbol{;}\AgdaSpace{}%
\AgdaSymbol{(}\AgdaBound{l}\AgdaSpace{}%
\AgdaSymbol{=}\AgdaSpace{}%
\AgdaInductiveConstructor{i0}\AgdaSymbol{)}\AgdaSpace{}%
\AgdaSymbol{→}\AgdaSpace{}%
\AgdaBound{x}\AgdaSpace{}%
\AgdaSymbol{\})}\AgdaSpace{}%
\AgdaSymbol{(}\AgdaBound{q}\AgdaSpace{}%
\AgdaInductiveConstructor{i0}\AgdaSymbol{)}\<%
\\
\>[25]\AgdaSymbol{;}\AgdaSpace{}%
\AgdaSymbol{(}\AgdaBound{k}\AgdaSpace{}%
\AgdaSymbol{=}\AgdaSpace{}%
\AgdaInductiveConstructor{i1}\AgdaSymbol{)}\AgdaSpace{}%
\AgdaSymbol{→}\AgdaSpace{}%
\AgdaBound{r}\AgdaSpace{}%
\AgdaBound{l}\AgdaSpace{}%
\AgdaSymbol{\})}\AgdaSpace{}%
\AgdaSymbol{(}\AgdaBound{q}\AgdaSpace{}%
\AgdaBound{k}\AgdaSymbol{)}\AgdaSpace{}%
\AgdaSymbol{\})}\<%
\\
\>[6]\AgdaSymbol{(}\AgdaPrimitive{hcomp}\AgdaSpace{}%
\AgdaSymbol{(λ}\AgdaSpace{}%
\AgdaBound{l}\AgdaSpace{}%
\AgdaSymbol{→}\<%
\\
\>[6][@{}l@{\AgdaIndent{0}}]%
\>[9]\AgdaSymbol{λ}%
\>[427I]\AgdaSymbol{\{}\AgdaSpace{}%
\AgdaSymbol{(}\AgdaBound{i}\AgdaSpace{}%
\AgdaSymbol{=}\AgdaSpace{}%
\AgdaInductiveConstructor{i1}\AgdaSymbol{)}\AgdaSpace{}%
\AgdaSymbol{→}\AgdaSpace{}%
\AgdaBound{p}\AgdaSpace{}%
\AgdaBound{j}\<%
\\
\>[.][@{}l@{}]\<[427I]%
\>[11]\AgdaSymbol{;}\AgdaSpace{}%
\AgdaSymbol{(}\AgdaBound{j}\AgdaSpace{}%
\AgdaSymbol{=}\AgdaSpace{}%
\AgdaInductiveConstructor{i0}\AgdaSymbol{)}\AgdaSpace{}%
\AgdaSymbol{→}\AgdaSpace{}%
\AgdaBound{w}\<%
\\
\>[11]\AgdaSymbol{;}\AgdaSpace{}%
\AgdaSymbol{(}\AgdaBound{j}\AgdaSpace{}%
\AgdaSymbol{=}\AgdaSpace{}%
\AgdaInductiveConstructor{i1}\AgdaSymbol{)}\AgdaSpace{}%
\AgdaSymbol{→}%
\>[443I]\AgdaPrimitive{hcomp}\AgdaSpace{}%
\AgdaSymbol{(λ}\AgdaSpace{}%
\AgdaBound{m}\AgdaSpace{}%
\AgdaSymbol{→}\<%
\\
\>[443I][@{}l@{\AgdaIndent{0}}]%
\>[26]\AgdaSymbol{λ}%
\>[447I]\AgdaSymbol{\{}\AgdaSpace{}%
\AgdaSymbol{(}\AgdaBound{i}\AgdaSpace{}%
\AgdaSymbol{=}\AgdaSpace{}%
\AgdaInductiveConstructor{i0}\AgdaSymbol{)}\AgdaSpace{}%
\AgdaSymbol{→}\AgdaSpace{}%
\AgdaBound{q}\AgdaSpace{}%
\AgdaSymbol{(}\AgdaBound{l}\AgdaSpace{}%
\AgdaOperator{\AgdaPrimitive{∧}}\AgdaSpace{}%
\AgdaBound{m}\AgdaSymbol{)}\<%
\\
\>[.][@{}l@{}]\<[447I]%
\>[28]\AgdaSymbol{;}\AgdaSpace{}%
\AgdaSymbol{(}\AgdaBound{i}\AgdaSpace{}%
\AgdaSymbol{=}\AgdaSpace{}%
\AgdaInductiveConstructor{i1}\AgdaSymbol{)}\AgdaSpace{}%
\AgdaSymbol{→}\AgdaSpace{}%
\AgdaBound{x}\<%
\\
\>[28]\AgdaSymbol{;}\AgdaSpace{}%
\AgdaSymbol{(}\AgdaBound{l}\AgdaSpace{}%
\AgdaSymbol{=}\AgdaSpace{}%
\AgdaInductiveConstructor{i0}\AgdaSymbol{)}\AgdaSpace{}%
\AgdaSymbol{→}\AgdaSpace{}%
\AgdaBound{x}\<%
\\
\>[28]\AgdaSymbol{\})}\AgdaSpace{}%
\AgdaSymbol{(}\AgdaBound{q}\AgdaSpace{}%
\AgdaInductiveConstructor{i0}\AgdaSymbol{)}\AgdaSpace{}%
\AgdaSymbol{\})}\AgdaSpace{}%
\AgdaSymbol{(}\AgdaBound{p}\AgdaSpace{}%
\AgdaBound{j}\AgdaSymbol{))}\<%
\end{code}}
\begin{code}[hide]%
\>[0]\<%
\\
\\[\AgdaEmptyExtraSkip]%
\\[\AgdaEmptyExtraSkip]%
\\[\AgdaEmptyExtraSkip]%
\\[\AgdaEmptyExtraSkip]%
\\[\AgdaEmptyExtraSkip]%
\\[\AgdaEmptyExtraSkip]%
\\[\AgdaEmptyExtraSkip]%
\\[\AgdaEmptyExtraSkip]%
\\[\AgdaEmptyExtraSkip]%
\\[\AgdaEmptyExtraSkip]%
\>[2]\AgdaFunction{∙assoc}\AgdaSpace{}%
\AgdaSymbol{:}\AgdaSpace{}%
\AgdaSymbol{(}\AgdaBound{p}\AgdaSpace{}%
\AgdaOperator{\AgdaFunction{∙}}\AgdaSpace{}%
\AgdaBound{q}\AgdaSymbol{)}\AgdaSpace{}%
\AgdaOperator{\AgdaFunction{∙}}\AgdaSpace{}%
\AgdaBound{r}\AgdaSpace{}%
\AgdaOperator{\AgdaFunction{≡}}\AgdaSpace{}%
\AgdaBound{p}\AgdaSpace{}%
\AgdaOperator{\AgdaFunction{∙}}\AgdaSpace{}%
\AgdaSymbol{(}\AgdaBound{q}\AgdaSpace{}%
\AgdaOperator{\AgdaFunction{∙}}\AgdaSpace{}%
\AgdaBound{r}\AgdaSymbol{)}\<%
\\
\>[2]\AgdaFunction{∙assoc}\AgdaSpace{}%
\AgdaSymbol{=}%
\>[14]\AgdaSymbol{λ}\AgdaSpace{}%
\AgdaBound{i}\AgdaSpace{}%
\AgdaBound{j}\AgdaSpace{}%
\AgdaSymbol{→}\AgdaSpace{}%
\AgdaPrimitive{hcomp}\AgdaSpace{}%
\AgdaSymbol{(λ}\AgdaSpace{}%
\AgdaBound{k}\AgdaSpace{}%
\AgdaSymbol{→}\AgdaSpace{}%
\AgdaSymbol{λ}\AgdaSpace{}%
\AgdaSymbol{\{}%
\>[41]\AgdaSymbol{(}\AgdaBound{i}\AgdaSpace{}%
\AgdaSymbol{=}\AgdaSpace{}%
\AgdaInductiveConstructor{i1}\AgdaSymbol{)}\AgdaSpace{}%
\AgdaSymbol{→}\AgdaSpace{}%
\AgdaFunction{hfill}\AgdaSpace{}%
\AgdaSymbol{(λ}\AgdaSpace{}%
\AgdaBound{l}\AgdaSpace{}%
\AgdaSymbol{→}\AgdaSpace{}%
\AgdaSymbol{λ}\AgdaSpace{}%
\AgdaSymbol{\{}\<%
\\
\>[2][@{}l@{\AgdaIndent{0}}]%
\>[8]\AgdaSymbol{(}\AgdaBound{j}\AgdaSpace{}%
\AgdaSymbol{=}\AgdaSpace{}%
\AgdaInductiveConstructor{i0}\AgdaSymbol{)}\AgdaSpace{}%
\AgdaSymbol{→}\AgdaSpace{}%
\AgdaBound{p}\AgdaSpace{}%
\AgdaInductiveConstructor{i0}\<%
\\
\>[2][@{}l@{\AgdaIndent{0}}]%
\>[6]\AgdaSymbol{;}%
\>[507I]\AgdaSymbol{(}\AgdaBound{j}\AgdaSpace{}%
\AgdaSymbol{=}\AgdaSpace{}%
\AgdaInductiveConstructor{i1}\AgdaSymbol{)}\AgdaSpace{}%
\AgdaSymbol{→}\AgdaSpace{}%
\AgdaPrimitive{hcomp}\AgdaSpace{}%
\AgdaSymbol{(λ}\AgdaSpace{}%
\AgdaBound{m}\AgdaSpace{}%
\AgdaSymbol{→}\AgdaSpace{}%
\AgdaSymbol{λ}\AgdaSpace{}%
\AgdaSymbol{\{}\<%
\\
\>[507I][@{}l@{\AgdaIndent{0}}]%
\>[10]\AgdaSymbol{(}\AgdaBound{l}\AgdaSpace{}%
\AgdaSymbol{=}\AgdaSpace{}%
\AgdaInductiveConstructor{i0}\AgdaSymbol{)}\AgdaSpace{}%
\AgdaSymbol{→}\AgdaSpace{}%
\AgdaBound{p}\AgdaSpace{}%
\AgdaInductiveConstructor{i1}\<%
\\
\>[.][@{}l@{}]\<[507I]%
\>[8]\AgdaSymbol{;}\AgdaSpace{}%
\AgdaSymbol{(}\AgdaBound{l}\AgdaSpace{}%
\AgdaSymbol{=}\AgdaSpace{}%
\AgdaInductiveConstructor{i1}\AgdaSymbol{)}\AgdaSpace{}%
\AgdaSymbol{→}\AgdaSpace{}%
\AgdaBound{r}\AgdaSpace{}%
\AgdaBound{m}\<%
\\
\>[8]\AgdaSymbol{\})}\AgdaSpace{}%
\AgdaSymbol{(}\AgdaBound{q}\AgdaSpace{}%
\AgdaBound{l}\AgdaSymbol{)}\<%
\\
\>[6]\AgdaSymbol{\})}\AgdaSpace{}%
\AgdaSymbol{(}\AgdaPostulate{inS}\AgdaSpace{}%
\AgdaSymbol{(}\AgdaBound{p}\AgdaSpace{}%
\AgdaBound{j}\AgdaSymbol{))}\AgdaSpace{}%
\AgdaBound{k}\<%
\\
\>[2][@{}l@{\AgdaIndent{0}}]%
\>[4]\AgdaSymbol{;}\AgdaSpace{}%
\AgdaSymbol{(}\AgdaBound{j}\AgdaSpace{}%
\AgdaSymbol{=}\AgdaSpace{}%
\AgdaInductiveConstructor{i0}\AgdaSymbol{)}\AgdaSpace{}%
\AgdaSymbol{→}\AgdaSpace{}%
\AgdaBound{p}\AgdaSpace{}%
\AgdaInductiveConstructor{i0}\<%
\\
\>[4]\AgdaSymbol{;}%
\>[540I]\AgdaSymbol{(}\AgdaBound{j}\AgdaSpace{}%
\AgdaSymbol{=}\AgdaSpace{}%
\AgdaInductiveConstructor{i1}\AgdaSymbol{)}\AgdaSpace{}%
\AgdaSymbol{→}\AgdaSpace{}%
\AgdaPrimitive{hcomp}\AgdaSpace{}%
\AgdaSymbol{(λ}\AgdaSpace{}%
\AgdaBound{l}\AgdaSpace{}%
\AgdaSymbol{→}\AgdaSpace{}%
\AgdaSymbol{λ}\AgdaSpace{}%
\AgdaSymbol{\{}\<%
\\
\>[540I][@{}l@{\AgdaIndent{0}}]%
\>[8]\AgdaSymbol{(}\AgdaBound{i}\AgdaSpace{}%
\AgdaSymbol{=}\AgdaSpace{}%
\AgdaInductiveConstructor{i1}\AgdaSymbol{)}\AgdaSpace{}%
\AgdaSymbol{→}\AgdaSpace{}%
\AgdaFunction{hfill}\AgdaSpace{}%
\AgdaSymbol{(λ}\AgdaSpace{}%
\AgdaBound{m}\AgdaSpace{}%
\AgdaSymbol{→}\AgdaSpace{}%
\AgdaSymbol{λ}\AgdaSpace{}%
\AgdaSymbol{\{}\<%
\\
\>[8][@{}l@{\AgdaIndent{0}}]%
\>[10]\AgdaSymbol{(}\AgdaBound{k}\AgdaSpace{}%
\AgdaSymbol{=}\AgdaSpace{}%
\AgdaInductiveConstructor{i0}\AgdaSymbol{)}\AgdaSpace{}%
\AgdaSymbol{→}\AgdaSpace{}%
\AgdaBound{p}\AgdaSpace{}%
\AgdaInductiveConstructor{i1}\<%
\\
\>[8]\AgdaSymbol{;}\AgdaSpace{}%
\AgdaSymbol{(}\AgdaBound{k}\AgdaSpace{}%
\AgdaSymbol{=}\AgdaSpace{}%
\AgdaInductiveConstructor{i1}\AgdaSymbol{)}\AgdaSpace{}%
\AgdaSymbol{→}\AgdaSpace{}%
\AgdaBound{r}\AgdaSpace{}%
\AgdaBound{m}\<%
\\
\>[8]\AgdaSymbol{\})}\AgdaSpace{}%
\AgdaSymbol{(}\AgdaPostulate{inS}\AgdaSpace{}%
\AgdaSymbol{(}\AgdaBound{q}\AgdaSpace{}%
\AgdaBound{k}\AgdaSymbol{))}\AgdaSpace{}%
\AgdaBound{l}\<%
\\
\>[.][@{}l@{}]\<[540I]%
\>[6]\AgdaSymbol{;}\AgdaSpace{}%
\AgdaSymbol{(}\AgdaBound{k}\AgdaSpace{}%
\AgdaSymbol{=}\AgdaSpace{}%
\AgdaInductiveConstructor{i0}\AgdaSymbol{)}\AgdaSpace{}%
\AgdaSymbol{→}\AgdaSpace{}%
\AgdaBound{p}\AgdaSpace{}%
\AgdaInductiveConstructor{i1}\<%
\\
\>[6]\AgdaSymbol{;}\AgdaSpace{}%
\AgdaSymbol{(}\AgdaBound{k}\AgdaSpace{}%
\AgdaSymbol{=}\AgdaSpace{}%
\AgdaInductiveConstructor{i1}\AgdaSymbol{)}\AgdaSpace{}%
\AgdaSymbol{→}\AgdaSpace{}%
\AgdaBound{r}\AgdaSpace{}%
\AgdaBound{l}\<%
\\
\>[6]\AgdaSymbol{;}%
\>[586I]\AgdaSymbol{(}\AgdaBound{i}\AgdaSpace{}%
\AgdaSymbol{=}\AgdaSpace{}%
\AgdaInductiveConstructor{i0}\AgdaSymbol{)}\AgdaSpace{}%
\AgdaSymbol{→}\AgdaSpace{}%
\AgdaFunction{hfill}\AgdaSpace{}%
\AgdaSymbol{(λ}\AgdaSpace{}%
\AgdaBound{m}\AgdaSpace{}%
\AgdaSymbol{→}\AgdaSpace{}%
\AgdaSymbol{λ}\AgdaSpace{}%
\AgdaSymbol{\{}\<%
\\
\>[586I][@{}l@{\AgdaIndent{0}}]%
\>[10]\AgdaSymbol{(}\AgdaBound{k}\AgdaSpace{}%
\AgdaSymbol{=}\AgdaSpace{}%
\AgdaInductiveConstructor{i0}\AgdaSymbol{)}\AgdaSpace{}%
\AgdaSymbol{→}\AgdaSpace{}%
\AgdaBound{p}\AgdaSpace{}%
\AgdaInductiveConstructor{i1}\<%
\\
\>[.][@{}l@{}]\<[586I]%
\>[8]\AgdaSymbol{;}\AgdaSpace{}%
\AgdaSymbol{(}\AgdaBound{k}\AgdaSpace{}%
\AgdaSymbol{=}\AgdaSpace{}%
\AgdaInductiveConstructor{i1}\AgdaSymbol{)}\AgdaSpace{}%
\AgdaSymbol{→}\AgdaSpace{}%
\AgdaBound{r}\AgdaSpace{}%
\AgdaBound{m}\<%
\\
\>[8]\AgdaSymbol{\})}\AgdaSpace{}%
\AgdaSymbol{(}\AgdaPostulate{inS}\AgdaSpace{}%
\AgdaSymbol{(}\AgdaBound{q}\AgdaSpace{}%
\AgdaBound{k}\AgdaSymbol{))}\AgdaSpace{}%
\AgdaBound{l}\<%
\\
\>[6]\AgdaSymbol{\})}\AgdaSpace{}%
\AgdaSymbol{(}\AgdaBound{q}\AgdaSpace{}%
\AgdaBound{k}\AgdaSymbol{)}\<%
\\
\>[4]\AgdaSymbol{;}%
\>[613I]\AgdaSymbol{(}\AgdaBound{i}\AgdaSpace{}%
\AgdaSymbol{=}\AgdaSpace{}%
\AgdaInductiveConstructor{i0}\AgdaSymbol{)}\AgdaSpace{}%
\AgdaSymbol{→}\AgdaSpace{}%
\AgdaPrimitive{hcomp}\AgdaSpace{}%
\AgdaSymbol{(λ}\AgdaSpace{}%
\AgdaBound{l}\AgdaSpace{}%
\AgdaSymbol{→}\AgdaSpace{}%
\AgdaSymbol{λ}\AgdaSpace{}%
\AgdaSymbol{\{}\<%
\\
\>[613I][@{}l@{\AgdaIndent{0}}]%
\>[8]\AgdaSymbol{(}\AgdaBound{j}\AgdaSpace{}%
\AgdaSymbol{=}\AgdaSpace{}%
\AgdaInductiveConstructor{i0}\AgdaSymbol{)}\AgdaSpace{}%
\AgdaSymbol{→}\AgdaSpace{}%
\AgdaBound{p}\AgdaSpace{}%
\AgdaInductiveConstructor{i0}\<%
\\
\>[.][@{}l@{}]\<[613I]%
\>[6]\AgdaSymbol{;}%
\>[628I]\AgdaSymbol{(}\AgdaBound{j}\AgdaSpace{}%
\AgdaSymbol{=}\AgdaSpace{}%
\AgdaInductiveConstructor{i1}\AgdaSymbol{)}\AgdaSpace{}%
\AgdaSymbol{→}\AgdaSpace{}%
\AgdaFunction{hfill}\AgdaSpace{}%
\AgdaSymbol{(λ}\AgdaSpace{}%
\AgdaBound{m}\AgdaSpace{}%
\AgdaSymbol{→}\AgdaSpace{}%
\AgdaSymbol{λ}\AgdaSpace{}%
\AgdaSymbol{\{}\<%
\\
\>[628I][@{}l@{\AgdaIndent{0}}]%
\>[10]\AgdaSymbol{(}\AgdaBound{k}\AgdaSpace{}%
\AgdaSymbol{=}\AgdaSpace{}%
\AgdaInductiveConstructor{i0}\AgdaSymbol{)}\AgdaSpace{}%
\AgdaSymbol{→}\AgdaSpace{}%
\AgdaBound{p}\AgdaSpace{}%
\AgdaInductiveConstructor{i1}\<%
\\
\>[.][@{}l@{}]\<[628I]%
\>[8]\AgdaSymbol{;}\AgdaSpace{}%
\AgdaSymbol{(}\AgdaBound{k}\AgdaSpace{}%
\AgdaSymbol{=}\AgdaSpace{}%
\AgdaInductiveConstructor{i1}\AgdaSymbol{)}\AgdaSpace{}%
\AgdaSymbol{→}\AgdaSpace{}%
\AgdaBound{r}\AgdaSpace{}%
\AgdaBound{m}\<%
\\
\>[8]\AgdaSymbol{\})}\AgdaSpace{}%
\AgdaSymbol{(}\AgdaPostulate{inS}\AgdaSpace{}%
\AgdaSymbol{(}\AgdaBound{q}\AgdaSpace{}%
\AgdaBound{k}\AgdaSymbol{))}\AgdaSpace{}%
\AgdaBound{l}\<%
\\
\>[6]\AgdaSymbol{;}\AgdaSpace{}%
\AgdaSymbol{(}\AgdaBound{k}\AgdaSpace{}%
\AgdaSymbol{=}\AgdaSpace{}%
\AgdaInductiveConstructor{i0}\AgdaSymbol{)}\AgdaSpace{}%
\AgdaSymbol{→}\AgdaSpace{}%
\AgdaBound{p}\AgdaSpace{}%
\AgdaBound{j}\<%
\\
\>[6]\AgdaSymbol{;}%
\>[659I]\AgdaSymbol{(}\AgdaBound{k}\AgdaSpace{}%
\AgdaSymbol{=}\AgdaSpace{}%
\AgdaInductiveConstructor{i1}\AgdaSymbol{)}\AgdaSpace{}%
\AgdaSymbol{→}\AgdaSpace{}%
\AgdaFunction{hfill}\AgdaSpace{}%
\AgdaSymbol{(λ}\AgdaSpace{}%
\AgdaBound{m}\AgdaSpace{}%
\AgdaSymbol{→}\AgdaSpace{}%
\AgdaSymbol{λ}\AgdaSpace{}%
\AgdaSymbol{\{}\<%
\\
\>[659I][@{}l@{\AgdaIndent{0}}]%
\>[10]\AgdaSymbol{(}\AgdaBound{j}\AgdaSpace{}%
\AgdaSymbol{=}\AgdaSpace{}%
\AgdaInductiveConstructor{i0}\AgdaSymbol{)}\AgdaSpace{}%
\AgdaSymbol{→}\AgdaSpace{}%
\AgdaBound{p}\AgdaSpace{}%
\AgdaInductiveConstructor{i0}\<%
\\
\>[.][@{}l@{}]\<[659I]%
\>[8]\AgdaSymbol{;}\AgdaSpace{}%
\AgdaSymbol{(}\AgdaBound{j}\AgdaSpace{}%
\AgdaSymbol{=}\AgdaSpace{}%
\AgdaInductiveConstructor{i1}\AgdaSymbol{)}\AgdaSpace{}%
\AgdaSymbol{→}\AgdaSpace{}%
\AgdaBound{r}\AgdaSpace{}%
\AgdaBound{m}\<%
\\
\>[8]\AgdaSymbol{\})}\AgdaSpace{}%
\AgdaSymbol{(}\AgdaPostulate{inS}\AgdaSpace{}%
\AgdaSymbol{(}\AgdaPrimitive{hcomp}\AgdaSpace{}%
\AgdaSymbol{(λ}\AgdaSpace{}%
\AgdaBound{m}\AgdaSpace{}%
\AgdaSymbol{→}\AgdaSpace{}%
\AgdaSymbol{λ}\AgdaSpace{}%
\AgdaSymbol{\{}\<%
\\
\>[8][@{}l@{\AgdaIndent{0}}]%
\>[10]\AgdaSymbol{(}\AgdaBound{j}\AgdaSpace{}%
\AgdaSymbol{=}\AgdaSpace{}%
\AgdaInductiveConstructor{i0}\AgdaSymbol{)}\AgdaSpace{}%
\AgdaSymbol{→}\AgdaSpace{}%
\AgdaBound{p}\AgdaSpace{}%
\AgdaInductiveConstructor{i0}\<%
\\
\>[8]\AgdaSymbol{;}\AgdaSpace{}%
\AgdaSymbol{(}\AgdaBound{j}\AgdaSpace{}%
\AgdaSymbol{=}\AgdaSpace{}%
\AgdaInductiveConstructor{i1}\AgdaSymbol{)}\AgdaSpace{}%
\AgdaSymbol{→}\AgdaSpace{}%
\AgdaBound{q}\AgdaSpace{}%
\AgdaBound{m}\<%
\\
\>[8]\AgdaSymbol{\})}\AgdaSpace{}%
\AgdaSymbol{(}\AgdaBound{p}\AgdaSpace{}%
\AgdaBound{j}\AgdaSymbol{)))}\AgdaSpace{}%
\AgdaBound{l}\<%
\\
\>[6]\AgdaSymbol{\})}\AgdaSpace{}%
\AgdaSymbol{(}\AgdaFunction{hfill}\AgdaSpace{}%
\AgdaSymbol{(λ}\AgdaSpace{}%
\AgdaBound{l}\AgdaSpace{}%
\AgdaSymbol{→}\AgdaSpace{}%
\AgdaSymbol{λ}\AgdaSpace{}%
\AgdaSymbol{\{}\<%
\\
\>[6][@{}l@{\AgdaIndent{0}}]%
\>[8]\AgdaSymbol{(}\AgdaBound{j}\AgdaSpace{}%
\AgdaSymbol{=}\AgdaSpace{}%
\AgdaInductiveConstructor{i0}\AgdaSymbol{)}\AgdaSpace{}%
\AgdaSymbol{→}\AgdaSpace{}%
\AgdaBound{p}\AgdaSpace{}%
\AgdaInductiveConstructor{i0}\<%
\\
\>[6]\AgdaSymbol{;}\AgdaSpace{}%
\AgdaSymbol{(}\AgdaBound{j}\AgdaSpace{}%
\AgdaSymbol{=}\AgdaSpace{}%
\AgdaInductiveConstructor{i1}\AgdaSymbol{)}\AgdaSpace{}%
\AgdaSymbol{→}\AgdaSpace{}%
\AgdaBound{q}\AgdaSpace{}%
\AgdaBound{l}\<%
\\
\>[6]\AgdaSymbol{\})}\AgdaSpace{}%
\AgdaSymbol{(}\AgdaPostulate{inS}\AgdaSpace{}%
\AgdaSymbol{(}\AgdaBound{p}\AgdaSpace{}%
\AgdaBound{j}\AgdaSymbol{))}\AgdaSpace{}%
\AgdaBound{k}\AgdaSymbol{)}\<%
\\
\>[4]\AgdaSymbol{\})}\AgdaSpace{}%
\AgdaSymbol{(}\AgdaBound{p}\AgdaSpace{}%
\AgdaBound{j}\AgdaSymbol{)}\<%
\\
\\[\AgdaEmptyExtraSkip]%
\\[\AgdaEmptyExtraSkip]%
\\[\AgdaEmptyExtraSkip]%
\\[\AgdaEmptyExtraSkip]%
\\[\AgdaEmptyExtraSkip]%
\\[\AgdaEmptyExtraSkip]%
\\[\AgdaEmptyExtraSkip]%
\\[\AgdaEmptyExtraSkip]%
\\[\AgdaEmptyExtraSkip]%
\\[\AgdaEmptyExtraSkip]%
\\[\AgdaEmptyExtraSkip]%
\\[\AgdaEmptyExtraSkip]%
\\[\AgdaEmptyExtraSkip]%
\\[\AgdaEmptyExtraSkip]%
\\[\AgdaEmptyExtraSkip]%
\\[\AgdaEmptyExtraSkip]%
\\[\AgdaEmptyExtraSkip]%
\\[\AgdaEmptyExtraSkip]%
\\[\AgdaEmptyExtraSkip]%
\\[\AgdaEmptyExtraSkip]%
\\[\AgdaEmptyExtraSkip]%
\\[\AgdaEmptyExtraSkip]%
\\[\AgdaEmptyExtraSkip]%
\\[\AgdaEmptyExtraSkip]%
\\[\AgdaEmptyExtraSkip]%
\\[\AgdaEmptyExtraSkip]%
\\[\AgdaEmptyExtraSkip]%
\\[\AgdaEmptyExtraSkip]%
\\[\AgdaEmptyExtraSkip]%
\\[\AgdaEmptyExtraSkip]%
\\[\AgdaEmptyExtraSkip]%
\\[\AgdaEmptyExtraSkip]%
\\[\AgdaEmptyExtraSkip]%
\>[0]\AgdaComment{--\ Only\ have\ them\ at\ level\ zero\ to\ suppress\ levels\ in\ code\ for\ the\ paper}\<%
\\
\>[0]\AgdaFunction{Pointed}\AgdaSpace{}%
\AgdaSymbol{:}\AgdaSpace{}%
\AgdaPrimitive{Type}\AgdaSpace{}%
\AgdaSymbol{\AgdaUnderscore{}}\<%
\\
\>[0]\AgdaFunction{Pointed}\AgdaSpace{}%
\AgdaSymbol{=}\AgdaSpace{}%
\AgdaRecord{Σ}\AgdaSpace{}%
\AgdaSymbol{(}\AgdaPrimitive{Type}\AgdaSpace{}%
\AgdaPrimitive{ℓ-zero}\AgdaSymbol{)}\AgdaSpace{}%
\AgdaSymbol{λ}\AgdaSpace{}%
\AgdaBound{A}\AgdaSpace{}%
\AgdaSymbol{→}\AgdaSpace{}%
\AgdaBound{A}\<%
\\
\\[\AgdaEmptyExtraSkip]%
\>[0]\<%
\end{code}
\newcommand{\agdaloopspace}{%
\begin{code}%
\>[0]\AgdaFunction{Ω}\AgdaSpace{}%
\AgdaSymbol{:}\AgdaSpace{}%
\AgdaFunction{Pointed}\AgdaSpace{}%
\AgdaSymbol{→}\AgdaSpace{}%
\AgdaFunction{Pointed}\<%
\\
\>[0]\AgdaFunction{Ω}\AgdaSpace{}%
\AgdaSymbol{(\AgdaUnderscore{}}\AgdaSpace{}%
\AgdaOperator{\AgdaInductiveConstructor{,}}\AgdaSpace{}%
\AgdaBound{a}\AgdaSymbol{)}\AgdaSpace{}%
\AgdaSymbol{=}\AgdaSpace{}%
\AgdaSymbol{((}\AgdaBound{a}\AgdaSpace{}%
\AgdaOperator{\AgdaFunction{≡}}\AgdaSpace{}%
\AgdaBound{a}\AgdaSymbol{)}\AgdaSpace{}%
\AgdaOperator{\AgdaInductiveConstructor{,}}\AgdaSpace{}%
\AgdaFunction{refl}\AgdaSymbol{)}\<%
\end{code}}
\begin{code}[hide]%
\>[0]\<%
\\
\>[0]\AgdaKeyword{module}\AgdaSpace{}%
\AgdaModule{\AgdaUnderscore{}}\AgdaSpace{}%
\AgdaSymbol{\{}\AgdaBound{A}\AgdaSpace{}%
\AgdaSymbol{:}\AgdaSpace{}%
\AgdaPrimitive{Type}\AgdaSpace{}%
\AgdaGeneralizable{ℓ}\AgdaSymbol{\}}\AgdaSpace{}%
\AgdaSymbol{\{}\AgdaBound{x}\AgdaSpace{}%
\AgdaSymbol{:}\AgdaSpace{}%
\AgdaBound{A}\AgdaSymbol{\}}\AgdaSpace{}%
\AgdaKeyword{where}\<%
\\
\\[\AgdaEmptyExtraSkip]%
\>[0]\AgdaComment{--\ Old\ version,\ a\ bit\ more\ general\ than\ then\ one\ in\ the\ paper}\<%
\\
\>[0]\AgdaComment{--\ (could\ be\ further\ generalised\ by\ varying\ x)}\<%
\\
\>[0][@{}l@{\AgdaIndent{0}}]%
\>[2]\AgdaComment{--\ Sq→Comp\ :\ (p₁\ p₂\ p₃\ p₄\ :\ x\ ≡\ x)\ →\ PathP\ (λ\ j\ →\ p₃\ j\ ≡\ p₂\ j)\ p₁\ p₄}\<%
\\
\>[2]\AgdaComment{--\ \ \ →\ p₁\ ∙\ p₂\ ≡\ p₃\ ∙\ p₄}\<%
\\
\>[2]\AgdaComment{--\ Sq→Comp\ p₁\ p₂\ p₃\ p₄\ α\ i\ j\ =}\<%
\\
\>[2]\AgdaComment{--\ \ \ \ \ hcomp\ (λ\ k\ →\ λ\ \{}\<%
\\
\>[2]\AgdaComment{--\ \ \ \ \ \ \ \ (j\ =\ i0)\ →\ x}\<%
\\
\>[2]\AgdaComment{--\ \ \ \ \ \ ;\ (j\ =\ i1)\ →\ hcomp\ (λ\ l\ →\ λ\ \{}\<%
\\
\>[2]\AgdaComment{--\ \ \ \ \ \ \ \ \ \ \ \ \ \ \ \ \ \ \ \ \ \ (i\ =\ i0)\ →\ α\ (k\ ∧\ l)\ l}\<%
\\
\>[2]\AgdaComment{--\ \ \ \ \ \ \ \ \ \ \ \ \ \ \ \ \ \ \ \ ;\ (i\ =\ i1)\ →\ α\ l\ (k\ ∧\ l)}\<%
\\
\>[2]\AgdaComment{--\ \ \ \ \ \ \ \ \ \ \ \ \ \ \ \ \ \ \ \ ;\ (k\ =\ i1)\ →\ α\ l\ l}\<%
\\
\>[2]\AgdaComment{--\ \ \ \ \ \ \ \ \ \ \ \ \ \ \ \ \ \ \ \ \})\ x}\<%
\\
\>[2]\AgdaComment{--\ \ \ \ \ \ \})\ (hcomp\ (λ\ m\ →\ λ\ \{}\<%
\\
\>[2]\AgdaComment{--\ \ \ \ \ \ \ \ \ \ \ \ \ (i\ =\ i0)\ →\ p₁\ (m\ ∧\ j)}\<%
\\
\>[2]\AgdaComment{--\ \ \ \ \ \ \ \ \ \ \ ;\ (i\ =\ i1)\ →\ p₃\ (m\ ∧\ j)}\<%
\\
\>[2]\AgdaComment{--\ \ \ \ \ \ \ \ \ \ \ ;\ (j\ =\ i0)\ →\ x}\<%
\\
\>[2]\AgdaComment{--\ \ \ \ \ \ \ \ \ \ \ \})\ x)}\<%
\\
\\[\AgdaEmptyExtraSkip]%
\\[\AgdaEmptyExtraSkip]%
\>[2]\AgdaKeyword{module}\AgdaSpace{}%
\AgdaModule{\AgdaUnderscore{}}\AgdaSpace{}%
\AgdaSymbol{(}\AgdaBound{p}\AgdaSpace{}%
\AgdaBound{q}\AgdaSpace{}%
\AgdaSymbol{:}\AgdaSpace{}%
\AgdaBound{x}\AgdaSpace{}%
\AgdaOperator{\AgdaFunction{≡}}\AgdaSpace{}%
\AgdaBound{x}\AgdaSymbol{)}\AgdaSpace{}%
\AgdaKeyword{where}\<%
\end{code}
\newcommand{\agdasqtocomp}{%
\begin{code}%
\>[2][@{}l@{\AgdaIndent{1}}]%
\>[4]\AgdaFunction{Sq→Comp}\AgdaSpace{}%
\AgdaSymbol{:}\AgdaSpace{}%
\AgdaPostulate{PathP}\AgdaSpace{}%
\AgdaSymbol{(λ}\AgdaSpace{}%
\AgdaBound{j}\AgdaSpace{}%
\AgdaSymbol{→}\AgdaSpace{}%
\AgdaBound{q}\AgdaSpace{}%
\AgdaBound{j}\AgdaSpace{}%
\AgdaOperator{\AgdaFunction{≡}}\AgdaSpace{}%
\AgdaBound{q}\AgdaSpace{}%
\AgdaBound{j}\AgdaSymbol{)}\AgdaSpace{}%
\AgdaBound{p}\AgdaSpace{}%
\AgdaBound{p}\AgdaSpace{}%
\AgdaSymbol{→}\AgdaSpace{}%
\AgdaBound{p}\AgdaSpace{}%
\AgdaOperator{\AgdaFunction{∙}}\AgdaSpace{}%
\AgdaBound{q}\AgdaSpace{}%
\AgdaOperator{\AgdaFunction{≡}}\AgdaSpace{}%
\AgdaBound{q}\AgdaSpace{}%
\AgdaOperator{\AgdaFunction{∙}}\AgdaSpace{}%
\AgdaBound{p}\<%
\\
\>[4]\AgdaFunction{Sq→Comp}\AgdaSpace{}%
\AgdaBound{α}\AgdaSpace{}%
\AgdaBound{i}\AgdaSpace{}%
\AgdaBound{j}\AgdaSpace{}%
\AgdaSymbol{=}\AgdaSpace{}%
\AgdaPrimitive{hcomp}\AgdaSpace{}%
\AgdaSymbol{(λ}\AgdaSpace{}%
\AgdaBound{k}\AgdaSpace{}%
\AgdaSymbol{→}\AgdaSpace{}%
\AgdaSymbol{λ}\AgdaSpace{}%
\AgdaSymbol{\{}\<%
\\
\>[4][@{}l@{\AgdaIndent{0}}]%
\>[6]\AgdaSymbol{(}\AgdaBound{i}\AgdaSpace{}%
\AgdaSymbol{=}\AgdaSpace{}%
\AgdaInductiveConstructor{i0}\AgdaSymbol{)}\AgdaSpace{}%
\AgdaSymbol{→}\AgdaSpace{}%
\AgdaPrimitive{hcomp}\AgdaSpace{}%
\AgdaSymbol{(λ}\AgdaSpace{}%
\AgdaBound{l}\AgdaSpace{}%
\AgdaSymbol{→}\AgdaSpace{}%
\AgdaSymbol{λ}\AgdaSpace{}%
\AgdaSymbol{\{}\<%
\\
\>[6][@{}l@{\AgdaIndent{0}}]%
\>[8]\AgdaSymbol{(}\AgdaBound{j}\AgdaSpace{}%
\AgdaSymbol{=}\AgdaSpace{}%
\AgdaInductiveConstructor{i0}\AgdaSymbol{)}\AgdaSpace{}%
\AgdaSymbol{→}\AgdaSpace{}%
\AgdaBound{x}\AgdaSpace{}%
\AgdaSymbol{;}\AgdaSpace{}%
\AgdaSymbol{(}\AgdaBound{k}\AgdaSpace{}%
\AgdaSymbol{=}\AgdaSpace{}%
\AgdaInductiveConstructor{i0}\AgdaSymbol{)}\AgdaSpace{}%
\AgdaSymbol{→}\AgdaSpace{}%
\AgdaBound{q}\AgdaSpace{}%
\AgdaSymbol{(}\AgdaBound{j}\AgdaSpace{}%
\AgdaOperator{\AgdaPrimitive{∧}}\AgdaSpace{}%
\AgdaBound{l}\AgdaSymbol{)}\AgdaSpace{}%
\AgdaSymbol{;}\AgdaSpace{}%
\AgdaSymbol{(}\AgdaBound{j}\AgdaSpace{}%
\AgdaSymbol{=}\AgdaSpace{}%
\AgdaInductiveConstructor{i1}\AgdaSymbol{)}\AgdaSpace{}%
\AgdaSymbol{→}\AgdaSpace{}%
\AgdaBound{α}\AgdaSpace{}%
\AgdaBound{l}\AgdaSpace{}%
\AgdaBound{k}\AgdaSpace{}%
\AgdaSymbol{;}\<%
\\
\>[8]\AgdaSymbol{(}\AgdaBound{k}\AgdaSpace{}%
\AgdaSymbol{=}\AgdaSpace{}%
\AgdaInductiveConstructor{i1}\AgdaSymbol{)}\AgdaSpace{}%
\AgdaSymbol{→}\AgdaSpace{}%
\AgdaFunction{hfill}\AgdaSpace{}%
\AgdaSymbol{(λ}\AgdaSpace{}%
\AgdaBound{m}\AgdaSpace{}%
\AgdaSymbol{→}\AgdaSpace{}%
\AgdaSymbol{λ}\AgdaSpace{}%
\AgdaSymbol{\{}\AgdaSpace{}%
\AgdaSymbol{(}\AgdaBound{j}\AgdaSpace{}%
\AgdaSymbol{=}\AgdaSpace{}%
\AgdaInductiveConstructor{i0}\AgdaSymbol{)}\AgdaSpace{}%
\AgdaSymbol{→}\AgdaSpace{}%
\AgdaBound{x}\AgdaSpace{}%
\AgdaSymbol{;}\AgdaSpace{}%
\AgdaSymbol{(}\AgdaBound{j}\AgdaSpace{}%
\AgdaSymbol{=}\AgdaSpace{}%
\AgdaInductiveConstructor{i1}\AgdaSymbol{)}\AgdaSpace{}%
\AgdaSymbol{→}\AgdaSpace{}%
\AgdaBound{q}\AgdaSpace{}%
\AgdaBound{m}\AgdaSpace{}%
\AgdaSymbol{\})}\AgdaSpace{}%
\AgdaSymbol{(}\AgdaPostulate{inS}\AgdaSpace{}%
\AgdaSymbol{(}\AgdaBound{p}\AgdaSpace{}%
\AgdaBound{j}\AgdaSymbol{))}\AgdaSpace{}%
\AgdaBound{l}\AgdaSpace{}%
\AgdaSymbol{\})}\<%
\\
\>[8]\AgdaSymbol{(}\AgdaBound{p}\AgdaSpace{}%
\AgdaSymbol{(}\AgdaBound{j}\AgdaSpace{}%
\AgdaOperator{\AgdaPrimitive{∧}}\AgdaSpace{}%
\AgdaBound{k}\AgdaSymbol{))}\AgdaSpace{}%
\AgdaSymbol{;}\<%
\\
\>[6]\AgdaSymbol{(}\AgdaBound{i}\AgdaSpace{}%
\AgdaSymbol{=}\AgdaSpace{}%
\AgdaInductiveConstructor{i1}\AgdaSymbol{)}\AgdaSpace{}%
\AgdaSymbol{→}\AgdaSpace{}%
\AgdaFunction{hfill}\AgdaSpace{}%
\AgdaSymbol{(λ}\AgdaSpace{}%
\AgdaBound{l}\AgdaSpace{}%
\AgdaSymbol{→}\AgdaSpace{}%
\AgdaSymbol{λ}\AgdaSpace{}%
\AgdaSymbol{\{}\AgdaSpace{}%
\AgdaSymbol{(}\AgdaBound{j}\AgdaSpace{}%
\AgdaSymbol{=}\AgdaSpace{}%
\AgdaInductiveConstructor{i0}\AgdaSymbol{)}\AgdaSpace{}%
\AgdaSymbol{→}\AgdaSpace{}%
\AgdaBound{x}\AgdaSpace{}%
\AgdaSymbol{;}\AgdaSpace{}%
\AgdaSymbol{(}\AgdaBound{j}\AgdaSpace{}%
\AgdaSymbol{=}\AgdaSpace{}%
\AgdaInductiveConstructor{i1}\AgdaSymbol{)}\AgdaSpace{}%
\AgdaSymbol{→}\AgdaSpace{}%
\AgdaBound{p}\AgdaSpace{}%
\AgdaBound{l}\AgdaSpace{}%
\AgdaSymbol{\})}\AgdaSpace{}%
\AgdaSymbol{(}\AgdaPostulate{inS}\AgdaSpace{}%
\AgdaSymbol{(}\AgdaBound{q}\AgdaSpace{}%
\AgdaBound{j}\AgdaSymbol{))}\AgdaSpace{}%
\AgdaBound{k}\AgdaSpace{}%
\AgdaSymbol{;}\<%
\\
\>[6]\AgdaSymbol{(}\AgdaBound{j}\AgdaSpace{}%
\AgdaSymbol{=}\AgdaSpace{}%
\AgdaInductiveConstructor{i0}\AgdaSymbol{)}\AgdaSpace{}%
\AgdaSymbol{→}\AgdaSpace{}%
\AgdaBound{x}\AgdaSpace{}%
\AgdaSymbol{;}\AgdaSpace{}%
\AgdaSymbol{(}\AgdaBound{j}\AgdaSpace{}%
\AgdaSymbol{=}\AgdaSpace{}%
\AgdaInductiveConstructor{i1}\AgdaSymbol{)}\AgdaSpace{}%
\AgdaSymbol{→}\AgdaSpace{}%
\AgdaBound{p}\AgdaSpace{}%
\AgdaBound{k}\AgdaSpace{}%
\AgdaSymbol{\})}\<%
\\
\>[6]\AgdaSymbol{(}\AgdaBound{q}\AgdaSpace{}%
\AgdaBound{j}\AgdaSymbol{)}\<%
\end{code}}
\begin{code}[hide]%
\>[0]\AgdaComment{--\ more\ whitespace}\<%
\\
\>[0][@{}l@{\AgdaIndent{0}}]%
\>[4]\AgdaComment{--\ Sq→Comp\ α\ i\ j\ =\ hcomp\ (λ\ k\ →\ λ\ \{\ (i\ =\ i0)\ →\ hcomp\ (λ\ l\ →\ λ\ \{\ (j\ =\ i0)\ →\ x}\<%
\\
\>[4]\AgdaComment{--\ \ \ \ \ \ \ \ \ \ \ \ \ \ \ \ \ \ \ \ \ \ \ \ \ \ \ \ \ \ \ \ \ \ \ \ \ \ \ \ \ \ \ \ \ \ \ \ \ \ \ \ \ \ \ \ \ \ \ ;\ (j\ =\ i1)\ →\ α\ l\ k}\<%
\\
\>[4]\AgdaComment{--\ \ \ \ \ \ \ \ \ \ \ \ \ \ \ \ \ \ \ \ \ \ \ \ \ \ \ \ \ \ \ \ \ \ \ \ \ \ \ \ \ \ \ \ \ \ \ \ \ \ \ \ \ \ \ \ \ \ \ ;\ (k\ =\ i0)\ →\ q\ (j\ ∧\ l)}\<%
\\
\>[4]\AgdaComment{--\ \ \ \ \ \ \ \ \ \ \ \ \ \ \ \ \ \ \ \ \ \ \ \ \ \ \ \ \ \ \ \ \ \ \ \ \ \ \ \ \ \ \ \ \ \ \ \ \ \ \ \ \ \ \ \})\ (p\ (j\ ∧\ k))}\<%
\\
\>[4]\AgdaComment{--\ \ \ \ \ \ \ \ \ \ \ \ \ \ \ \ \ \ \ \ \ \ \ \ \ \ \ \ \ \ \ ;\ (j\ =\ i0)\ →\ x}\<%
\\
\>[4]\AgdaComment{--\ \ \ \ \ \ \ \ \ \ \ \ \ \ \ \ \ \ \ \ \ \ \ \ \ \ \ \ \ \ \ ;\ (j\ =\ i1)\ →\ p\ k}\<%
\\
\>[4]\AgdaComment{--\ \ \ \ \ \ \ \ \ \ \ \ \ \ \ \ \ \ \ \ \ \ \ \ \ \ \ \ \ \ \ \})\ (q\ j)}\<%
\\
\\[\AgdaEmptyExtraSkip]%
\>[0]\AgdaKeyword{module}\AgdaSpace{}%
\AgdaModule{\AgdaUnderscore{}}\AgdaSpace{}%
\AgdaSymbol{\{}\AgdaBound{(}\AgdaBound{A}\AgdaSpace{}%
\AgdaOperator{\AgdaInductiveConstructor{,}}\AgdaSpace{}%
\AgdaBound{x}\AgdaBound{)}\AgdaSpace{}%
\AgdaSymbol{:}\AgdaSpace{}%
\AgdaFunction{Pointed}\AgdaSymbol{\}}\AgdaSpace{}%
\AgdaSymbol{(}\AgdaBound{p}\AgdaSpace{}%
\AgdaBound{q}\AgdaSpace{}%
\AgdaSymbol{:}\AgdaSpace{}%
\AgdaField{fst}\AgdaSpace{}%
\AgdaSymbol{(}\AgdaFunction{Ω}\AgdaSpace{}%
\AgdaSymbol{(}\AgdaFunction{Ω}\AgdaSpace{}%
\AgdaSymbol{(}\AgdaBound{A}\AgdaSpace{}%
\AgdaOperator{\AgdaInductiveConstructor{,}}\AgdaSpace{}%
\AgdaBound{x}\AgdaSymbol{))))}%
\>[58]\AgdaKeyword{where}\<%
\\
\\[\AgdaEmptyExtraSkip]%
\>[0][@{}l@{\AgdaIndent{0}}]%
\>[2]\AgdaComment{--\ EckmannHilton-Cube\ :\ PathP\ (λ\ i\ →\ q\ i\ ≡\ q\ i)\ p\ p}\<%
\\
\>[2]\AgdaKeyword{postulate}\<%
\\
\>[2][@{}l@{\AgdaIndent{0}}]%
\>[4]\AgdaPostulate{p'}\AgdaSpace{}%
\AgdaSymbol{:}\AgdaSpace{}%
\AgdaPostulate{PathP}\AgdaSpace{}%
\AgdaSymbol{(λ}\AgdaSpace{}%
\AgdaBound{j}\AgdaSpace{}%
\AgdaSymbol{→}\AgdaSpace{}%
\AgdaBound{x}\AgdaSpace{}%
\AgdaOperator{\AgdaFunction{≡}}\AgdaSpace{}%
\AgdaBound{x}\AgdaSymbol{)}\AgdaSpace{}%
\AgdaSymbol{(λ}\AgdaSpace{}%
\AgdaBound{i}\AgdaSpace{}%
\AgdaSymbol{→}\AgdaSpace{}%
\AgdaBound{x}\AgdaSymbol{)}\AgdaSpace{}%
\AgdaSymbol{(λ}\AgdaSpace{}%
\AgdaBound{i}\AgdaSpace{}%
\AgdaSymbol{→}\AgdaSpace{}%
\AgdaBound{x}\AgdaSymbol{)}\<%
\\
\\[\AgdaEmptyExtraSkip]%
\>[0]\AgdaComment{--\ Old\ type:\ EckmannHilton-Cube\ :\ PathP\ (λ\ i\ →\ PathP\ (λ\ j\ →\ a\ ≡\ a)\ (q\ i)\ (q\ i))\ p\ p}\<%
\\
\>[0]\AgdaComment{--\ Anders\ changed\ tot\ he\ one\ below\ as\ it\ is\ easier\ to\ understand\ and\ explain?}\<%
\end{code}
\newcommand{\agdaeckmannhiltoncube}{%
\begin{code}%
\>[0][@{}l@{\AgdaIndent{1}}]%
\>[2]\AgdaFunction{EckmannHilton-Cube}\AgdaSpace{}%
\AgdaSymbol{:}\AgdaSpace{}%
\AgdaPostulate{PathP}\AgdaSpace{}%
\AgdaSymbol{(λ}\AgdaSpace{}%
\AgdaBound{i}\AgdaSpace{}%
\AgdaSymbol{→}\AgdaSpace{}%
\AgdaBound{q}\AgdaSpace{}%
\AgdaBound{i}\AgdaSpace{}%
\AgdaOperator{\AgdaFunction{≡}}\AgdaSpace{}%
\AgdaBound{q}\AgdaSpace{}%
\AgdaBound{i}\AgdaSymbol{)}\AgdaSpace{}%
\AgdaBound{p}\AgdaSpace{}%
\AgdaBound{p}\<%
\\
\>[2]\AgdaFunction{EckmannHilton-Cube}\AgdaSpace{}%
\AgdaSymbol{=}\AgdaSpace{}%
\AgdaSymbol{λ}\AgdaSpace{}%
\AgdaBound{i}\AgdaSpace{}%
\AgdaBound{j}\AgdaSpace{}%
\AgdaBound{k}\AgdaSpace{}%
\AgdaSymbol{→}\AgdaSpace{}%
\AgdaPrimitive{hcomp}\AgdaSpace{}%
\AgdaSymbol{(λ}\AgdaSpace{}%
\AgdaBound{l}\AgdaSpace{}%
\AgdaSymbol{→}\AgdaSpace{}%
\AgdaSymbol{λ}\AgdaSpace{}%
\AgdaSymbol{\{}\<%
\\
\>[2][@{}l@{\AgdaIndent{0}}]%
\>[4]\AgdaSymbol{(}\AgdaBound{i}\AgdaSpace{}%
\AgdaSymbol{=}\AgdaSpace{}%
\AgdaInductiveConstructor{i0}\AgdaSymbol{)}\AgdaSpace{}%
\AgdaSymbol{→}\AgdaSpace{}%
\AgdaBound{p}\AgdaSpace{}%
\AgdaBound{j}\AgdaSpace{}%
\AgdaSymbol{(}\AgdaBound{k}\AgdaSpace{}%
\AgdaOperator{\AgdaPrimitive{∧}}\AgdaSpace{}%
\AgdaBound{l}\AgdaSymbol{)}\AgdaSpace{}%
\AgdaSymbol{;}\AgdaSpace{}%
\AgdaSymbol{(}\AgdaBound{j}\AgdaSpace{}%
\AgdaSymbol{=}\AgdaSpace{}%
\AgdaInductiveConstructor{i0}\AgdaSymbol{)}\AgdaSpace{}%
\AgdaSymbol{→}\AgdaSpace{}%
\AgdaBound{q}\AgdaSpace{}%
\AgdaBound{i}\AgdaSpace{}%
\AgdaBound{k}\AgdaSpace{}%
\AgdaSymbol{;}\AgdaSpace{}%
\AgdaSymbol{(}\AgdaBound{k}\AgdaSpace{}%
\AgdaSymbol{=}\AgdaSpace{}%
\AgdaInductiveConstructor{i0}\AgdaSymbol{)}\AgdaSpace{}%
\AgdaSymbol{→}\AgdaSpace{}%
\AgdaBound{x}\AgdaSpace{}%
\AgdaSymbol{;}\<%
\\
\>[4]\AgdaSymbol{(}\AgdaBound{i}\AgdaSpace{}%
\AgdaSymbol{=}\AgdaSpace{}%
\AgdaInductiveConstructor{i1}\AgdaSymbol{)}\AgdaSpace{}%
\AgdaSymbol{→}\AgdaSpace{}%
\AgdaBound{p}\AgdaSpace{}%
\AgdaBound{j}\AgdaSpace{}%
\AgdaSymbol{(}\AgdaBound{k}\AgdaSpace{}%
\AgdaOperator{\AgdaPrimitive{∧}}\AgdaSpace{}%
\AgdaBound{l}\AgdaSymbol{)}\AgdaSpace{}%
\AgdaSymbol{;}\AgdaSpace{}%
\AgdaSymbol{(}\AgdaBound{j}\AgdaSpace{}%
\AgdaSymbol{=}\AgdaSpace{}%
\AgdaInductiveConstructor{i1}\AgdaSymbol{)}\AgdaSpace{}%
\AgdaSymbol{→}\AgdaSpace{}%
\AgdaBound{q}\AgdaSpace{}%
\AgdaBound{i}\AgdaSpace{}%
\AgdaBound{k}\AgdaSpace{}%
\AgdaSymbol{;}\AgdaSpace{}%
\AgdaSymbol{(}\AgdaBound{k}\AgdaSpace{}%
\AgdaSymbol{=}\AgdaSpace{}%
\AgdaInductiveConstructor{i1}\AgdaSymbol{)}\AgdaSpace{}%
\AgdaSymbol{→}\AgdaSpace{}%
\AgdaBound{p}\AgdaSpace{}%
\AgdaBound{j}\AgdaSpace{}%
\AgdaBound{l}\AgdaSpace{}%
\AgdaSymbol{\})}\AgdaSpace{}%
\AgdaSymbol{(}\AgdaBound{q}\AgdaSpace{}%
\AgdaBound{i}\AgdaSpace{}%
\AgdaBound{k}\AgdaSymbol{)}\<%
\end{code}}
\begin{code}[hide]%
\>[2]\AgdaComment{--\ EckmannHilton-Cube\ i\ j\ k\ =\ hcomp\ (λ\ l\ →\ λ\ \{}\<%
\\
\>[2][@{}l@{\AgdaIndent{0}}]%
\>[6]\AgdaComment{--\ \ \ (i\ =\ i0)\ →\ p\ j\ (k\ ∧\ l)}\<%
\\
\>[6]\AgdaComment{--\ ;\ (i\ =\ i1)\ →\ p\ j\ (k\ ∧\ l)}\<%
\\
\>[6]\AgdaComment{--\ ;\ (j\ =\ i0)\ →\ q\ i\ k}\<%
\\
\>[6]\AgdaComment{--\ ;\ (j\ =\ i1)\ →\ q\ i\ k}\<%
\\
\>[6]\AgdaComment{--\ ;\ (k\ =\ i0)\ →\ x}\<%
\\
\>[6]\AgdaComment{--\ ;\ (k\ =\ i1)\ →\ p\ j\ l\ \})}\<%
\\
\>[6]\AgdaComment{--\ \ \ (q\ i\ k)}\<%
\\
\>[0]\<%
\end{code}
\newcommand{\agdaeckmannhilton}{%
\begin{code}%
\>[0][@{}l@{\AgdaIndent{1}}]%
\>[2]\AgdaFunction{EckmannHilton}\AgdaSpace{}%
\AgdaSymbol{:}\AgdaSpace{}%
\AgdaBound{p}\AgdaSpace{}%
\AgdaOperator{\AgdaFunction{∙}}\AgdaSpace{}%
\AgdaBound{q}\AgdaSpace{}%
\AgdaOperator{\AgdaFunction{≡}}\AgdaSpace{}%
\AgdaBound{q}\AgdaSpace{}%
\AgdaOperator{\AgdaFunction{∙}}\AgdaSpace{}%
\AgdaBound{p}\<%
\\
\>[2]\AgdaFunction{EckmannHilton}\AgdaSpace{}%
\AgdaSymbol{=}\AgdaSpace{}%
\AgdaFunction{Sq→Comp}\AgdaSpace{}%
\AgdaBound{p}\AgdaSpace{}%
\AgdaBound{q}\AgdaSpace{}%
\AgdaFunction{EckmannHilton-Cube}\<%
\end{code}}
\begin{code}[hide]%
\>[0]\<%
\\
\\[\AgdaEmptyExtraSkip]%
\>[0]\AgdaKeyword{module}\AgdaSpace{}%
\AgdaModule{\AgdaUnderscore{}}\AgdaSpace{}%
\AgdaSymbol{\{}\AgdaBound{(}\AgdaBound{A}\AgdaSpace{}%
\AgdaOperator{\AgdaInductiveConstructor{,}}\AgdaSpace{}%
\AgdaBound{a}\AgdaBound{)}\AgdaSpace{}%
\AgdaSymbol{:}\AgdaSpace{}%
\AgdaFunction{Pointed}\AgdaSymbol{\}}%
\>[31]\AgdaKeyword{where}\<%
\\
\\[\AgdaEmptyExtraSkip]%
\>[0]\AgdaComment{--\ Adapted\ from\ https://github.com/pi3js2/pi4s3/blob/main/Syllepsis.agda}\<%
\\
\>[0][@{}l@{\AgdaIndent{0}}]%
\>[2]\AgdaFunction{2+2-thingy}\AgdaSpace{}%
\AgdaSymbol{:}\AgdaSpace{}%
\AgdaSymbol{(}\AgdaBound{p}\AgdaSpace{}%
\AgdaBound{q}\AgdaSpace{}%
\AgdaSymbol{:}\AgdaSpace{}%
\AgdaField{fst}\AgdaSpace{}%
\AgdaSymbol{(}\AgdaFunction{Ω}\AgdaSpace{}%
\AgdaSymbol{(}\AgdaFunction{Ω}\AgdaSpace{}%
\AgdaSymbol{(}\AgdaFunction{Ω}\AgdaSpace{}%
\AgdaSymbol{(}\AgdaBound{A}\AgdaSpace{}%
\AgdaOperator{\AgdaInductiveConstructor{,}}\AgdaSpace{}%
\AgdaBound{a}\AgdaSymbol{)))))}\AgdaSpace{}%
\AgdaSymbol{→}\<%
\\
\>[2][@{}l@{\AgdaIndent{0}}]%
\>[4]\AgdaPostulate{PathP}\AgdaSpace{}%
\AgdaSymbol{(λ}\AgdaSpace{}%
\AgdaBound{i}\AgdaSpace{}%
\AgdaSymbol{→}\AgdaSpace{}%
\AgdaPostulate{PathP}\AgdaSpace{}%
\AgdaSymbol{(λ}\AgdaSpace{}%
\AgdaBound{j}\AgdaSpace{}%
\AgdaSymbol{→}\AgdaSpace{}%
\AgdaPostulate{PathP}\AgdaSpace{}%
\AgdaSymbol{(λ}\AgdaSpace{}%
\AgdaBound{k}\AgdaSpace{}%
\AgdaSymbol{→}\AgdaSpace{}%
\AgdaFunction{Path}\AgdaSpace{}%
\AgdaSymbol{(}\AgdaField{fst}\AgdaSpace{}%
\AgdaSymbol{(}\AgdaFunction{Ω}\AgdaSpace{}%
\AgdaSymbol{(}\AgdaBound{A}\AgdaSpace{}%
\AgdaOperator{\AgdaInductiveConstructor{,}}\AgdaSpace{}%
\AgdaBound{a}\AgdaSymbol{)))}\<%
\\
\>[4][@{}l@{\AgdaIndent{0}}]%
\>[6]\AgdaSymbol{(}\AgdaBound{q}\AgdaSpace{}%
\AgdaBound{i}\AgdaSpace{}%
\AgdaBound{j}\AgdaSymbol{)}\AgdaSpace{}%
\AgdaSymbol{(}\AgdaBound{q}\AgdaSpace{}%
\AgdaBound{i}\AgdaSpace{}%
\AgdaBound{j}\AgdaSymbol{))}\<%
\\
\>[6]\AgdaSymbol{(λ}\AgdaSpace{}%
\AgdaSymbol{\AgdaUnderscore{}}\AgdaSpace{}%
\AgdaSymbol{→}\AgdaSpace{}%
\AgdaBound{q}\AgdaSpace{}%
\AgdaBound{i}\AgdaSpace{}%
\AgdaBound{j}\AgdaSymbol{)}\AgdaSpace{}%
\AgdaSymbol{(λ}\AgdaSpace{}%
\AgdaSymbol{\AgdaUnderscore{}}\AgdaSpace{}%
\AgdaSymbol{→}\AgdaSpace{}%
\AgdaBound{q}\AgdaSpace{}%
\AgdaBound{i}\AgdaSpace{}%
\AgdaBound{j}\AgdaSymbol{))}\<%
\\
\>[6]\AgdaBound{p}\AgdaSpace{}%
\AgdaBound{p}\AgdaSymbol{)}\<%
\\
\>[6]\AgdaSymbol{(λ}\AgdaSpace{}%
\AgdaSymbol{\AgdaUnderscore{}}\AgdaSpace{}%
\AgdaSymbol{→}\AgdaSpace{}%
\AgdaBound{p}\AgdaSymbol{)}\AgdaSpace{}%
\AgdaSymbol{(λ}\AgdaSpace{}%
\AgdaSymbol{\AgdaUnderscore{}}\AgdaSpace{}%
\AgdaSymbol{→}\AgdaSpace{}%
\AgdaBound{p}\AgdaSymbol{)}\<%
\\
\>[2]\AgdaFunction{2+2-thingy}\AgdaSpace{}%
\AgdaBound{p}\AgdaSpace{}%
\AgdaBound{q}\AgdaSpace{}%
\AgdaBound{i}\AgdaSpace{}%
\AgdaBound{j}\AgdaSpace{}%
\AgdaBound{k}\AgdaSpace{}%
\AgdaBound{l}\AgdaSpace{}%
\AgdaBound{m}\AgdaSpace{}%
\AgdaSymbol{=}\AgdaSpace{}%
\AgdaPrimitive{hcomp}\AgdaSpace{}%
\AgdaSymbol{(λ}\AgdaSpace{}%
\AgdaBound{n}\AgdaSpace{}%
\AgdaSymbol{→}\AgdaSpace{}%
\AgdaSymbol{λ}\AgdaSpace{}%
\AgdaSymbol{\{}\<%
\\
\>[2][@{}l@{\AgdaIndent{0}}]%
\>[6]\AgdaSymbol{(}\AgdaBound{i}\AgdaSpace{}%
\AgdaSymbol{=}\AgdaSpace{}%
\AgdaInductiveConstructor{i0}\AgdaSymbol{)}\AgdaSpace{}%
\AgdaSymbol{→}\AgdaSpace{}%
\AgdaBound{p}\AgdaSpace{}%
\AgdaBound{k}\AgdaSpace{}%
\AgdaBound{l}\AgdaSpace{}%
\AgdaSymbol{(}\AgdaBound{m}\AgdaSpace{}%
\AgdaOperator{\AgdaPrimitive{∨}}\AgdaSpace{}%
\AgdaOperator{\AgdaPrimitive{\textasciitilde{}}}\AgdaSpace{}%
\AgdaBound{n}\AgdaSymbol{)}\<%
\\
\>[2][@{}l@{\AgdaIndent{0}}]%
\>[4]\AgdaSymbol{;}\AgdaSpace{}%
\AgdaSymbol{(}\AgdaBound{i}\AgdaSpace{}%
\AgdaSymbol{=}\AgdaSpace{}%
\AgdaInductiveConstructor{i1}\AgdaSymbol{)}\AgdaSpace{}%
\AgdaSymbol{→}\AgdaSpace{}%
\AgdaBound{p}\AgdaSpace{}%
\AgdaBound{k}\AgdaSpace{}%
\AgdaBound{l}\AgdaSpace{}%
\AgdaSymbol{(}\AgdaBound{m}\AgdaSpace{}%
\AgdaOperator{\AgdaPrimitive{∨}}\AgdaSpace{}%
\AgdaOperator{\AgdaPrimitive{\textasciitilde{}}}\AgdaSpace{}%
\AgdaBound{n}\AgdaSymbol{)}\<%
\\
\>[4]\AgdaSymbol{;}\AgdaSpace{}%
\AgdaSymbol{(}\AgdaBound{j}\AgdaSpace{}%
\AgdaSymbol{=}\AgdaSpace{}%
\AgdaInductiveConstructor{i0}\AgdaSymbol{)}\AgdaSpace{}%
\AgdaSymbol{→}\AgdaSpace{}%
\AgdaBound{p}\AgdaSpace{}%
\AgdaBound{k}\AgdaSpace{}%
\AgdaBound{l}\AgdaSpace{}%
\AgdaSymbol{(}\AgdaBound{m}\AgdaSpace{}%
\AgdaOperator{\AgdaPrimitive{∨}}\AgdaSpace{}%
\AgdaOperator{\AgdaPrimitive{\textasciitilde{}}}\AgdaSpace{}%
\AgdaBound{n}\AgdaSymbol{)}\<%
\\
\>[4]\AgdaSymbol{;}\AgdaSpace{}%
\AgdaSymbol{(}\AgdaBound{j}\AgdaSpace{}%
\AgdaSymbol{=}\AgdaSpace{}%
\AgdaInductiveConstructor{i1}\AgdaSymbol{)}\AgdaSpace{}%
\AgdaSymbol{→}\AgdaSpace{}%
\AgdaBound{p}\AgdaSpace{}%
\AgdaBound{k}\AgdaSpace{}%
\AgdaBound{l}\AgdaSpace{}%
\AgdaSymbol{(}\AgdaBound{m}\AgdaSpace{}%
\AgdaOperator{\AgdaPrimitive{∨}}\AgdaSpace{}%
\AgdaOperator{\AgdaPrimitive{\textasciitilde{}}}\AgdaSpace{}%
\AgdaBound{n}\AgdaSymbol{)}\<%
\\
\>[4]\AgdaSymbol{;}\AgdaSpace{}%
\AgdaSymbol{(}\AgdaBound{k}\AgdaSpace{}%
\AgdaSymbol{=}\AgdaSpace{}%
\AgdaInductiveConstructor{i0}\AgdaSymbol{)}\AgdaSpace{}%
\AgdaSymbol{→}\AgdaSpace{}%
\AgdaBound{q}\AgdaSpace{}%
\AgdaBound{i}\AgdaSpace{}%
\AgdaBound{j}\AgdaSpace{}%
\AgdaSymbol{(}\AgdaBound{m}\AgdaSpace{}%
\AgdaOperator{\AgdaPrimitive{∧}}\AgdaSpace{}%
\AgdaBound{n}\AgdaSymbol{)}\<%
\\
\>[4]\AgdaSymbol{;}\AgdaSpace{}%
\AgdaSymbol{(}\AgdaBound{k}\AgdaSpace{}%
\AgdaSymbol{=}\AgdaSpace{}%
\AgdaInductiveConstructor{i1}\AgdaSymbol{)}\AgdaSpace{}%
\AgdaSymbol{→}\AgdaSpace{}%
\AgdaBound{q}\AgdaSpace{}%
\AgdaBound{i}\AgdaSpace{}%
\AgdaBound{j}\AgdaSpace{}%
\AgdaSymbol{(}\AgdaBound{m}\AgdaSpace{}%
\AgdaOperator{\AgdaPrimitive{∧}}\AgdaSpace{}%
\AgdaBound{n}\AgdaSymbol{)}\<%
\\
\>[4]\AgdaSymbol{;}\AgdaSpace{}%
\AgdaSymbol{(}\AgdaBound{l}\AgdaSpace{}%
\AgdaSymbol{=}\AgdaSpace{}%
\AgdaInductiveConstructor{i0}\AgdaSymbol{)}\AgdaSpace{}%
\AgdaSymbol{→}\AgdaSpace{}%
\AgdaBound{q}\AgdaSpace{}%
\AgdaBound{i}\AgdaSpace{}%
\AgdaBound{j}\AgdaSpace{}%
\AgdaSymbol{(}\AgdaBound{m}\AgdaSpace{}%
\AgdaOperator{\AgdaPrimitive{∧}}\AgdaSpace{}%
\AgdaBound{n}\AgdaSymbol{)}\<%
\\
\>[4]\AgdaSymbol{;}\AgdaSpace{}%
\AgdaSymbol{(}\AgdaBound{l}\AgdaSpace{}%
\AgdaSymbol{=}\AgdaSpace{}%
\AgdaInductiveConstructor{i1}\AgdaSymbol{)}\AgdaSpace{}%
\AgdaSymbol{→}\AgdaSpace{}%
\AgdaBound{q}\AgdaSpace{}%
\AgdaBound{i}\AgdaSpace{}%
\AgdaBound{j}\AgdaSpace{}%
\AgdaSymbol{(}\AgdaBound{m}\AgdaSpace{}%
\AgdaOperator{\AgdaPrimitive{∧}}\AgdaSpace{}%
\AgdaBound{n}\AgdaSymbol{)}\<%
\\
\>[4]\AgdaSymbol{;}\AgdaSpace{}%
\AgdaSymbol{(}\AgdaBound{m}\AgdaSpace{}%
\AgdaSymbol{=}\AgdaSpace{}%
\AgdaInductiveConstructor{i0}\AgdaSymbol{)}\AgdaSpace{}%
\AgdaSymbol{→}\AgdaSpace{}%
\AgdaBound{p}\AgdaSpace{}%
\AgdaBound{k}\AgdaSpace{}%
\AgdaBound{l}\AgdaSpace{}%
\AgdaSymbol{(}\AgdaOperator{\AgdaPrimitive{\textasciitilde{}}}\AgdaSpace{}%
\AgdaBound{n}\AgdaSymbol{)}\<%
\\
\>[4]\AgdaSymbol{;}\AgdaSpace{}%
\AgdaSymbol{(}\AgdaBound{m}\AgdaSpace{}%
\AgdaSymbol{=}\AgdaSpace{}%
\AgdaInductiveConstructor{i1}\AgdaSymbol{)}\AgdaSpace{}%
\AgdaSymbol{→}\AgdaSpace{}%
\AgdaBound{q}\AgdaSpace{}%
\AgdaBound{i}\AgdaSpace{}%
\AgdaBound{j}\AgdaSpace{}%
\AgdaBound{n}\<%
\\
\>[4]\AgdaSymbol{\})}\AgdaSpace{}%
\AgdaBound{a}\<%
\\
\\[\AgdaEmptyExtraSkip]%
\>[2]\AgdaFunction{cubicalSyllepsis}\AgdaSpace{}%
\AgdaSymbol{:}\AgdaSpace{}%
\AgdaSymbol{(}\AgdaBound{p}\AgdaSpace{}%
\AgdaBound{q}\AgdaSpace{}%
\AgdaSymbol{:}\AgdaSpace{}%
\AgdaField{fst}\AgdaSpace{}%
\AgdaSymbol{(}\AgdaFunction{Ω}\AgdaSpace{}%
\AgdaSymbol{(}\AgdaFunction{Ω}\AgdaSpace{}%
\AgdaSymbol{(}\AgdaFunction{Ω}\AgdaSpace{}%
\AgdaSymbol{(}\AgdaBound{A}\AgdaSpace{}%
\AgdaOperator{\AgdaInductiveConstructor{,}}\AgdaSpace{}%
\AgdaBound{a}\AgdaSymbol{)))))}\<%
\\
\>[2][@{}l@{\AgdaIndent{0}}]%
\>[4]\AgdaSymbol{→}\AgdaSpace{}%
\AgdaFunction{EckmannHilton-Cube}\AgdaSpace{}%
\AgdaBound{p}\AgdaSpace{}%
\AgdaBound{q}\AgdaSpace{}%
\AgdaOperator{\AgdaFunction{≡}}\AgdaSpace{}%
\AgdaSymbol{(λ}\AgdaSpace{}%
\AgdaBound{i}\AgdaSpace{}%
\AgdaBound{j}\AgdaSpace{}%
\AgdaSymbol{→}\AgdaSpace{}%
\AgdaFunction{EckmannHilton-Cube}\AgdaSpace{}%
\AgdaBound{q}\AgdaSpace{}%
\AgdaBound{p}\AgdaSpace{}%
\AgdaBound{j}\AgdaSpace{}%
\AgdaBound{i}\AgdaSymbol{)}\<%
\\
\>[2]\AgdaFunction{cubicalSyllepsis}\AgdaSpace{}%
\AgdaBound{p}\AgdaSpace{}%
\AgdaBound{q}\AgdaSpace{}%
\AgdaBound{i}\AgdaSpace{}%
\AgdaBound{j}\AgdaSpace{}%
\AgdaBound{k}\AgdaSpace{}%
\AgdaBound{l}\AgdaSpace{}%
\AgdaSymbol{=}\<%
\\
\>[2][@{}l@{\AgdaIndent{0}}]%
\>[4]\AgdaPrimitive{hcomp}\AgdaSpace{}%
\AgdaSymbol{(λ}\AgdaSpace{}%
\AgdaBound{m}\AgdaSpace{}%
\AgdaSymbol{→}%
\>[1224I]\AgdaSymbol{λ}\AgdaSpace{}%
\AgdaSymbol{\{}\AgdaSpace{}%
\AgdaSymbol{(}\AgdaBound{j}\AgdaSpace{}%
\AgdaSymbol{=}\AgdaSpace{}%
\AgdaInductiveConstructor{i0}\AgdaSymbol{)}\AgdaSpace{}%
\AgdaSymbol{→}\AgdaSpace{}%
\AgdaBound{p}\AgdaSpace{}%
\AgdaBound{k}\AgdaSpace{}%
\AgdaSymbol{(}\AgdaBound{l}\AgdaSpace{}%
\AgdaOperator{\AgdaPrimitive{∧}}\AgdaSpace{}%
\AgdaSymbol{(}\AgdaBound{m}\AgdaSpace{}%
\AgdaOperator{\AgdaPrimitive{∨}}\AgdaSpace{}%
\AgdaBound{i}\AgdaSymbol{))}\<%
\\
\>[1224I][@{}l@{\AgdaIndent{0}}]%
\>[18]\AgdaSymbol{;}\AgdaSpace{}%
\AgdaSymbol{(}\AgdaBound{j}\AgdaSpace{}%
\AgdaSymbol{=}\AgdaSpace{}%
\AgdaInductiveConstructor{i1}\AgdaSymbol{)}\AgdaSpace{}%
\AgdaSymbol{→}\AgdaSpace{}%
\AgdaBound{p}\AgdaSpace{}%
\AgdaBound{k}\AgdaSpace{}%
\AgdaSymbol{(}\AgdaBound{l}\AgdaSpace{}%
\AgdaOperator{\AgdaPrimitive{∧}}\AgdaSpace{}%
\AgdaSymbol{(}\AgdaBound{m}\AgdaSpace{}%
\AgdaOperator{\AgdaPrimitive{∨}}\AgdaSpace{}%
\AgdaBound{i}\AgdaSymbol{))}\<%
\\
\>[18]\AgdaSymbol{;}\AgdaSpace{}%
\AgdaSymbol{(}\AgdaBound{k}\AgdaSpace{}%
\AgdaSymbol{=}\AgdaSpace{}%
\AgdaInductiveConstructor{i0}\AgdaSymbol{)}\AgdaSpace{}%
\AgdaSymbol{→}\AgdaSpace{}%
\AgdaBound{q}\AgdaSpace{}%
\AgdaBound{j}\AgdaSpace{}%
\AgdaSymbol{(}\AgdaBound{l}\AgdaSpace{}%
\AgdaOperator{\AgdaPrimitive{∧}}\AgdaSpace{}%
\AgdaSymbol{(}\AgdaBound{m}\AgdaSpace{}%
\AgdaOperator{\AgdaPrimitive{∨}}\AgdaSpace{}%
\AgdaOperator{\AgdaPrimitive{\textasciitilde{}}}\AgdaSpace{}%
\AgdaBound{i}\AgdaSymbol{))}\<%
\\
\>[18]\AgdaSymbol{;}\AgdaSpace{}%
\AgdaSymbol{(}\AgdaBound{k}\AgdaSpace{}%
\AgdaSymbol{=}\AgdaSpace{}%
\AgdaInductiveConstructor{i1}\AgdaSymbol{)}\AgdaSpace{}%
\AgdaSymbol{→}\AgdaSpace{}%
\AgdaBound{q}\AgdaSpace{}%
\AgdaBound{j}\AgdaSpace{}%
\AgdaSymbol{(}\AgdaBound{l}\AgdaSpace{}%
\AgdaOperator{\AgdaPrimitive{∧}}\AgdaSpace{}%
\AgdaSymbol{(}\AgdaBound{m}\AgdaSpace{}%
\AgdaOperator{\AgdaPrimitive{∨}}\AgdaSpace{}%
\AgdaOperator{\AgdaPrimitive{\textasciitilde{}}}\AgdaSpace{}%
\AgdaBound{i}\AgdaSymbol{))}\<%
\\
\>[18]\AgdaSymbol{;}\AgdaSpace{}%
\AgdaSymbol{(}\AgdaBound{l}\AgdaSpace{}%
\AgdaSymbol{=}\AgdaSpace{}%
\AgdaInductiveConstructor{i0}\AgdaSymbol{)}\AgdaSpace{}%
\AgdaSymbol{→}\AgdaSpace{}%
\AgdaFunction{refl}\<%
\\
\>[18]\AgdaSymbol{;}\AgdaSpace{}%
\AgdaSymbol{(}\AgdaBound{l}\AgdaSpace{}%
\AgdaSymbol{=}\AgdaSpace{}%
\AgdaInductiveConstructor{i1}\AgdaSymbol{)}\AgdaSpace{}%
\AgdaSymbol{→}\AgdaSpace{}%
\AgdaFunction{2+2-thingy}\AgdaSpace{}%
\AgdaBound{p}\AgdaSpace{}%
\AgdaBound{q}\AgdaSpace{}%
\AgdaBound{j}\AgdaSpace{}%
\AgdaSymbol{(}\AgdaBound{m}\AgdaSpace{}%
\AgdaOperator{\AgdaPrimitive{∨}}\AgdaSpace{}%
\AgdaOperator{\AgdaPrimitive{\textasciitilde{}}}\AgdaSpace{}%
\AgdaBound{i}\AgdaSymbol{)}\AgdaSpace{}%
\AgdaBound{k}\AgdaSpace{}%
\AgdaSymbol{(}\AgdaBound{m}\AgdaSpace{}%
\AgdaOperator{\AgdaPrimitive{∨}}\AgdaSpace{}%
\AgdaBound{i}\AgdaSymbol{)}\<%
\\
\>[18]\AgdaSymbol{\})}\AgdaSpace{}%
\AgdaSymbol{(}\AgdaFunction{2+2-thingy}\AgdaSpace{}%
\AgdaBound{p}\AgdaSpace{}%
\AgdaBound{q}\AgdaSpace{}%
\AgdaBound{j}\AgdaSpace{}%
\AgdaSymbol{(}\AgdaBound{l}\AgdaSpace{}%
\AgdaOperator{\AgdaPrimitive{∧}}\AgdaSpace{}%
\AgdaOperator{\AgdaPrimitive{\textasciitilde{}}}\AgdaSpace{}%
\AgdaBound{i}\AgdaSymbol{)}\AgdaSpace{}%
\AgdaBound{k}\AgdaSpace{}%
\AgdaSymbol{(}\AgdaBound{l}\AgdaSpace{}%
\AgdaOperator{\AgdaPrimitive{∧}}\AgdaSpace{}%
\AgdaBound{i}\AgdaSymbol{))}\<%
\\
\\[\AgdaEmptyExtraSkip]%
\\[\AgdaEmptyExtraSkip]%
\>[2]\AgdaComment{--\ syllepsis\ :\ EckmannHilton\ p\ q\ ≡\ sym\ (EckmannHilton\ q\ p)}\<%
\\
\>[2]\AgdaComment{--\ syllepsis\ =\ \{!\ TODO\ !\}}\<%
\\
\\[\AgdaEmptyExtraSkip]%
\>[2]\AgdaComment{--\ Example\ discussed\ on\ the\ Univalent\ Agda\ Discord}\<%
\\
\>[2]\AgdaKeyword{module}\AgdaSpace{}%
\AgdaModule{\AgdaUnderscore{}}\AgdaSpace{}%
\AgdaSymbol{\{}\AgdaBound{a}\AgdaSpace{}%
\AgdaBound{b}\AgdaSpace{}%
\AgdaBound{c}\AgdaSpace{}%
\AgdaSymbol{:}\AgdaSpace{}%
\AgdaBound{A}\AgdaSymbol{\}}\AgdaSpace{}%
\AgdaSymbol{\{}\AgdaBound{p}\AgdaSpace{}%
\AgdaBound{p'}\AgdaSpace{}%
\AgdaSymbol{:}\AgdaSpace{}%
\AgdaBound{a}\AgdaSpace{}%
\AgdaOperator{\AgdaFunction{≡}}\AgdaSpace{}%
\AgdaBound{b}\AgdaSymbol{\}}\AgdaSpace{}%
\AgdaSymbol{\{}\AgdaBound{q}\AgdaSpace{}%
\AgdaBound{q'}\AgdaSpace{}%
\AgdaSymbol{:}\AgdaSpace{}%
\AgdaBound{b}\AgdaSpace{}%
\AgdaOperator{\AgdaFunction{≡}}\AgdaSpace{}%
\AgdaBound{c}\AgdaSymbol{\}}\AgdaSpace{}%
\AgdaKeyword{where}\<%
\\
\>[2][@{}l@{\AgdaIndent{0}}]%
\>[4]\AgdaFunction{double-connection'}\AgdaSpace{}%
\AgdaSymbol{:}\AgdaSpace{}%
\AgdaBound{p}\AgdaSpace{}%
\AgdaOperator{\AgdaFunction{≡}}\AgdaSpace{}%
\AgdaBound{p'}\AgdaSpace{}%
\AgdaSymbol{→}\AgdaSpace{}%
\AgdaBound{q}\AgdaSpace{}%
\AgdaOperator{\AgdaFunction{≡}}\AgdaSpace{}%
\AgdaBound{q'}\AgdaSpace{}%
\AgdaSymbol{→}\AgdaSpace{}%
\AgdaPostulate{PathP}\AgdaSpace{}%
\AgdaSymbol{(λ}\AgdaSpace{}%
\AgdaBound{i}\AgdaSpace{}%
\AgdaSymbol{→}\AgdaSpace{}%
\AgdaBound{p'}\AgdaSpace{}%
\AgdaBound{i}\AgdaSpace{}%
\AgdaOperator{\AgdaFunction{≡}}\AgdaSpace{}%
\AgdaBound{q'}\AgdaSpace{}%
\AgdaBound{i}\AgdaSymbol{)}\AgdaSpace{}%
\AgdaBound{p}\AgdaSpace{}%
\AgdaBound{q}\<%
\\
\>[4]\AgdaFunction{double-connection'}\AgdaSpace{}%
\AgdaBound{alpha}\AgdaSpace{}%
\AgdaBound{beta}\AgdaSpace{}%
\AgdaSymbol{=}\AgdaSpace{}%
\AgdaSymbol{λ}\AgdaSpace{}%
\AgdaBound{i}\AgdaSpace{}%
\AgdaBound{j}\AgdaSpace{}%
\AgdaSymbol{→}\AgdaSpace{}%
\AgdaPrimitive{hcomp}\AgdaSpace{}%
\AgdaSymbol{(λ}\AgdaSpace{}%
\AgdaBound{k}\AgdaSpace{}%
\AgdaSymbol{→}\AgdaSpace{}%
\AgdaSymbol{λ}\AgdaSpace{}%
\AgdaSymbol{\{}\<%
\\
\>[4][@{}l@{\AgdaIndent{0}}]%
\>[10]\AgdaSymbol{(}\AgdaBound{i}\AgdaSpace{}%
\AgdaSymbol{=}\AgdaSpace{}%
\AgdaInductiveConstructor{i0}\AgdaSymbol{)}\AgdaSpace{}%
\AgdaSymbol{→}\AgdaSpace{}%
\AgdaBound{p}\AgdaSpace{}%
\AgdaBound{j}\<%
\\
\>[4][@{}l@{\AgdaIndent{0}}]%
\>[8]\AgdaSymbol{;}\AgdaSpace{}%
\AgdaSymbol{(}\AgdaBound{i}\AgdaSpace{}%
\AgdaSymbol{=}\AgdaSpace{}%
\AgdaInductiveConstructor{i1}\AgdaSymbol{)}\AgdaSpace{}%
\AgdaSymbol{→}\AgdaSpace{}%
\AgdaBound{q}\AgdaSpace{}%
\AgdaBound{j}\<%
\\
\>[8]\AgdaSymbol{;}\AgdaSpace{}%
\AgdaSymbol{(}\AgdaBound{j}\AgdaSpace{}%
\AgdaSymbol{=}\AgdaSpace{}%
\AgdaInductiveConstructor{i0}\AgdaSymbol{)}\AgdaSpace{}%
\AgdaSymbol{→}\AgdaSpace{}%
\AgdaBound{p'}\AgdaSpace{}%
\AgdaBound{i}\<%
\\
\>[8]\AgdaSymbol{;}\AgdaSpace{}%
\AgdaSymbol{(}\AgdaBound{j}\AgdaSpace{}%
\AgdaSymbol{=}\AgdaSpace{}%
\AgdaInductiveConstructor{i1}\AgdaSymbol{)}\AgdaSpace{}%
\AgdaSymbol{→}\AgdaSpace{}%
\AgdaBound{beta}\AgdaSpace{}%
\AgdaBound{k}\AgdaSpace{}%
\AgdaBound{i}\<%
\\
\>[4][@{}l@{\AgdaIndent{0}}]%
\>[6]\AgdaSymbol{\})}\AgdaSpace{}%
\AgdaSymbol{(}\AgdaPrimitive{hcomp}\AgdaSpace{}%
\AgdaSymbol{(λ}\AgdaSpace{}%
\AgdaBound{k}\AgdaSpace{}%
\AgdaSymbol{→}\AgdaSpace{}%
\AgdaSymbol{λ}\AgdaSpace{}%
\AgdaSymbol{\{}\<%
\\
\>[6][@{}l@{\AgdaIndent{0}}]%
\>[8]\AgdaSymbol{(}\AgdaBound{i}\AgdaSpace{}%
\AgdaSymbol{=}\AgdaSpace{}%
\AgdaInductiveConstructor{i0}\AgdaSymbol{)}\AgdaSpace{}%
\AgdaSymbol{→}\AgdaSpace{}%
\AgdaBound{p}\AgdaSpace{}%
\AgdaBound{j}\<%
\\
\>[6]\AgdaSymbol{;}\AgdaSpace{}%
\AgdaSymbol{(}\AgdaBound{i}\AgdaSpace{}%
\AgdaSymbol{=}\AgdaSpace{}%
\AgdaInductiveConstructor{i1}\AgdaSymbol{)}\AgdaSpace{}%
\AgdaSymbol{→}\AgdaSpace{}%
\AgdaBound{q}\AgdaSpace{}%
\AgdaSymbol{(}\AgdaBound{j}\AgdaSpace{}%
\AgdaOperator{\AgdaPrimitive{∧}}\AgdaSpace{}%
\AgdaBound{k}\AgdaSymbol{)}\<%
\\
\>[6]\AgdaSymbol{;}\AgdaSpace{}%
\AgdaSymbol{(}\AgdaBound{j}\AgdaSpace{}%
\AgdaSymbol{=}\AgdaSpace{}%
\AgdaInductiveConstructor{i0}\AgdaSymbol{)}\AgdaSpace{}%
\AgdaSymbol{→}\AgdaSpace{}%
\AgdaBound{alpha}\AgdaSpace{}%
\AgdaBound{k}\AgdaSpace{}%
\AgdaBound{i}\<%
\\
\>[6]\AgdaSymbol{;}\AgdaSpace{}%
\AgdaSymbol{(}\AgdaBound{j}\AgdaSpace{}%
\AgdaSymbol{=}\AgdaSpace{}%
\AgdaInductiveConstructor{i1}\AgdaSymbol{)}\AgdaSpace{}%
\AgdaSymbol{→}\AgdaSpace{}%
\AgdaBound{q}\AgdaSpace{}%
\AgdaSymbol{(}\AgdaBound{i}\AgdaSpace{}%
\AgdaOperator{\AgdaPrimitive{∧}}\AgdaSpace{}%
\AgdaBound{k}\AgdaSymbol{)}\<%
\\
\>[6]\AgdaSymbol{\})}\AgdaSpace{}%
\AgdaSymbol{(}\AgdaBound{p}\AgdaSpace{}%
\AgdaSymbol{(}\AgdaBound{i}\AgdaSpace{}%
\AgdaOperator{\AgdaPrimitive{∨}}\AgdaSpace{}%
\AgdaBound{j}\AgdaSymbol{)))}\<%
\\
\\[\AgdaEmptyExtraSkip]%
\\[\AgdaEmptyExtraSkip]%
\>[2]\AgdaFunction{mmh}\AgdaSpace{}%
\AgdaSymbol{:}\AgdaSpace{}%
\AgdaSymbol{\{}\AgdaBound{x}\AgdaSpace{}%
\AgdaBound{y}\AgdaSpace{}%
\AgdaBound{z}\AgdaSpace{}%
\AgdaSymbol{:}\AgdaSpace{}%
\AgdaBound{A}\AgdaSymbol{\}}\AgdaSpace{}%
\AgdaSymbol{(}\AgdaBound{p}\AgdaSpace{}%
\AgdaSymbol{:}\AgdaSpace{}%
\AgdaBound{x}\AgdaSpace{}%
\AgdaOperator{\AgdaFunction{≡}}\AgdaSpace{}%
\AgdaBound{y}\AgdaSymbol{)}\<%
\\
\>[2][@{}l@{\AgdaIndent{0}}]%
\>[4]\AgdaComment{--\ →\ q\ ≡\ (q\ ∙\ sym\ p)\ ∙\ p}\<%
\\
\>[4]\AgdaSymbol{→}\AgdaSpace{}%
\AgdaPostulate{PathP}\AgdaSpace{}%
\AgdaSymbol{(λ}\AgdaSpace{}%
\AgdaBound{i}\AgdaSpace{}%
\AgdaSymbol{→}\AgdaSpace{}%
\AgdaBound{p}\AgdaSpace{}%
\AgdaBound{i}\AgdaSpace{}%
\AgdaOperator{\AgdaFunction{≡}}\AgdaSpace{}%
\AgdaBound{x}\AgdaSymbol{)}\AgdaSpace{}%
\AgdaSymbol{(λ}\AgdaSpace{}%
\AgdaSymbol{\AgdaUnderscore{}}\AgdaSpace{}%
\AgdaSymbol{→}\AgdaSpace{}%
\AgdaBound{x}\AgdaSymbol{)}\AgdaSpace{}%
\AgdaSymbol{(λ}\AgdaSpace{}%
\AgdaBound{i}\AgdaSpace{}%
\AgdaSymbol{→}\AgdaSpace{}%
\AgdaPrimitive{hcomp}\AgdaSpace{}%
\AgdaSymbol{(λ}\AgdaSpace{}%
\AgdaBound{j}\AgdaSpace{}%
\AgdaSymbol{→}\AgdaSpace{}%
\AgdaSymbol{λ}\AgdaSpace{}%
\AgdaSymbol{\{}\AgdaSpace{}%
\AgdaSymbol{(}\AgdaBound{i}\AgdaSpace{}%
\AgdaSymbol{=}\AgdaSpace{}%
\AgdaInductiveConstructor{i0}\AgdaSymbol{)}\AgdaSpace{}%
\AgdaSymbol{→}\AgdaSpace{}%
\AgdaBound{p}\AgdaSpace{}%
\AgdaBound{j}\AgdaSpace{}%
\AgdaSymbol{;}\AgdaSpace{}%
\AgdaSymbol{(}\AgdaBound{i}\AgdaSpace{}%
\AgdaSymbol{=}\AgdaSpace{}%
\AgdaInductiveConstructor{i1}\AgdaSymbol{)}\AgdaSpace{}%
\AgdaSymbol{→}\AgdaSpace{}%
\AgdaBound{p}\AgdaSpace{}%
\AgdaInductiveConstructor{i0}\AgdaSpace{}%
\AgdaSymbol{\})}\AgdaSpace{}%
\AgdaSymbol{(}\AgdaBound{p}\AgdaSpace{}%
\AgdaInductiveConstructor{i0}\AgdaSymbol{))}\<%
\\
\>[2]\AgdaFunction{mmh}%
\>[1467I]\AgdaBound{p}\AgdaSpace{}%
\AgdaBound{j}\AgdaSpace{}%
\AgdaBound{i}\AgdaSpace{}%
\AgdaSymbol{=}\AgdaSpace{}%
\AgdaFunction{hfill}\<%
\\
\\[\AgdaEmptyExtraSkip]%
\>[.][@{}l@{}]\<[1467I]%
\>[6]\AgdaSymbol{(λ}\AgdaSpace{}%
\AgdaBound{k}\AgdaSpace{}%
\AgdaSymbol{→}\AgdaSpace{}%
\AgdaSymbol{λ}\AgdaSpace{}%
\AgdaSymbol{\{}\<%
\\
\>[6][@{}l@{\AgdaIndent{0}}]%
\>[8]\AgdaSymbol{(}\AgdaBound{i}\AgdaSpace{}%
\AgdaSymbol{=}\AgdaSpace{}%
\AgdaInductiveConstructor{i0}\AgdaSymbol{)}\AgdaSpace{}%
\AgdaSymbol{→}\AgdaSpace{}%
\AgdaBound{p}\AgdaSpace{}%
\AgdaBound{k}\<%
\\
\>[6]\AgdaSymbol{;}\AgdaSpace{}%
\AgdaSymbol{(}\AgdaBound{i}\AgdaSpace{}%
\AgdaSymbol{=}\AgdaSpace{}%
\AgdaInductiveConstructor{i1}\AgdaSymbol{)}\AgdaSpace{}%
\AgdaSymbol{→}\AgdaSpace{}%
\AgdaBound{p}\AgdaSpace{}%
\AgdaInductiveConstructor{i0}\<%
\\
\>[6]\AgdaSymbol{\})}\<%
\\
\>[6]\AgdaSymbol{(}\AgdaPostulate{inS}\AgdaSpace{}%
\AgdaSymbol{(}\AgdaBound{p}\AgdaSpace{}%
\AgdaInductiveConstructor{i0}\AgdaSymbol{))}\AgdaSpace{}%
\AgdaBound{j}\<%
\\
\\[\AgdaEmptyExtraSkip]%
\\[\AgdaEmptyExtraSkip]%
\>[6][@{}l@{\AgdaIndent{0}}]%
\>[10]\AgdaComment{--\ hfill\ \ (inS\ (p\ i0))\ j}\<%
\\
\\[\AgdaEmptyExtraSkip]%
\\[\AgdaEmptyExtraSkip]%
\>[2]\AgdaFunction{compPathl-cancel}\AgdaSpace{}%
\AgdaSymbol{:}\AgdaSpace{}%
\AgdaSymbol{\{}\AgdaBound{x}\AgdaSpace{}%
\AgdaBound{y}\AgdaSpace{}%
\AgdaBound{z}\AgdaSpace{}%
\AgdaSymbol{:}\AgdaSpace{}%
\AgdaBound{A}\AgdaSymbol{\}}\AgdaSpace{}%
\AgdaSymbol{(}\AgdaBound{p}\AgdaSpace{}%
\AgdaSymbol{:}\AgdaSpace{}%
\AgdaBound{x}\AgdaSpace{}%
\AgdaOperator{\AgdaFunction{≡}}\AgdaSpace{}%
\AgdaBound{y}\AgdaSymbol{)}\AgdaSpace{}%
\AgdaSymbol{(}\AgdaBound{q}\AgdaSpace{}%
\AgdaSymbol{:}\AgdaSpace{}%
\AgdaBound{x}\AgdaSpace{}%
\AgdaOperator{\AgdaFunction{≡}}\AgdaSpace{}%
\AgdaBound{z}\AgdaSymbol{)}\<%
\\
\>[2][@{}l@{\AgdaIndent{0}}]%
\>[4]\AgdaComment{--\ →\ q\ ≡\ (q\ ∙\ sym\ p)\ ∙\ p}\<%
\\
\>[4]\AgdaSymbol{→}\AgdaSpace{}%
\AgdaPostulate{PathP}\AgdaSpace{}%
\AgdaSymbol{(λ}\AgdaSpace{}%
\AgdaBound{i}\AgdaSpace{}%
\AgdaSymbol{→}\AgdaSpace{}%
\AgdaBound{p}\AgdaSpace{}%
\AgdaBound{i}\AgdaSpace{}%
\AgdaOperator{\AgdaFunction{≡}}\AgdaSpace{}%
\AgdaBound{q}\AgdaSpace{}%
\AgdaInductiveConstructor{i1}\AgdaSymbol{)}\AgdaSpace{}%
\AgdaBound{q}\AgdaSpace{}%
\AgdaSymbol{((}\AgdaFunction{sym}\AgdaSpace{}%
\AgdaBound{p}\AgdaSymbol{)}\AgdaSpace{}%
\AgdaOperator{\AgdaFunction{∙}}\AgdaSpace{}%
\AgdaBound{q}\AgdaSymbol{)}\<%
\\
\>[2]\AgdaFunction{compPathl-cancel}\AgdaSpace{}%
\AgdaBound{p}\AgdaSpace{}%
\AgdaBound{q}\AgdaSpace{}%
\AgdaSymbol{=}\<%
\\
\>[2][@{}l@{\AgdaIndent{0}}]%
\>[6]\AgdaSymbol{λ}%
\>[1523I]\AgdaBound{i}\AgdaSpace{}%
\AgdaBound{j}\AgdaSpace{}%
\AgdaSymbol{→}\AgdaSpace{}%
\AgdaPrimitive{hcomp}\AgdaSpace{}%
\AgdaSymbol{(λ}\AgdaSpace{}%
\AgdaBound{k}\AgdaSpace{}%
\AgdaSymbol{→}\AgdaSpace{}%
\AgdaSymbol{λ}\AgdaSpace{}%
\AgdaSymbol{\{}\<%
\\
\>[.][@{}l@{}]\<[1523I]%
\>[8]\AgdaSymbol{(}\AgdaBound{i}\AgdaSpace{}%
\AgdaSymbol{=}\AgdaSpace{}%
\AgdaInductiveConstructor{i0}\AgdaSymbol{)}\AgdaSpace{}%
\AgdaSymbol{→}\AgdaSpace{}%
\AgdaBound{q}\AgdaSpace{}%
\AgdaSymbol{(}\AgdaBound{j}\AgdaSpace{}%
\AgdaOperator{\AgdaPrimitive{∧}}\AgdaSpace{}%
\AgdaBound{k}\AgdaSymbol{)}\<%
\\
\>[6]\AgdaSymbol{;}%
\>[1539I]\AgdaSymbol{(}\AgdaBound{i}%
\>[1540I]\AgdaSymbol{=}\AgdaSpace{}%
\AgdaInductiveConstructor{i1}\AgdaSymbol{)}\AgdaSpace{}%
\AgdaSymbol{→}\AgdaSpace{}%
\AgdaFunction{hfill}\AgdaSpace{}%
\AgdaSymbol{(λ}\AgdaSpace{}%
\AgdaBound{l}\AgdaSpace{}%
\AgdaSymbol{→}\AgdaSpace{}%
\AgdaSymbol{λ}\AgdaSpace{}%
\AgdaSymbol{\{}\<%
\\
\>[1540I][@{}l@{\AgdaIndent{0}}]%
\>[12]\AgdaSymbol{(}\AgdaBound{j}\AgdaSpace{}%
\AgdaSymbol{=}\AgdaSpace{}%
\AgdaInductiveConstructor{i0}\AgdaSymbol{)}\AgdaSpace{}%
\AgdaSymbol{→}\AgdaSpace{}%
\AgdaBound{p}\AgdaSpace{}%
\AgdaInductiveConstructor{i1}\<%
\\
\>[1539I][@{}l@{\AgdaIndent{0}}]%
\>[10]\AgdaSymbol{;}\AgdaSpace{}%
\AgdaSymbol{(}\AgdaBound{j}\AgdaSpace{}%
\AgdaSymbol{=}\AgdaSpace{}%
\AgdaInductiveConstructor{i1}\AgdaSymbol{)}\AgdaSpace{}%
\AgdaSymbol{→}\AgdaSpace{}%
\AgdaBound{q}\AgdaSpace{}%
\AgdaBound{l}\<%
\\
\>[10]\AgdaSymbol{\})}%
\>[1560I]\AgdaSymbol{(}\AgdaPostulate{inS}\AgdaSpace{}%
\AgdaSymbol{(}\AgdaPrimitive{hcomp}\AgdaSpace{}%
\AgdaSymbol{(λ}\AgdaSpace{}%
\AgdaBound{l}\AgdaSpace{}%
\AgdaSymbol{→}\AgdaSpace{}%
\AgdaSymbol{λ}\AgdaSpace{}%
\AgdaSymbol{\{}\<%
\\
\>[1560I][@{}l@{\AgdaIndent{0}}]%
\>[16]\AgdaSymbol{(}\AgdaBound{j}\AgdaSpace{}%
\AgdaSymbol{=}\AgdaSpace{}%
\AgdaInductiveConstructor{i0}\AgdaSymbol{)}\AgdaSpace{}%
\AgdaSymbol{→}\AgdaSpace{}%
\AgdaBound{p}\AgdaSpace{}%
\AgdaBound{l}\<%
\\
\>[1560I][@{}l@{\AgdaIndent{0}}]%
\>[14]\AgdaSymbol{;}%
\>[1572I]\AgdaSymbol{(}\AgdaBound{j}\AgdaSpace{}%
\AgdaSymbol{=}\AgdaSpace{}%
\AgdaInductiveConstructor{i1}\AgdaSymbol{)}\AgdaSpace{}%
\AgdaSymbol{→}\AgdaSpace{}%
\AgdaBound{p}\AgdaSpace{}%
\AgdaInductiveConstructor{i0}\<%
\\
\>[1572I][@{}l@{\AgdaIndent{0}}]%
\>[18]\AgdaSymbol{\})}\AgdaSpace{}%
\AgdaSymbol{(}\AgdaBound{p}\AgdaSpace{}%
\AgdaInductiveConstructor{i0}\AgdaSymbol{)))}\AgdaSpace{}%
\AgdaBound{k}\<%
\\
\>[6]\AgdaSymbol{;}\AgdaSpace{}%
\AgdaSymbol{(}\AgdaBound{j}\AgdaSpace{}%
\AgdaSymbol{=}\AgdaSpace{}%
\AgdaInductiveConstructor{i0}\AgdaSymbol{)}\AgdaSpace{}%
\AgdaSymbol{→}\AgdaSpace{}%
\AgdaBound{p}\AgdaSpace{}%
\AgdaBound{i}\<%
\\
\>[6]\AgdaSymbol{;}\AgdaSpace{}%
\AgdaSymbol{(}\AgdaBound{j}\AgdaSpace{}%
\AgdaSymbol{=}\AgdaSpace{}%
\AgdaInductiveConstructor{i1}\AgdaSymbol{)}\AgdaSpace{}%
\AgdaSymbol{→}\AgdaSpace{}%
\AgdaBound{q}\AgdaSpace{}%
\AgdaBound{k}\<%
\\
\>[6]\AgdaSymbol{\})}%
\>[1593I]\AgdaSymbol{(}\AgdaFunction{hfill}\AgdaSpace{}%
\AgdaSymbol{(λ}\AgdaSpace{}%
\AgdaBound{k}\AgdaSpace{}%
\AgdaSymbol{→}\AgdaSpace{}%
\AgdaSymbol{λ}\AgdaSpace{}%
\AgdaSymbol{\{}\<%
\\
\>[1593I][@{}l@{\AgdaIndent{0}}]%
\>[12]\AgdaSymbol{(}\AgdaBound{j}\AgdaSpace{}%
\AgdaSymbol{=}\AgdaSpace{}%
\AgdaInductiveConstructor{i0}\AgdaSymbol{)}\AgdaSpace{}%
\AgdaSymbol{→}\AgdaSpace{}%
\AgdaBound{p}\AgdaSpace{}%
\AgdaBound{k}\<%
\\
\>[1593I][@{}l@{\AgdaIndent{0}}]%
\>[10]\AgdaSymbol{;}\AgdaSpace{}%
\AgdaSymbol{(}\AgdaBound{j}\AgdaSpace{}%
\AgdaSymbol{=}\AgdaSpace{}%
\AgdaInductiveConstructor{i1}\AgdaSymbol{)}\AgdaSpace{}%
\AgdaSymbol{→}\AgdaSpace{}%
\AgdaBound{p}\AgdaSpace{}%
\AgdaInductiveConstructor{i0}\<%
\\
\>[10]\AgdaSymbol{\})}\AgdaSpace{}%
\AgdaSymbol{(}\AgdaPostulate{inS}\AgdaSpace{}%
\AgdaSymbol{(}\AgdaBound{p}\AgdaSpace{}%
\AgdaInductiveConstructor{i0}\AgdaSymbol{))}\AgdaSpace{}%
\AgdaBound{i}\AgdaSymbol{)}\<%
\\
\\[\AgdaEmptyExtraSkip]%
\\[\AgdaEmptyExtraSkip]%
\>[2]\AgdaComment{--\ compPathr-cancel\ :\ \{x\ y\ z\ :\ A\}\ (p\ :\ z\ ≡\ y)\ (q\ :\ x\ ≡\ y)\ →\ q\ ≡\ (q\ ∙\ sym\ p)\ ∙\ p}\<%
\\
\>[2]\AgdaComment{--\ compPathr-cancel\ p\ q\ i\ j\ =}\<%
\\
\>[2]\AgdaComment{--\ \ \ hcomp\ (λ\ k\ →\ λ\ \{}\<%
\\
\>[2]\AgdaComment{--\ \ \ \ \ \ \ \ \ (i\ =\ i0)\ →\ q\ j}\<%
\\
\>[2]\AgdaComment{--\ \ \ \ \ \ \ ;\ (i\ =\ i1)\ →\ \{!!\}}\<%
\\
\>[2]\AgdaComment{--\ \ \ \ \ \ \ --\ hfill\ (λ\ m\ →\ λ\ \{}\<%
\\
\>[2]\AgdaComment{--\ \ \ \ \ \ \ --\ \ \ \ \ \ \ \ \ \ \ (j\ =\ i0)\ →\ q\ i0}\<%
\\
\>[2]\AgdaComment{--\ \ \ \ \ \ \ --\ \ \ \ \ \ \ \ \ ;\ (j\ =\ i1)\ →\ p\ i}\<%
\\
\>[2]\AgdaComment{--\ \ \ \ \ \ \ --\ \ \ \ \ \ \ \ \ \ \ \ \ \ \ \ \ \})\ (hcomp\ (λ\ o\ →\ λ\ \{}\<%
\\
\>[2]\AgdaComment{--\ \ \ \ \ \ \ --\ \ \ \ \ \ \ \ \ \ \ \ \ \ \ \ \ (j\ =\ i0)\ →\ q\ i0}\<%
\\
\>[2]\AgdaComment{--\ \ \ \ \ \ \ --\ \ \ \ \ \ \ \ \ \ \ \ \ \ \ \ \ ;\ (j\ =\ i1)\ →\ hcomp\ (λ\ q\ →\ λ\ \{}\<%
\\
\>[2]\AgdaComment{--\ \ \ \ \ \ \ --\ \ \ \ \ \ \ \ \ \ \ \ \ \ \ \ \ (j\ =\ i0)\ →\ p\ i}\<%
\\
\>[2]\AgdaComment{--\ \ \ \ \ \ \ --\ \ \ \ \ \ \ \ \ \ \ \ \ \ \ \ \ ;\ (j\ =\ i1)\ →\ p\ i0}\<%
\\
\>[2]\AgdaComment{--\ \ \ \ \ \ \ --\ \ \ \ \ \ \ \ \ \ \ \ \ \ \ \ \ \})\ (p\ i0)}\<%
\\
\>[2]\AgdaComment{--\ \ \ \ \ \ \ --\ \ \ \ \ \ \ \ \ \ \ \ \ \ \ \ \ \})\ (q\ i))}\<%
\\
\>[2]\AgdaComment{--\ \ \ \ \ \ \ ;\ (j\ =\ i0)\ →\ q\ i0}\<%
\\
\>[2]\AgdaComment{--\ \ \ \ \ \ \ ;\ (j\ =\ i1)\ →}\<%
\\
\>[2]\AgdaComment{--\ \ \ \ \ \ \ \ \ \ \ hcomp\ (λ\ l\ →\ λ\ \{}\<%
\\
\>[2]\AgdaComment{--\ \ \ \ \ \ \ \ \ \ \ \ \ (i\ =\ i0)\ →\ p\ l}\<%
\\
\>[2]\AgdaComment{--\ \ \ \ \ \ \ \ \ \ \ ;\ (i\ =\ i1)\ →\ p\ (i\ ∧\ l)}\<%
\\
\>[2]\AgdaComment{--\ \ \ \ \ \ \ \ \ \ \ ;\ (k\ =\ i1)\ →\ p\ l}\<%
\\
\>[2]\AgdaComment{--\ \ \ \ \ \ \ \ \ \ \ ;\ (k\ =\ i0)\ →\ \{!!\}}\<%
\\
\>[2]\AgdaComment{--\ \ \ \ \ \ \ \ \ \ \ \ \ --\ hfill\ (λ\ o\ →\ λ\ \{}\<%
\\
\>[2]\AgdaComment{--\ \ \ \ \ \ \ \ \ \ \ \ \ --\ \ \ \ \ (j\ =\ i0)\ →\ p\ i}\<%
\\
\>[2]\AgdaComment{--\ \ \ \ \ \ \ \ \ \ \ \ \ --\ \ \ ;\ (j\ =\ i1)\ →\ p\ i0}\<%
\\
\>[2]\AgdaComment{--\ \ \ \ \ \ \ \ \ \ \ \ \ --\ \})\ (p\ i0)}\<%
\\
\>[2]\AgdaComment{--\ \ \ \ \ \ \ \ \ \ \ \})\ (p\ i0)}\<%
\\
\>[2]\AgdaComment{--\ \ \ \ \ \ \ \})\ (hfill}\<%
\\
\>[2]\AgdaComment{--\ \ \ \ \ \ \ \ \ \ \ \ (λ\ k\ →\ λ\ \{\ (j\ =\ i0)\ →\ q\ i0\ ;\ (j\ =\ i1)\ →\ sym\ p\ k\ \})}\<%
\\
\>[2]\AgdaComment{--\ \ \ \ \ \ \ \ \ \ \ \ (inS\ (q\ j))\ i)}\<%
\\
\\[\AgdaEmptyExtraSkip]%
\>[2][@{}l@{\AgdaIndent{0}}]%
\>[10]\AgdaComment{--\ (hfill\ (λ\ m\ →\ λ\ \{\ (i\ =\ i0)\ →\ q\ i0\ ;\ (i\ =\ i1)\ →\ sym\ p\ m\ \})\ (inS\ ?)\ i)}\<%
\\
\>[2][@{}l@{\AgdaIndent{0}}]%
\>[8]\AgdaComment{--\ (hfill\ (λ\ m\ →\ λ\ \{}\<%
\\
\>[8]\AgdaComment{--\ (j\ =\ i0)\ →\ q\ i0}\<%
\\
\>[8]\AgdaComment{--\ ;\ (j\ =\ i1)\ →\ hcomp\ (λ\ o\ →\ λ\ \{}\<%
\\
\>[8]\AgdaComment{--\ (j\ =\ i0)\ →\ p\ i}\<%
\\
\>[8]\AgdaComment{--\ ;\ (j\ =\ i1)\ →\ p\ i0}\<%
\\
\>[8]\AgdaComment{--\ \})\ (p\ i0)}\<%
\\
\>[8]\AgdaComment{--\ \})\ (q\ i))}\<%
\\
\>[0]\<%
\end{code}

\title{Automating Boundary Filling in Cubical Type Theories}
\titlecomment{{\lsuper*} This paper is an extended version of
  \emph{Automating Boundary Filling in Cubical Agda}
  \cite{dore24_autom_bound_fillin_cubic_agda}.}  \thanks{Maximilian
  Doré was supported by COST Action EuroProofNet, supported by COST
  (European Cooperation in Science and Technology,
  \url{www.cost.eu}). Evan Cavallo was supported by the Knut and Alice
  Wallenberg Foundation through the Foundation's program for
  mathematics. Anders Mörtberg was supported by the Swedish Research
  Council (Vetenskapsrådet) under Grant No. 2019-04545.}

\author[M.~Doré]{Maximilian Doré\lmcsorcid{0000-0003-4012-6401}}[a]
\author[E.~Cavallo]{Evan Cavallo\lmcsorcid{0000-0001-8174-7496}}[b]
\author[A.~Mörtberg]{Anders Mörtberg\lmcsorcid{0000-0001-9558-6080}}[c]

\address{Department of Computer Science, University of Oxford, United
  Kingdom}
\email{maximilian.dore@cs.ox.ac.uk}

\address{Department of Computer Science and Engineering,
  University of Gothenburg and Chalmers University of Technology, Sweden}
\email{evan.cavallo@gu.se}

\address{Department of Mathematics, Stockholm University,
  Sweden}
\email{anders.mortberg@math.su.se}





\begin{abstract}
  \noindent When working in a proof assistant, automation is
  key to discharging routine proof goals such as equations
  between algebraic expressions. Homotopy type theory allows
  the user to reason about higher structures, such as
  topological spaces, using higher inductive types~(HITs)
  and univalence. Cubical type theory provides computational
  support for HITs and univalence. A difficulty when working
  in cubical type theory is dealing with the complex
  combinatorics of higher structures, an
  infinite-dimensional generalisation of equational
  reasoning. To solve these higher-dimensional equations
  consists in constructing cubes with specified boundaries.

  We develop a simplified cubical language in which we isolate and
  study two automation problems: contortion solving, where we attempt
  to ``contort'' a cube to fit a given boundary, and the more general
  Kan solving, where we search for solutions that involve pasting
  multiple cubes together. Both problems are difficult in the general
  case---Kan solving is even undecidable---so we focus on heuristics
  that perform well on practical examples. Our language encompasses
  different variations of cubical type theory which differ in their
  ``contortion theory'', i.e., the class of contortions they support.
  We provide a solver for the contortion problem for the most complex
  contortion theories currently being researched, namely the Dedekind
  and De~Morgan contortions, by utilising a reformulation of
  contortions in terms of poset maps. We solve Kan problems using
  constraint satisfaction programming, which is applicable
  independently of the underlying contortion theory. We have
  implemented our algorithms in an experimental Haskell solver that
  can be used to automatically solve many goals a user of cubical type
  theory might face. We illustrate this with a case study establishing
  the Eckmann-Hilton theorem using our solver, as well as various
  benchmarks---providing the ground for further study of proof
  automation in cubical type theories.
\end{abstract}

\maketitle

\section{Introduction} \label{sec:intro}

Homotopy type theory (HoTT) \cite{hott} adds new constructs to intensional
dependent type theory \cite{martin-loef75_intuit_theor_types} reflecting an
interpretation of types as homotopy types of topological spaces. This allows
homotopy theory to be developed \emph{synthetically} inside HoTT; many classical
results have been reconstructed this way, such as the Hopf fibration
\cite{hott}, Blakers-Massey
theorem~\cite{favonia16_mechan_blaker_massey_connec_theor}, Seifert-van Kampen
theorem~\cite{FavoniaShulman16}, Atiyah-Hirzebruch and Serre spectral
sequences~\cite{FlorisPhd}, Hurewicz theorem \cite{christensen2020hurewicz},
etc.  However, as originally formulated, HoTT postulates both the univalence
axiom \cite{Voevodsky10cmu} and the existence of HITs \cite{hott} without proper
computational content---to rectify this, \emph{cubical type theories}
\cite{cohen18_cubic_type_theor,angiuli18} replace the identity type with a
primitive \emph{path} type, yielding a computationally well-behaved theory which
validates the axioms of HoTT.

Inspired by Daniel Kan's cubical sets \cite{kan55}, cubical type theory
represents elements of iterated identity types as higher-dimensional cubes.
Synthetic homotopy theory in cubical type theory thereby attains a particular
``cubical'' flavour \cite{cubicalsynthetic}. A path in a type $A$ connecting
elements $a$ and $b$ can be thought of as a function
$p \colon [\izero,\ione] \to A$ from the unit interval into the ``space'' $A$
such that $p(\izero) = a$ and $p(\ione) = b$. Paths play the role of equalities
in the theory, and operations on paths encode familiar laws of equality:
reflexivity is a constant path, transitivity is concatenation of paths, and
symmetry is following a path in reverse.

Paths can also be studied in their own right. In particular, we can
consider equalities \emph{between} paths in $A$, which as functions
$[\izero,\ione] \to ([\izero,\ione] \to A)$ can be read as maps from the unit
\emph{square} (or \emph{2-cube}) $[\izero,\ione]^2$ to $A$; iterating, we find
ourselves considering $n$-cubes in $A$.  Algebraic laws such as the
associativity of path concatenation or identity laws
are
represented as squares with certain boundaries.

\begin{figure}[h!]
  \centering
  \begin{subfigure}{0.2\textwidth}
    \begin{tikzpicture}[scale=0.8]
      \filldraw[pattern=north west lines] (0,1,0) -- (0,1,2) -- (0,2,2) -- (0,2,0) -- cycle;
      \filldraw[color=gray] (0,0,0) -- (0,0,2) -- (0,1,2) -- (0,1,0) -- cycle;

      \draw[thick] (0,0,0) -- (0,2,0) -- (2,2,0) -- (2,0,0) -- cycle;
      \draw[thick] (0,0,0) -- (0,0,2) -- (0,2,2) -- (0,2,0) -- cycle;
      \draw[thick] (0,0,0) -- (2,0,0) -- (2,0,2) -- (0,0,2) -- cycle;
      \draw[thick] (0,1,0) -- (0,1,2);

      \filldraw[color=gray] (2,1,0) -- (2,1,2) -- (2,2,2) -- (2,2,0) -- cycle;

      \draw[thick,fill=white,fill opacity=0.7] (0,2,0) -- (0,2,2) -- (2,2,2) -- (2,2,0) -- cycle;
      \draw[thick,fill=white,fill opacity=0.7] (0,0,2) -- (2,0,2) -- (2,2,2) -- (0,2,2) -- cycle;
      \draw[thick,fill=white,fill opacity=0.7] (2,0,0) -- (2,0,2) -- (2,2,2) -- (2,2,0) -- cycle;
      \draw[thick] (2,1,0) -- (2,1,2);
      \draw[thick,preaction={fill, white, fill opacity=0.8},pattern=north west lines] (2,0,0) -- (2,0,2) -- (2,1,2) -- (2,1,0) -- cycle;
    \end{tikzpicture}
    \caption{The theorem \label{eh-main}}
  \end{subfigure}
  \begin{subfigure}{0.2\textwidth}
    \begin{tikzpicture}[scale=0.8]
      \filldraw[gray] (0,0,0) -- (0,0,2) -- (0,2,2) -- (0,2,0) -- cycle;
      \filldraw[pattern=north west lines] (0,0,0) -- (2,0,0) -- (2,0,2) -- (0,0,2) -- cycle;

      \draw[thick] (0,0,0) -- (0,2,0) -- (2,2,0) -- (2,0,0) -- cycle;
      \draw[thick] (0,0,0) -- (0,0,2) -- (0,2,2) -- (0,2,0) -- cycle;
      \draw[thick] (0,0,0) -- (2,0,0) -- (2,0,2) -- (0,0,2) -- cycle;

      \filldraw[color=gray] (2,0,0) -- (2,0,2) -- (2,2,2) -- (2,2,0) -- cycle;

      \draw[thick,preaction={fill, white, fill opacity=0.8},pattern=north west lines] (0,2,0) -- (0,2,2) -- (2,2,2) -- (2,2,0) -- cycle;
      \draw[thick,fill=white,fill opacity=0.7] (0,0,2) -- (2,0,2) -- (2,2,2) -- (0,2,2) -- cycle;
      \draw[thick,fill=white,fill opacity=0.7] (2,0,0) -- (2,0,2) -- (2,2,2) -- (2,2,0) -- cycle;
    \end{tikzpicture}
    \caption{\label{eh-cubical}}
  \end{subfigure}
  \begin{subfigure}{0.2\textwidth}
    \begin{tikzpicture}[scale=0.8]
      \filldraw[color=gray] (0,0,0) -- (0,0,2) -- (0,2,2) -- (0,2,0) -- cycle;
      \filldraw[pattern=north west lines] (0,0,0) -- (0,1,0) -- (2,1,0) -- (2,0,0) -- cycle;
      \filldraw[pattern=north east lines] (0,1,0) -- (0,2,0) -- (2,2,0) -- (2,1,0) -- cycle;

      \draw[thick] (0,0,0) -- (0,2,0) -- (2,2,0) -- (2,0,0) -- cycle;
      \draw[thick] (0,0,0) -- (0,0,2) -- (0,2,2) -- (0,2,0) -- cycle;
      \draw[thick] (0,0,0) -- (2,0,0) -- (2,0,2) -- (0,0,2) -- cycle;
      \draw[thick] (0,1,0) -- (2,1,0);

      \filldraw[color=gray] (2,0,0) -- (2,0,2) -- (2,2,2) -- (2,2,0) -- cycle;

      \draw[thick,fill=white,fill opacity=0.7] (0,2,0) -- (0,2,2) -- (2,2,2) -- (2,2,0) -- cycle;
      \draw[thick,fill=white,fill opacity=0.7] (0,0,2) -- (2,0,2) -- (2,2,2) -- (0,2,2) -- cycle;
      \draw[thick,fill=white,fill opacity=0.7] (2,0,0) -- (2,0,2) -- (2,2,2) -- (2,2,0) -- cycle;
    \end{tikzpicture}
    \caption{\label{eh-inverses}}
  \end{subfigure}
  \begin{subfigure}{0.2\textwidth}
    \begin{tikzpicture}[scale=0.8]

      \filldraw[thick,gray] (0,0,0) -- (0,0,2) -- (0,2,2) -- (0,2,0) -- cycle;

      \draw[thick] (0,0,0) -- (0,2,0) -- (2,2,0) -- (2,0,0) -- cycle;
      \draw[thick] (0,0,0) -- (0,0,2) -- (0,2,2) -- (0,2,0) -- cycle;
      \draw[thick] (0,0,0) -- (2,0,0) -- (2,0,2) -- (0,0,2) -- cycle;

      \filldraw[color=gray] (2,0,0) -- (2,0,2) -- (2,2,2) -- (2,2,0) -- cycle;

      \draw[thick,fill=white,fill opacity=0.7] (0,2,0) -- (0,2,2) -- (2,2,2) -- (2,2,0) -- cycle;
      \draw[thick,fill=white,fill opacity=0.7] (0,0,2) -- (2,0,2) -- (2,2,2) -- (0,2,2) -- cycle;
      \draw[thick,fill=white,fill opacity=0.7] (2,0,0) -- (2,0,2) -- (2,2,2) -- (2,2,0) -- cycle;
    \end{tikzpicture}
    \caption{\label{eh-degeneracy}}
  \end{subfigure}
  \vspace{-.6em}
  \caption{A cubical Eckmann-Hilton argument in four steps.}
\label{fig:ehcube}
\end{figure}
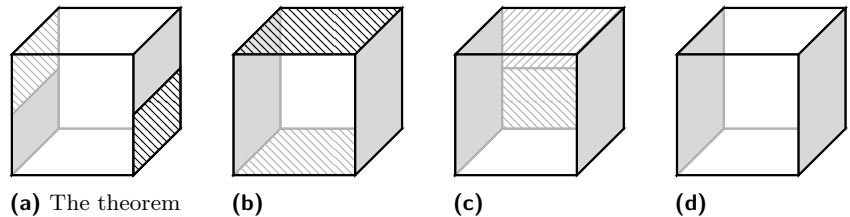

For instance, a foundational result in algebraic topology is
the \emph{Eckmann-Hilton argument}~\cite{EH62}, which states that concatenation
of 2-spheres, i.e., 2-cubes with constant boundaries, is commutative up to a path. As
a path between 2-cubes, the theorem is a 3-cube as shown
in
\autoref{eh-main}: on the left
we have a grey 2-cube concatenated with a
hatched 2-cube, on the right
they are concatenated in the opposite order,
and the interior is the path between them.

\vspace{-.7em}

In cubical type theory, we can construct such an interior by starting from some
3-cube we know can be filled, then deforming its boundary via certain basic
operations until it has the desired form. It can be more intuitive to work
backwards: deform the ``goal'' boundary until we reach a boundary we can
fill. \autoref{fig:ehcube} shows one solution: we \subref{eh-cubical} shift the
copies of the hatched 2-cube to the top and bottom faces, \subref{eh-inverses}
further shift them both to the back face, whereupon they face each other in
opposite directions, and then \subref{eh-degeneracy} cancel the concatenation of
the hatched 2-cube with its inverse. The boundary in \subref{eh-degeneracy} can
be filled immediately by the constant homotopy---i.e., reflexive equality---from
the gray 2-cube to itself.

This example illustrates the two main principles we use to build cubes in type
theory, which we call contortion and Kan filling.%
\footnote{Because we only reason within individual types in this paper, we
  encounter only so-called \emph{homogeneous} Kan filling. General Kan filling
  also incorporates \emph{transport} (or \emph{coercion}) between different
  indices of a dependent type family, but we leave this aspect to future work.
}
To \emph{contort} a cube is to reparameterise it, stretching it into a higher
dimension or projecting a face. For example, we fill the cube
\subref{eh-degeneracy} by taking the gray 2-cube and stretching it into a
degenerate 3-cube, reparameterising by a projection
$[\izero,\ione]^3 \to [\izero,\ione]^2$. Different cubical type theories come
equipped with different kinds of reparameterisations, which we call their
``contortion theories''.
None of the contortion theories we will consider in this paper allow us to
derive the cube in \autoref{eh-main} only by contorting the gray or hatched
2-cube in isolation, however. Thus the role of \emph{Kan filling}, which lets us
modify a cube by a continuous deformation of its boundary. Each of the
reductions \subref{eh-main} to \subref{eh-cubical} to \subref{eh-inverses} to
\subref{eh-degeneracy} above is an instance of Kan filling.%
\footnote{The fact that a concatenation of a 2-cube with its inverse can be
  deformed away, which we use in the step \subref{eh-inverses} to
  \subref{eh-degeneracy}, is a lemma that can itself be proven with contortion
  and Kan filling.}
Kan filling admits a second geometric
reading: it states that for every \emph{open box}, i.e., the boundary of a cube with
one face unspecified, there is a lid for the box for which an interior
(``filler'') exists. The two readings agree because a continuous deformation of
the boundary of an $n$-cube over ``time'' $t \in [\izero,\ione]$ can also be
seen as all but two opposing faces of an $(n+1)$-cube; the cube to be deformed
fits into one of the missing faces, and the lid produced by box filling is then
the deformed cube.

Reasoning with contortions and Kan fillings can pose a challenge when
formalising mathematics or computer science in cubical type theories. It is the
essence of standalone theorems such as Eckmann-Hilton, but cubical puzzles also
often appear as routine lemmas in more complex proofs. One may need to relate
one arrangement of concatenations and inverses of paths to another, for example;
such coherence conditions often appear in definitions by pattern-matching on
HITs. Just as it is difficult to anticipate all types of equations between
algebraic expressions that one might need in a large formalisation project, it
is infeasible to enumerate every routine cubical lemma in a standard
library. The purpose of this paper is to instead devise an algorithm which can
automatically prove such lemmas as needed.

Different underlying contortion theories for cubical types theories
have been considered, ranging from simple contortion theories which
form the basis of the \redtt proof assistant \cite{redtt} to more expressive contortion
theories which provide the basis of the \CubicalAgda proof assistant \cite{vezzosi19}. Our
language supports all contortion theories currently under study and is
thereby applicable to all variations of cubical type theory in the literature. In the paper we will, in particular,
focus on the more complex contortion theories as
\CubicalAgda is currently the most widely used cubical system. With
\agdaCubical \cite{agda-cubical} and \onelab \cite{1lab} there are
extensive libraries for \CubicalAgda which contain ad-hoc collections
of cubical reasoning combinators that we aim to automate with our
solver.

\medskip

\noindent \textbf{Contributions:} The work presented in this paper constitutes
one of the first systematic studies of automated reasoning for cubical type
theories. In it we
\begin{itemize}
\item formulate a general cubical language which contains a Kan filling operator
  and is parameterised over the class of ``contortions'' the language possesses,
  which allows us to precisely state important classes of automation problems in
  cubical type theories (\S\ref{sec:bdryproblems}),
\item map out the computational complexity landscape for different cubical type theories and
  show Kan filling undecidable for all theories under investigation
  (\S\ref{sec:complexity}),
\item formulate an algorithm based on poset maps for solving problems
  with the ``Dedekind'' and ``De Morgan'' contortions, thereby making
  these computationally hard problems more tractable
  (\S\ref{sec:contortions}),
\item formulate an algorithm based on constraint satisfaction programming for
  solving problems using Kan filling (\S\ref{sec:kan}), and
\item provide a practical \Haskell implementation of our algorithms and exhibit
  its effectiveness on a selection of theorems and lemmas taken from libraries
  for \CubicalAgda (\S\ref{sec:casestudy}).
\end{itemize}

\noindent 
This paper extends \cite{dore24_autom_bound_fillin_cubic_agda}, which
only considered Dedekind contortions, by generalising our framework to
work with any kind contortion theory that is considered in the
literature (cartesian \cite{angiuli18, FootballHockeyLeague19,
  accrs24}, disjunctive \cite{cavallo23}, Dedekind
\cite{awodey2023cartesian} and De Morgan
\cite{cohen18_cubic_type_theor, vezzosi19}). We give additional
complexity results for these newly arisen problems and extend our
efficient representation of Dedekind contortions to De Morgan
contortions. Furthermore, we spell out the proof that finding Kan
filling is undecidable, which was only sketched in
\cite{dore24_autom_bound_fillin_cubic_agda}. Making precise the
reduction from the word problem for groups to Kan filling requires a
considerable amount of care.
Lastly, we have rectified an error in
the definition of our language given in
\cite{dore24_autom_bound_fillin_cubic_agda}; see \autoref{rem:kan-error} below.

\section{Filling cubes in cubical type theories}
\label{sec:bdryproblems}

Cubical type theories are complex systems. Besides path types, one has the usual type
formers of type theory---dependent functions, products, inductive types,
etc.---not to mention univalence and HITs. To make automation
tractable, we restrict attention to a fragment including only basic operations
on cubical cells in a single type.\footnote{This is similar to the fragments of
  type theory used to axiomatise higher structures such as weak
  $\omega$-groupoids in e.g.\ \cite[Appendix A]{brunerie16} and
  \cite{finster17}.}

Rather than use path \emph{types} to encode cubical cells, as one does in a
fully-featured cubical type theory, we take cells as a primitive notion. A cell
is a term parameterised by one or more \emph{dimension variables}, which we
think of as ranging in the interval $[\izero,\ione]$; intuitively, a cell of
type $A$ in $n$ variables is a function $[\izero,\ione]^n \to
A$. \emph{Contexts} are lists of cells each of which can have a specified
\emph{boundary}. For example, an entry
$\isof{p}{i}[\assign{i}{\izero}{a} \se \assign{i}{\ione}{b}]$ specifies a
1-dimensional cell $p$ varying in $i \in [\izero,\ione]$ such that
$p(\izero) = a$ and $p(\ione) = b$, i.e., a path from $a$ to $b$. In general, an
entry in a context has the form $\isof{q}{\Psi}[\phi]$ where $\Psi$ is a list of
variables and $\phi$ is a list of values at faces ($i = \izero$ and $i = \ione$
in the example above). A cell hypothesis is thus a judgemental analogue of a
hypothesis of \emph{extension type} à la Riehl and Shulman
\cite[\S2.2]{riehl17}.

The problems we aim to solve are \emph{boundary problems}: given a context of
cells $\Gamma$, a list of dimension variables $\Psi$, and a boundary $\phi$, can
we use the cells in $\Gamma$ to build a cell varying in~$\Psi$ with boundary
$\phi$? We write such a problem as
``$\isfterm[\Gamma][\Psi]{\hole}<\phi>$''. For example, if we want to prove
that paths are invertible,
then we could pose the boundary problem
\begin{equation}\label{inverse}
  \isfterm[\isof{a},\isof{b},\isof{p}{i}[\assign{i}{\izero}{a} \se \assign{i}{\ione}{b}]][j]{\hole}<\assign{j}{\izero}{b} \se \assign{j}{\ione}{a}>
\end{equation}
Here $\Gamma$
has three cells: points $a$ and $b$, and a path $p$.
Our goal is a path from $b$ to $a$, written as a function of
$j \in [\izero,\ione]$ with fixed endpoints. We allow leaving the boundary of cells partially or
completely unspecified, so that we can formulate the same problem more compactly as
\begin{equation}\label{short-inverse}
  \isfterm[\isof{p}{i}][j]{\hole}<\assign{j}{\izero}{p(\ione)} \se \assign{j}{\ione}{p(\izero)}>
\end{equation}
This variant assumes a path $p$ without naming its endpoints and seeks a path from $p(\ione)$ to $p(\izero)$.
The format extends gracefully to higher cells; for example, the
diagonal of a square can be requested by posing
\begin{equation}\label{diagonal}
  \isfterm[\isof{s}{i,j}][k]{\hole}<\assign{k}{\izero}{s(\izero,\izero)} \se \assign{k}{\ione}{s(\ione,\ione)}>
\end{equation}
Here we assume a 2-dimensional cell $s$ with unspecified boundary and seek a
path from $s(\izero,\izero)$ to $s(\ione,\ione)$. In the remaining section, we
introduce two ways to produce \emph{solutions} to boundary problems: contortions
and Kan filling.

\subsection{Contorting cubes}
\label{ssec:contorting-cubes}

Intuitively, the problem (\ref{diagonal}) has a simple solution:
$\hole \coloneqq s(k,k)$.  That is, we take the hypothesised 2-cube $s$ and
apply a reparameterisation $k \mapsto (k,k)$. We call such reparameterisations
\emph{contortions}. Different cubical type theories offer different kinds of
contortions. The only contortions of cartesian cubical type theory of Angiuli et
al.\ \cite{angiuli18,FootballHockeyLeague19} are variables and the constants
$\izero$,$\ione$, whereas the theory of Cohen et al.\ (CCHM)
\cite{cohen18_cubic_type_theor} includes binary operators $\lor$ and $\land$,
conventionally called \emph{connections} \cite{brown81}, as well as a unary
\emph{involution} operator $\inv$. We think of $\lor$ as taking the
\emph{maximum} of two parameters and $\land$ as taking the minimum, whereas
$\inv$ is thought of as negation sending $i \in [\izero,\ione]$ to $\ione - i$.
For example, given a cell context containing a path $p$, the operator $\lor$ can
be used to define a square whose value at coordinate $(j,k)$ is the value of $p$
at the maximum of $j$ and $k$:
\begin{equation}
  \label{or-connection-cell}
  \isfterm*[\isof{p}{i}][j,k]{p(j \lor k)}<\begin{array}{lll}
                                  \assign{j}{\izero}{p(k)} &\se&
                                  \assign{k}{\izero}{p(j)} \\
                                  \assign{j}{\ione}{p(\ione)} &\se&
                                  \assign{k}{\ione}{p(\ione)}
                                \end{array}>
\end{equation}

We will study both the cartesian and CCHM theories in the following, as well as
two theories which lie in between the two in terms of expressiveness. If we
remove the involution operation of CCHM, leaving $\lor$ and $\land$, we have a
distributive lattice which we call the \emph{Dedekind} contortion
theory. Removing moreover one of the connections yields the \emph{disjunctive}
contortion theory, which is used by Cavallo and
Sattler~\cite{cavallo23}. Choosing a more expressive contortion theory naturally
means more problems can be solved by contortion.  For example, the path
inversion problem \eqref{short-inverse} above is immediately solved with an
involution:
\begin{equation*}
  \isfterm[\isof{p}{i}][j]{p(\inv i)}<\assign{j}{\izero}{p(\ione)} \se \assign{j}{\ione}{p(\izero)}>
\end{equation*}
Without an involution, this problem instead requires Kan filling (which will be introduced in \S\ref{ssec:kan-fillers}). On the other
hand, adding more contortions makes contortion solving more
complex.\footnote{It is also unclear whether cubical type theories with more complex contortion theories admit semantics in standard homotopy types; see discussion in \cite{cavallo23,accrs24}.} There is hence a trade-off for which class of
contortions are allowed for proof search.

We now formally introduce the language of boundary problems, starting with the
fragment needed to formulate solutions by contortion.

\begin{defi}
  A \textbf{dimension context} $\Psi$ is either a list of (unique) dimension variables
  $(i_1,\ldots,i_n)$ or the \textbf{inconsistent context} $\bot$.
\end{defi}

We think of a dimension context with $n$ variables as a topological unit
$n$-cube, each axis being labelled with one variable, while $\bot$ is the empty
space; note that the ``empty'' context $()$ is the unit 0-cube, which does have
a unique point.  We write $\Psi, i$ for the extension of $\Psi$ by a fresh
variable $i$, which is $(i_1,\ldots,i_n,i)$ when $\Psi = (i_1,\ldots,i_n)$ and
$\bot$ when $\Psi = \bot$.

\begin{defi}
  The \textbf{dimension terms} $\iscdim[\Psi]{r}$ over a dimension context
  $\Psi$ are generated by variables in $\Psi$, the constants $\izero$ and
  $\ione$, the unary operator $\inv$, and binary operators $\lor$ and $\land$,
  subject to the equations
  \begin{mathpar}
    \inv \izero = \ione \and
    \inv \ione = \izero \and
    \inv \inv x = x \and
    r \lor s = s \lor r \and
    r \lor \izero = r \and
    r \lor \ione = \ione \and
    r \lor r = r \and
    r \land (s \lor t) = (r \land s) \lor (s \land t) \and
    \inv (r \lor s) = \inv r \land \inv s \and
    \inv (r \land s) = \inv r \lor \inv s
  \end{mathpar}
  When $\Psi$ is the inconsistent context $\bot$, we consider all dimension
  terms to be equal.

  In other words, a dimension term over $\Psi = (i_1,\ldots,i_n)$ is an element of the free De
  Morgan algebra on $n$ variables.
  We call terms in the full language \textbf{De Morgan} dimension terms.
  We also consider several sublanguages of \textbf{De Morgan}: we say that a dimension term is
  \begin{itemize}
  \item \textbf{cartesian} if it is an element of $\Psi$ or
    $\izero$ or $\ione$.
  \item \textbf{disjunctive} if it mentions only variables, $\izero$, $\ione$,
    and $\lor$, i.e., is an element of the free bounded semilattice with join
    $\lor$ over $\Psi$.
  \item \textbf{Dedekind} if it mentions only variables, $\izero$, $\ione$, and
    $\lor$ and $\land$, i.e., is an element of the free bounded distributive
    lattice over $\Psi$.
  \end{itemize}
  The cartesian dimension terms play a special role in the definition of
  our theory. In this context we will instead call them \textbf{atomic} terms
  for emphasis; we use the judgement $\isdim[\Psi]{r}$ to denote atomic dimension
  terms.
\end{defi}

\begin{defi}
  We write $\negI{e}$ for the opposite of an endpoint
  $e$, so $\negI{\izero} \coloneq \ione$ and $\negI{\ione} \coloneq \izero$.
\end{defi}

Note that any cartesian term is also a disjunctive term, etc., as the language
for dimension terms gets increasingly more expressive from cartesian to De
Morgan.

\begin{defi}
  A \textbf{contortion} $\isisubst{\psi}{\Psi'}{\Psi}$ when
  $\Psi = (i_1,\ldots,i_n)$ is a list
  \[
    \psi = (i_1 \mapsto r_1,\ldots, i_n \mapsto r_n)
  \]
  consisting of a dimension term $\iscdim[\Psi']{r_k}$ for each $i_k \in \Psi$.
  When $\Psi = \bot$, there is a contortion
  $\isisubst{\psi}{\Psi'}{\Psi}$ only when $\Psi' = \bot$, in which case there
  is a unique one. A contortion is called
  \textbf{cartesian}/\textbf{disjunctive}/\textbf{Dedekind}/\textbf{De Morgan}
  if all of its dimension terms are cartesian/disjunctive/Dedekind/De Morgan.

  A \textbf{substitution} $\isincl{\psi}{\Psi'}{\Psi}$ is a contortion whose
  terms are atomic.
\end{defi}

\begin{rem}
  When $\Psi$ is clear from context, we will write $(r_1,\ldots, r_n)$ for a
  substitution as shorthand for $(i_1 \mapsto r_1,\ldots, i_n \mapsto r_n)$.  We
  also write, e.g., $\isincl{(i \mapsto r)}{\Psi}{(\Psi,i)}$ as shorthand for
  $(i_1 \mapsto i_1,\ldots, i_n \mapsto i_n, i \mapsto r)$ when
  $\Psi = (i_1,\ldots,i_n)$.
\end{rem}

A contortion $\isisubst{\psi}{\Psi'}{\Psi}$ can be seen as a continuous mapping from the domain cube to the codomain cube.
For example, $\isisubst{(i \mapsto i, j \mapsto \izero)}{(i)}{(i,j)}$ is the inclusion of a 1-dimensional face of the 2-cube, specifically the face where the second coordinate is $\izero$.
Another example is, $\isisubst{(i \mapsto i)}{(i,j)}{(i)}$, which maps a 2-cube to a 1-cube by collapsing the second coordinate.

We will need the following operation on dimension contexts to define boundaries.

\begin{defi}
  \label{def:constrained-context}
  When $\Psi$ is a dimension context, $r$ is an atomic dimension term, and $e$
  is an endpoint, we define the \textbf{constrained} dimension context
  $\ictxtcon{\Psi}{r}{e}$ by cases:
  \begin{mathpar}
    \ictxtcon{(\Psi,i,\Psi')}{i}{e} \coloneq \Psi,\Psi' \and
    \ictxtcon{\Psi}{\negI{e}}{e} \coloneq \bot \and
    \ictxtcon{\Psi}{r}{e} \coloneq \Psi\text{, otherwise}
  \end{mathpar}
  We have a \textbf{constraining
    substitution} $(\inclcon{r}{e}) \colon \ictxtcon{\Psi}{r}{e} \to \Psi$ that
  sends $r$ to $e$ if $r$ is a variable, is the unique substitution from $\bot$
  when $r$ is $\negI{e}$, and the identity substitution otherwise.
\end{defi}

For example, $(\inclcon{j}{\izero}) \colon \ictxtcon{(i,j)}{j}{\izero} \to (i,j)$ is the inclusion $\isisubst{(i \mapsto i, j \mapsto \izero)}{(i)}{(i,j)}$ of the face where $j = \izero$ into the 2-cube $(i,j)$.

The \emph{cell contexts} ($\isctxt{\Gamma}$), \emph{contorted boundaries}
($\iscbdy[\Gamma][\Psi][\Psi']{\phi}$), and \emph{contorted cells}
($\iscterm[\Gamma][\Psi]{t}$ and $\iscterm[\Gamma][\Psi]{t}<\phi>$) are mutually
inductively defined as follows.
The subscript $c$ on $\vdash_c$ indicate that these judgements concern contortions and we leave the contortion theory implicit as the judgements are the same for all theories.
Substitutions act on each of these judgements in
the usual way: given some kind of term $t$ and a substitution
$\isincl{\psi}{\Psi'}{\Psi}$ where $\Psi = (i_1,\ldots,i_n)$ and
$\psi = (r_1,\ldots,r_n)$, we write $\face{t}{\psi}$ for the result of replacing
each $i_k$ by $r_k$ in $t$. General contortions act only on some of our
syntactic sorts, namely dimension terms and contorted cells
(\autoref{def:contorted-cell}); for those sorts we write $\cont{t}{\psi}$ for
application of a contortion. As above, we say a context/boundary/term is
\textbf{cartesian}/\textbf{disjunctive}/\textbf{Dedekind}/\textbf{De Morgan}
when it only mentions contortion operations from that sublanguage.

\begin{defi}
  The \textbf{cell contexts} $\isctxt{\Gamma}$ are inductively defined by the
  rules
  \vspace{-.7em}
  \begin{mathpar}
    \inferrule
    { }
    {\isctxt{()}}
    \and
    \inferrule
    {\isctxt{\Gamma} \\ \iscbdy[\Gamma][\Psi][()]{\phi}}
    {\isctxt{(\Gamma, \isof{a}{\Psi}[\phi])}}
  \end{mathpar}
\end{defi}
\noindent 
where in the second rule, $a$ is a fresh variable name standing for a cell
over dimension variables $\Psi$ and with boundary $\phi$.

That is, a cell context is a list of variables each paired with a dimension
context and boundary over that context; the boundary for one variable may
mention preceding variables.
The list of inputs to a boundary problem, such as $\isof{a},\isof{b},\isof{p}{i}[{\assign{i}{\izero}{a} \se \assign{i}{\ione}{b}}]$ from \eqref{inverse}, is a cell context.

\begin{defi}
  The \textbf{contorted boundaries} $\iscbdy[\Gamma][\Psi][\Psi']{\phi}$ are inductively
  defined by the rules
  \vspace{-.7em}
  \begin{mathpar}
    \inferrule
    { }
    {\iscbdy[\Gamma][\Psi][\Psi']{()}}
    \\
    \inferrule
    {\iscbdy[\Gamma][\Psi][()]{\phi} \\
      \isdim[\Psi]{r} \\
      e \in \{\izero,\ione\} \\
      \iscterm[\Gamma][\ictxtcon{\Psi}{r}{e},\Psi']{t}<\face{\phi}{\inclcon{r}{e}}>}
    {\iscbdy[\Gamma][\Psi][\Psi']{(\phi \se \assign{r}{e}{t})}}
  \end{mathpar}
  Here $\face{\phi}{\inclcon{r}{e}}$ is the application of the
  constraining substitution $(r = e)$ to the boundary $\phi$.
\end{defi}

A contorted boundary is thus a list of entries $\assign{r}{e}{t}$, where each
$t$ is a contorted cell over $\ictxtcon{\Psi}{r}{e},\Psi'$, such that each entry
agrees with the previous entries when their constraints overlap.
The constraints
$r = e$ can only refer to variables in $\Psi$, while the constrained terms $t$
can also refer to variables in $\Psi'$. We will only use the cases where $\Psi'$
is empty or a singleton, the latter being used in the definition of Kan cells
(\autoref{defi:kan-cells}).
We write $\iscbdy[\Gamma][\Psi]{\phi}$ as shorthand for $\iscbdy[\Gamma][\Psi][()]{\phi}$, and use implicitly that any $\iscbdy[\Gamma][\Psi][\Psi']{\phi}$ can be viewed as a $\iscbdy[\Gamma][\Psi,\Psi']{\phi}$ (but not vice versa).

In \eqref{or-connection-cell}, for example, we saw the contorted boundary
\begin{equation}\label{or-connection-boundary}
  \iscbdy[\isof{p}{i}][j,k]{(\assign{j}{\izero}{p(k)} \se \assign{k}{\izero}{p(j)} \se \assign{j}{\ione}{p(\ione)} \se \assign{k}{\ione}{p(\ione)})}
\end{equation}
which we picture as the following unfilled square:
\begin{mathpar}
  \dimsquare{j}{k}
  \quad
  \begin{tikzpicture}[xscale=.3,yscale=.3]
    \draw[->,>=stealth] (0,0) -- (4,0) node [midway,below, fill=none] {$p(k)$};
    \draw[->,>=stealth] (0,0) -- (0,4) node [midway,left, fill=none] {$p(j)$};
    \draw[->,>=stealth] (4,0) -- (4,4) node [midway,right, fill=none] {$p(\ione)$};
    \draw[->,>=stealth] (0,4) -- (4,4) node [midway,above, fill=none] {$p(\ione)$};
  \end{tikzpicture}
\end{mathpar}
The compatibility condition in the formation rule ensures that we can only form this boundary when the edges of the squares actually line up on the intersections; for example, when we add the final face $\assign{k}{\ione}{p(\ione)}$, we must check that $p(k) = p(\ione)$ when both $j = \izero$ and $k = \ione$.

\begin{rem}
  The requirement that the term $r$ in a constraint $r = e$ is atomic is absent
  in \CubicalAgda. Imposing it simplifies the constrained context operation
  (\autoref{def:constrained-context}), while relaxing it is not particularly useful
  for practical boundary solving. We distinguish between substitutions and
  contortions to make this requirement sensible.
\end{rem}

Finally, we can introduce the main protagonists of our contortion theory: the
cells that stem from contorting some generating cell of the cell context.

\begin{defi}
  \label{def:contorted-cell}
  A \textbf{contorted cell} $\iscterm[\Gamma][\Psi]{t}$ is a contortion $\psi$
  applied to a variable $a$ from the cell context:
  \begin{mathpar}
    \inferrule
    {(\isof{a}{\Psi'}[\phi]) \in \Gamma \\
      \isisubst{\psi}{\Psi}{\Psi'}}
    {\iscterm[\Gamma][\Psi]{a(\psi)}}
  \end{mathpar}
  We write $a$ instead of $a()$ when $\Psi' = ()$.
  Equality of contorted cells is generated by
  the rule
  \begin{mathpar}
    \inferrule
    {(\isof{a}{\Psi'}[\phi]) \in \Gamma \\
      (\assign{r}{e}{t}) \in \phi \\
      \isisubst{\psi}{\Psi}{\Psi'} \\
      \iscdim[\Psi]{\cont{r}{\psi}}[\cont{e}{\psi}]}
    {\iscterm[\Gamma][\Psi]{a(\psi)}[\cont{t}{\psi}]}
  \end{mathpar}
  which is to say that $a(\psi)$ has exactly the boundary assigned to it by the
  context. Additionally, all contorted cells are considered equal in the
  inconsistent dimension context $\bot$.

  We write $\iscterm[\Gamma][\Psi]{t}<\phi>$ when $t$ is a cell agreeing with
  $\phi$, i.e., such that $\iscterm{\face{t}{\inclcon{r}{e}}}[t']$ for each
  $(\assign{r}{e}{t'}) \in \phi$. The two rules above can then be restated as
  the single rule
  \begin{mathpar}
    \inferrule
    {(\isof{a}{\Psi'}[\phi]) \in \Gamma \\
      \isisubst{\psi}{\Psi}{\Psi'}}
    {\iscterm[\Gamma][\Psi]{a(\psi)}<\cont{\phi}{\psi}>}
  \end{mathpar}
\end{defi}

For example, \eqref{or-connection-cell} applies the contortion $\isisubst{(j \lor k)}{(j,k)}{(i)}$ to a 1-cell $p(i)$ to define a contorted cell $\iscterm[\isof{p}{i}][j,k]{p(j \lor k)}$ matching the boundary \eqref{or-connection-boundary}.

\begin{rem}
  \label{rem:explicit-substitutions}
  For ease of reading, we have described the action of substitutions and
  contortions above as a meta-operation on raw terms, replacing dimension
  variables by terms in the expected way. Formally, however, we shall consider
  the theory to come with a calculus of explicit substitutions as described by
  Abadi et al.\ \cite{abadi91}. That is, for example, the action of
  substitutions on terms is given by a term former
  \begin{mathpar}
    \inferrule
    {\iscterm[\Gamma][\Psi]{t} \and
      \isincl{\psi}{\Psi'}{\Psi}}
    {\iscterm[\Gamma][\Psi']{\face{t}{\psi}}}
  \end{mathpar}
  whose behaviour is specified by equations such as
  $\face{a(\psi)}{\psi'} = a(\psi\psi')$. We make use of this formal description
  in \S\ref{sec:undecidability}, where we prove some results by induction on
  syntax.
\end{rem}

\subsection{Kan filling}
\label{ssec:kan-fillers}

Paths in spaces can be \emph{concatenated}: if there is a path from $a$ to $b$
and a path from $b$ to $c$, then there is a path from $a$ to $c$. Concatenation
generalises to higher cells; for example, we can attach several surfaces at
their boundaries to form a new surface. Kan \cite{kan55} devised a single
property that encompasses all of these operations in the context of cubical
sets. In cubical type theory, it is embodied by the \emph{Kan filling}
(sometimes \emph{Kan composition}) operator.

We write an application of the filling operator as $\ffill{e}{r}{j}{\phi}{u}$,
where $u$ is a cell, $j.\boundary{\phi}$ is a boundary varying in a dimension
variable $j$, $e$ is an endpoint, and $r$ is an atomic dimension term. For the
operator to be well-formed $u$ must have boundary $\phi[\inclface{j}{e}]$, while
the resulting cell has boundary $\phi[\inclface{j}{r}]$; thus we think of
$\ffillop$ as deforming the boundary of $u$ from $\phi[\inclface{j}{e}]$ to
$\phi[\inclface{j}{r}]$.
The fact that we fill to a dimension term $r$ means
that the operation unifies $\mathsf{hcomp}$/$\mathsf{hfill}$ of
\cite{CoquandHuberMortberg18}, while being a special case of the more general
$\mathsf{hcom}$ of \cite{angiuli18,FootballHockeyLeague19}.

As an example, consider the context
$\isof{p}{i},\isof{q}{j}[\assign{j}{\izero}{p(\ione)}]$ with $p$ and $q$ such
that the $\ione$-endpoint of $p$ lines up with the $\izero$-endpoint of
$q$. Suppose we want to concatenate them and produce a cell with boundary
$\boundary{\assign{i}{\izero}{p(\izero)} \se
  \assign{i}{\ione}{q(\ione)}}$. Observe that the boundary
$\boundary{\assign{i}{\izero}{p(\izero)} \se \assign{i}{\ione}{q(j)}}$ varying
in $j$ is the boundary of $p$ at ``time'' $j = \izero$ and of our desired
concatenation at ``time'' $j = \ione$. Thus, deforming $p$ with $\ffillop$ can
give us our goal:
\begin{mathpar}
  \dimsquare{i}{j}
  \and
  \begin{tikzpicture}[scale=.5]
    \draw[->,>=stealth] (0,0) -- (3,0) node [midway,below, fill=none] {$p(\izero)$};
    \draw[->,>=stealth] (0,0) -- (0,3) node [midway,left, fill=none] {$p(i)$};
    \draw[->,>=stealth,dashed] (3,0) -- (3,3) node [midway,right, fill=none] {$\ffill{\izero}{\ione}{j}{\assign{i}{\izero}{p(\izero)} \se \assign{i}{\ione}{q(j)}}{p(i)}$};
    \draw[->,>=stealth] (0,3) -- (3,3) node [midway,above, fill=none] {$q(j)$};
  \end{tikzpicture}
\end{mathpar}
We write $(\pcomp{p}{q})(i)$ for the filler above.  If we replace the target
$\ione$ of the $\ffillop$ with a variable, we get a 2-dimensional cell serving
as an interior for the depicted square. One says that this cell \emph{fills} the
open box formed by the solid paths, hence the name. We define a \emph{Kan cell}
to be one formed by iterated applications of the filling operator to contorted cells.

\begin{defi}
  \label{defi:kan-cells}
  The \textbf{Kan cells} $\isfterm[\Gamma][\Psi]{t}$ are inductively generated by
  the rules
  \begin{mathpar}
    \inferrule
    {\iscterm[\Gamma][\Psi]{t}}
    {\isfterm[\Gamma][\Psi]{t}}
    \\
    \inferrule
    {e \in \{\izero,\ione\} \\
      \isdim[\Psi]{r} \\
      \isfbdy[\Gamma][\Psi][i]{\phi} \\
      \isfterm[\Gamma][\Psi]{u}<\face{\phi}{\inclface{i}{e}}>}
    {\isfterm[\Gamma][\Psi]{\ffill{e}{r}{i}{\phi}{u}}<\face{\phi}{\inclface{i}{r}}, \assign{r}{e}{u}>}
  \end{mathpar}
  where, as in \autoref{def:contorted-cell}, we write
  $\isfterm[\Gamma][\Psi]{t}<\phi>$ as shorthand to mean that
  $\isfterm[\Gamma][\Psi]{t}$ and $t$ agrees with $\isfbdy[\Gamma][\Psi]{\phi}$.
  All Kan cells are considered equal in the inconsistent dimension context
  $\bot$.  The \textbf{Kan boundaries} $\isfbdy[\Gamma][\Psi][\Psi']{\phi}$ are
  defined analogously to the contorted boundaries
  $\iscbdy[\Gamma][\Psi][\Psi']{\phi}$.
\end{defi}

\begin{rem}
  Note that the $\assign{r}{e}{u}$ constraint makes filling in direction $e \to
  e$ the identity function. This ensures that the face opposite the missing side
  in the filler agrees with the input $u$ to the filling operator.
\end{rem}

\begin{rem}
  \label{rem:kan-error}
  In \cite{dore24_autom_bound_fillin_cubic_agda}, the rule we gave for the
  $\ffillop$ constructor had the premise $\isfbdy[\Gamma][\Psi,i]{\phi}$ in
  place of $\isfbdy[\Gamma][\Psi][i]{\phi}$. This is incorrect: it would allow
  us for every $\isfterm[\Gamma][\Psi]{t,u}$ to construct a path
  \[
    \isfterm[\Gamma][\Psi,i]{\ffill{\izero}{i}{j}{\assign{j}{\ione}{u}}{t}}<\assign{i}{\izero}{t} \se \assign{i}{\ione}{u}>
  \]
  between them.  Fortunately, our results and implementation from
  \cite{dore24_autom_bound_fillin_cubic_agda} did not actually make use of this
  error.
\end{rem}

\section{Complexity of Contortion Solving and Undecidability of Kan Filling}
\label{sec:complexity}

After having specified what solutions to boundary problems look like, we will
now classify the different kinds of problems that we study in this paper. We
will first look at the complexities of contortion solving for the different
contortion theories that we introduced, and then show that the problem of
finding Kan fillers is undecidable.
\vspace{-.5em}

\subsection{Complexity of contortion solving}

Let us formally introduce the contortion problem for the different contortion
theories that we study.
\vspace{-.5em}

\begin{prob}[\Cartesian/\Disjunctive/\Dedekind/\DeMorgan] Given
  $\iscbdy[\Gamma][\Psi]{\phi}$,

  \begin{itemize}
  \item the problem $\Cartesian(\Gamma,\Psi,\phi)$ is to determine if there
    exists a cartesian contortion $\isisubst{\psi}{\Psi}{\Psi'}$ such that
    $\iscterm[\Gamma][\Psi]{a(\psi)}<\phi>$ for some variable
    $\isof{a}{\Psi'}[\phi']$ in $\Gamma$.
  \item the problem $\Disjunctive(\Gamma,\Psi,\phi)$ is to determine if there
    exists a disjunctive contortion $\isisubst{\psi}{\Psi}{\Psi'}$ such that
    $\iscterm[\Gamma][\Psi]{a(\psi)}<\phi>$ for some variable
    $\isof{a}{\Psi'}[\phi']$ in $\Gamma$.
  \item the problem $\Dedekind(\Gamma,\Psi,\phi)$ is to determine if there
    exists a Dedekind contortion $\isisubst{\psi}{\Psi}{\Psi'}$ such that
    $\iscterm[\Gamma][\Psi]{a(\psi)}<\phi>$ for some variable
    $\isof{a}{\Psi'}[\phi']$ in $\Gamma$.
  \item the problem $\DeMorgan(\Gamma,\Psi,\phi)$ is to determine if there
    exists a De Morgan contortion $\isisubst{\psi}{\Psi}{\Psi'}$ such that
    $\iscterm[\Gamma][\Psi]{a(\psi)}<\phi>$ for some variable
    $\isof{a}{\Psi'}[\phi']$ in $\Gamma$.
  \end{itemize}
\end{prob}
\noindent 
All four problems are decidable: there are finitely many cell variables in
$\Gamma$ and all contortion theories that we consider are finite, so we can try
all possible contortions of each cell variable by brute-force.

Moreover, we can efficiently recognise a solution if we are given one. For this,
we need to decide equality between contorted cells, for which we need to
normalise a contorted cell by looking at its boundary, if it is specified. For
example, in context (\ref{inverse}) the cell $p(\izero)$ normalises to $a$. The
number of such normalisation steps is bounded by the length of the context.

\begin{prop} \label{prop:np} $\Cartesian(\Gamma,\Psi,\phi)$,
  $\Disjunctive(\Gamma,\Psi,\phi)$, $\Dedekind(\Gamma,\Psi,\phi)$ and
  $\DeMorgan(\Gamma,\Psi,\phi)$ are in NP as functions of $\Gamma$, $\Psi$, and
  $\phi$.
\end{prop}
\begin{proof}
  For any contortion theory, we can determine in polynomial time if a given
  variable $\isof{a}{\Psi'}[\phi']$ and contortion
  $\isisubst{\psi}{\Psi}{\Psi'}$ form a solution to a boundary problem, i.e.,
  that $\iscterm[\Gamma][\Psi]{a(\psi)}<\phi>$, by normalising $a(\psi)$ and
  $\phi$ and comparing. Note that the complexity of our check increases
  polynomially with the number of cells in $\Gamma$; the dimension variables in
  $\Psi$; and the specified faces of $\phi$.
\end{proof}

Boundary problems have multiple parameters; for more detailed analysis of the
complexity of the different contortion problems we will in the following fix
some of them. In our contortion solver in \S\ref{sec:contortions}, we will
primarily study boundary problems over some small fixed cell context $\Gamma$,
and try to contort a given cell into ever higher dimensions. It hence makes
sense to study the contortion problems only with respect to the number of
variables in $\Psi$, which in turn determines the size of the goal boundary
$\phi$.

\begin{prop}
  For any $\Gamma$, $\Cartesian(\Gamma,\Psi,\phi)$ is in P as a function of
  $\Psi$ and $\phi$.
\end{prop}
\begin{proof}
  For any $\isof{a}{\Psi'}[\phi']$ in $\Gamma$, there are $(n + 2)^m$ many
  cartesian contortions for $m \coloneq \card{\Psi'}$ and $n = \card{\Psi}$.
  Since we treat the cell context (and therefore also the dimension $m$ of each
  of its cells) as constant, there are polynomially many contortions that we
  need to check.
\end{proof}

If we include disjunctions into our dimension terms, we have $(2^n + 1)^{m}$
contortions of an $m$-dimensional cell into $n$ dimensions, which means
brute-force is not polynomial for \Disjunctive, even for fixed cell contexts.
However, enumerating all \Disjunctive contortions can still feasibly be done
also in higher dimensions---in contrast to the Dedekind and De Morgan theories,
whose sizes explode with growing $n$. In the case of $\Dedekind$, the number of
ways to contort an $m$-cube to fit an $n$-dimensional goal is $D(n)^m$ where
$D(n)$ is the $n$-th \emph{Dedekind number}~\cite[App.~B]{awodey2023cartesian}.
The Dedekind numbers grow extremely quickly: there are $D(6) = 7\,828\,354$ many
ways to contort a $1$-cube into a $6$-dimensional cube; the $42$-digit $D(9)$
was only recently computed using supercomputing~\cite{vanhirtum,jaekel}. The
problems \Dedekind and \DeMorgan thus seem to be computationally very hard, and
our focus in \S\ref{sec:contortions} will be on heuristics that quickly yield
solutions to boundary problems that appear in practice, rather than on
worst-case asymptotics. The following results give some indication of this
difficulty.

\begin{prop}\label{prop:dedekindconp}
  There exist $\Gamma$ for which $\Dedekind(\Gamma,\Psi,\phi)$ is coNP-hard as a
  function of $\Psi$ and $\phi$.
\end{prop}
\begin{proof}
  We give a reduction from the entailment problem for monotone Boolean formulas,
  which is equivalent to the equivalence problem for monotone Boolean formulas,
  which is known to be coNP-complete \cite[Theorem 15]{reith03}.

  Given two monotone formulas $\varphi_0$ and $\varphi_1$ over variables
  $\vec{x} = x_1,\ldots,x_n$, we want to decide whether
  $\varphi_0 \vDash \varphi_1$. Note that we can treat each $\varphi_i$ as a
  Dedekind dimension term $\iscdim[\vec{x}]{\varphi_i}$ by reading $\bot$ and
  $\top$ as $\izero$ and $\ione$ and disjunction and conjunction as $\lor$ and
  $\land$. We claim that $\varphi_0 \vDash \varphi_1$ iff
  the following boundary problem is solvable:
  \[
  \iscterm[\isof{s}{j}][l,\vec{x}]{\hole}< \assign{l}{\izero}{
    s(\varphi_0) } \se \assign{l}{\ione}{ s(\varphi_1)} >.
  \]

  Suppose $\varphi_0 \vDash \varphi_1$. Then we can define a Dedekind contortion
  $\isisubst{\psi}{(l,\vec{x})}{(j)}$ as
  $\psi(l,\vec{x}) = (\varphi_0(\vec x) \vee (l \land \varphi_1(\vec x)))$.  The
  contorted cell $s(\psi)$ then solves the boundary problem.

  Conversely, if $\varphi_0 \nvDash \varphi_1$, then there is an assignment
  $\vec{e}$ to the variables $\vec{x}$ such that $\varphi_0(\vec{e}) = \ione$
  but $\varphi_1(\vec{e}) = \izero$. But then any contortion
  $\isisubst{\psi}{(l,\vec{x})}{(j)}$ which would contort $s$ into the goal
  would be non-monotone, with
  $\psi(\izero,\vec{e}) = \varphi_0(\vec{e}) > \varphi_1(\vec{e}) =
  \psi(\ione,\vec{e})$, which is impossible with a Dedekind contortion.
\end{proof}

Another perspective on contortion problems is given by considering not the cell
context as constant, but instead the dimension of the goal boundary. Even
restricting to 1-dimensional goals, contortion problems are NP-hard as soon as
the contortion language includes two connections, which underlines that the
contortion problems \Dedekind and \DeMorgan are very complex.

\begin{prop}
  For $\Psi$ containing at least one variable, $\Dedekind(\Gamma,\Psi,\phi)$ is
  NP-complete as a function of $\Psi$ and $\phi$.
\end{prop}
\begin{proof}
  Membership in NP follows from Proposition~\ref{prop:np}. For completeness, we
  give a reduction from SAT. Suppose we have a Boolean CNF formula $\varphi$
  over $\vec{x} = x_1,\ldots,x_n$. Replace each $\lnot x_i$ in $\varphi$ by a
  variable $y_i$ to obtain a dimension term $r$ in variables $\vec{x},\vec{y}$.
  Then $\varphi$ is satisfiable if and only if there is
  $\isisubst{\psi}{()}{(\vec{x},\vec{y})}$ such that $\cont{r}{\psi} = \ione$
  and $\cont{(x_k \land y_k)}{\psi} = \izero$ and $\cont{(x_k \lor y_k)}{\psi} =
  \ione$ for each $k$. Take $\Gamma_\varphi$ to be the context
  \[
    \textstyle
    \isof{a},
    \isof{p}{z,j_0,j_1},
    \isof{q}{\vec{x},\vec{y},i}[
      \assign{i}{\izero}{a} \se
      \assign{i}{\ione}{p\left(r,\bigvee_k (x_k \land y_k), \bigwedge_k (x_k \lor y_k)\right)}
    ]
  \]
  and consider the boundary problem
  $
    \iscterm[\Gamma_\varphi][i]{\hole}<
      \assign{i}{\izero}{a} \se
      \assign{i}{\ione}{p(\ione,\izero,\ione)}
    >
  $.
  Any $\isisubst{\psi}{()}{(\vec{x},\vec{y})}$ such that
  $\cont{r}{\psi} = \ione$ and $\cont{(x_k \land y_k)}{\psi} = \izero$ and
  $\cont{(x_k \lor y_k)}{\psi} = \ione$ for each $k$ yields a solution
  $\iscterm[\Gamma_\varphi][i]{q(\psi, i)}$. Conversely, any
  solution to the problem will be of the form
  $\iscterm[\Gamma_\varphi][i]{q(\psi',r)}$ for some
  $\isisubst{\psi'}{i}{(\vec{x},\vec{y})}$ and $\iscdim[i]{r}$, in which case
  $\isisubst{\psi'(\ione)}{()}{(\vec{x},\vec{y})}$ induces a satisfying
  assignment for $\varphi$.
\end{proof}

The same reduction also works to establish that $\DeMorgan(\Gamma,\Psi,\phi)$ is
NP-hard. In fact, the proof can be slightly simplified as the negated variables do not have to be replaced first.
\begin{cor}
  For $\Psi$ with at least one variable, $\DeMorgan(\Gamma,\Psi,\phi)$ is NP-complete as a function of $\Psi$ and $\phi$.
\end{cor}
\noindent 
In summary, contortion problems are, even if decidable, not necessarily
tractable for the more complicated contortion theories that we consider.

\subsection{Undecidability of Kan solving}
\label{sec:undecidability}

The Kan filling problem has---in contrast to the contortion problems---an
infinite search space, and we will in the following establish that it is
undecidable. This result is independent of which underlying contortion theory
one considers. Let us formally introduce the problem of finding Kan cells.

\begin{prob}[\KanCubicalCell]
  Given a Kan boundary $\isfbdy[\Gamma][\Psi]{\phi}$, the problem
  $\KanCubicalCell(\Gamma,\Psi,\phi)$ is to determine if there exists a Kan cell $t$ such that
  $\isfterm[\Gamma][\Psi]{t}<\phi>$.
\end{prob}

For example, the problem (\ref{short-inverse}) of inverting a path does not have a
solution in \Dedekind but does have solutions in \KanCubicalCell, such as
$\ffill{\izero}{\ione}{i}{\assign{j}{\izero}{p(i)} \se \assign{j}{\ione}{p(\izero)}}{p(\izero})$.

Unlike contortion solving, deciding whether a Kan solution to a problem exists
is not only difficult but actually impossible in general. Intuitively, Kan
solving is a higher-dimensional generalisation of a more familiar undecidable
problem: the word problem for a finitely presented group. This is the problem of
deciding, for finite sets $X$ of generators and $R$ of equations, whether two
words on $X$ are equal in the free group on $X$ modulo $R$. In Kan solving, the
context $\Gamma$ can be thought of as a collection of generators in which each
$(n+1)$-dimensional cell serves as an ``equation'' between $n$-dimensional
cells, while Kan filling generalises the multiplication and inverse operations
available in a group.

We now make this precise by giving a reduction from the word problem for a given
finitely presented group to Kan solving over a corresponding context.  This
argument applies to Kan solving relative to any of the sublanguages of
contortions (cartesian, disjunctive, Dedekind, De Morgan) we have introduced.
As a side effect, we will get to see some more complex constructions in the
cubical type theory we have defined.

A group presentation $\present{X}{R}$ consists of a finite set $X$, the
\emph{generators}, and a finite set $R$ of equations of the form $w = 1$ where
$w$ is a word on $X$, i.e., a finite list of the form
$x_{0}^{\alpha_{0}},\ldots,x^{\alpha_{k}}_{k}$ where each $x_i$ is in $X$ and
each $\alpha_i$ is $-1$ or $1$. The group $G$ presented by $\present{X}{R}$ is
the free group on $X$ modulo the equations in $R$; given words $w,v$ over $X$,
we write $w \equiv_G v$ to mean that they represent equal elements of $G$.
Before beginning the reduction, we first show that we can assume the equations
in $R$ are of a more restricted form.

\begin{defi}
  Say a finite presentation $\present{X}{R}$ of a group $G$ is \emph{convenient} when
  \begin{itemize}
  \item $X$ is closed under inverses;
  \item every equation in $R$ is of the form $abc^{-1} = 1$ for some $a,b,c \in X$.
  \end{itemize}
\end{defi}

\begin{prop}
  Let $G$ be a finitely presented group. Then $G$ has a convenient presentation.
\end{prop}
\begin{proof}
  Suppose that $G$ is presented by a finite set of generators $X$ and a finite set of equations
  $y_{i,0}^{\alpha_{i,0}},\ldots,y^{\alpha_{i,k_i}}_{i,k_i} = 1$ for $0 \le i < n$, where for each $i$ we have $k_i \in \mathbb{N}$ and then $y_{i,j} \in X$ and $\alpha_{i,j} \in \braces{-1,1}$ for $0 \le j \le k_i$.

  For each $0 \le i < n$ and $0 \le j \le k_i + 1$, define $z_{i,j} \coloneq y_{i,0}^{\alpha_{i,0}},\ldots,y^{\alpha_{i,k_i}}_{i,j-1} \in G$. Then $G$ is presented by the set of generators $X \cup \set{z_{i,j}}{0 \le i < n, 0 \le j \le k_i}$ and equations
  \[
    \begin{array}{lcrl}
      z_{i,0}z_{i,0} &=& z_{i,0} &\text{for $0 \le i < m$} \\
      z_{i,j}y_{i,j} &=& z_{i,j+1} &\text{for $0 \le i < m$ and $0 \le j \le k_i$ with $\alpha_{i,j+1} = 1$} \\
      z_{i,j+1}y_{i,j} &=& z_{i,j} &\text{for $0 \le i < m$ and $0 \le j \le k_i$ with $\alpha_{i,j+1} = -1$} \\
      z_{i,k_i}z_{i,k_i} &=& z_{i,k_i} &\text{for $0 \le i < m$}
    \end{array}
  \]
  Note that the first and last equations encode that $z_{i,0} = 1$ and $z_{i,k_i} = 1$ for $0 \le i < m$ respectively.
\end{proof}

We encode a (conveniently) finitely presented group $\present{X}{R}$ as a
context with a single point $\star$. Each generator is encoded as a 1-cell, namely
a path from $\star$ to itself, and each equation as a 2-cell.

\begin{defi}
  In this section, we use $\assignbdy{i}{t}$ as a shorthand for the pair of boundary entries $\assign{i}{\izero}{t} \se \assign{i}{\ione}{t}$.
\end{defi}

\begin{defi}
  Given a convenient presentation $\present{X}{R}$, define the context
  $\isctxt{\presentctxt{X}{R}}$ to consist of
  \begin{itemize}
  \item a point $\isof{\star}$,
  \item a loop $\isof{\hat a}{i}[\assignbdy{i}{\star}]$ for each $a \in X$,
  \item a square
    \[
      \isof{s_{a,b,c}}{j,k}[\assign{k}{\izero}{\hat a(j)} \se \assign{k}{\ione}{\hat c(j)} \se \assign{j}{\izero}{\star} \se \assign{j}{\ione}{\hat b(k)}]
    \]
    for each equation $abc^{-1} = 1$ in $R$:
    \[
      \dimsquare{j}{k}
      \begin{tikzpicture}[xscale=.7,yscale=.4]
        \draw[->,>=stealth] (0,0) -- (4,0) node [midway,below, fill=none] {$\star$};
        \draw[->,>=stealth] (0,0) -- (0,4) node [midway,left, fill=none] {$\hat a(j)$};
        \draw[->,>=stealth] (4,0) -- (4,4) node [midway,right, fill=none] {$\hat c(j)$};
        \draw[->,>=stealth] (0,4) -- (4,4) node [midway,above, fill=none] {$\hat b(k)$};
        \node at (2,2) {$s_{a,b,c}(j,k)$};
      \end{tikzpicture}
    \]
  \end{itemize}
\end{defi}
\noindent 
Any word on $X$ can then be encoded as a path from $\star$ to $\star$ in the context
$\isctxt{\presentctxt{X}{R}}$.

\begin{defi}
  Let $\present{X}{R}$ be a convenient presentation of a group, $a \in X$ be a generator, and
  $\isfterm[\presentctxt{X}{R}][i]{t}<\assignbdy{i}{\star}>$ be a cell.
  For $e \in \braces{\izero,\ione}$, define cells
  \begin{mathpar}
    \isfterm[\presentctxt{X}{R}][i]{t \rhd^e_i a}<\assignbdy{i}{\star}>
    \and
    \isfterm[\presentctxt{X}{R}][i,\ell]{t \blacktriangleright^e_{i,\ell} a}<\assign{i}{\izero}{\star} \se \assign{i}{\ione}{\hat a(\ell)} \se \assign{\ell}{\izero}{t} \se \assign{\ell}{\ione}{t \rhd^e_i a}>
  \end{mathpar}
  by
  \begin{align*}
    t \blacktriangleright^e_{i,\ell} a &\coloneq \ffill{\negI{e}}{\ell}{j}{\assign{i}{\izero}{\star} \se \assign{i}{\ione}{\hat a(j)}}{t} \\
    t \rhd^e_i a &\coloneq t \blacktriangleright^e_{i,e} a
  \end{align*}
  as pictured below.
  \begin{mathpar}
    \dimsquare{i}{\ell}
    \quad
    \begin{tikzpicture}[xscale=.5,yscale=.4]
      \draw[->,>=stealth] (0,0) -- (4,0) node [midway,below, fill=none] {$\star$};
      \draw[->,>=stealth] (0,0) -- (0,4) node [midway,left, fill=none] {$t$};
      \draw[->,>=stealth,dashed] (4,0) -- (4,4) node [midway,right, fill=none] {$t \rhd^\ione_i a$};
      \draw[->,>=stealth] (0,4) -- (4,4) node [midway,above, fill=none] {$\hat a(\ell)$};
      \node at (2,2) {$t \blacktriangleright^\ione_{i,\ell} a$};
    \end{tikzpicture}
    \quad\quad
    \begin{tikzpicture}[xscale=.5,yscale=.4]
      \draw[->,>=stealth] (0,0) -- (4,0) node [midway,below, fill=none] {$\star$};
      \draw[->,>=stealth,dashed] (0,0) -- (0,4) node [midway,left, fill=none] {$t \rhd^\izero_i a$};
      \draw[->,>=stealth] (4,0) -- (4,4) node [midway,right, fill=none] {$t$};
      \draw[->,>=stealth] (0,4) -- (4,4) node [midway,above, fill=none] {$\hat a(\ell)$};
      \node at (2,2) {$t \blacktriangleright^\ione_{i,\ell} a$};
    \end{tikzpicture}
  \end{mathpar}
\end{defi}

\begin{defi}
  Let $\present{X}{R}$ be a finite presentation of a group.
  For each word $w$ on $X$, define a cell $\isfterm[\presentctxt{X}{R}][i]{\tmword{w}{i}}<\assignbdy{i}{\star}>$ by recursion on $w$ as follows.
  \begin{align*}
    \tmword{\epsilon}{i} &= \star \\
    \tmword{wa}{i} &= \tmword{w}{i} \rhd^\ione_i a \\
    \tmword{wa^{-1}}{i} &= \tmword{w}{i} \rhd^\izero_i a
  \end{align*}
\end{defi}

Now we show that when two words represent the same element of the presented group, their encodings as paths are related by a 2-cell.
First, we prove a lemma corresponding to cancellation of inverses.

\begin{defi}
  \label{defi:equation-word-cancel}
  Let $\present{X}{R}$ be a convenient presentation of a group, $a \in X$ be a generator, and
  $\isfterm[\presentctxt{X}{R}][i]{t}<\assignbdy{i}{\star}>$ be a cell.
  For $e \in \braces{\izero,\ione}$, define the cell
  \[
    \isfterm[\presentctxt{X}{R}][i,k]{\tmcancel{t}{e}{i}{a}{k}}<\assignbdy{i}{\star} \se \assign{k}{\izero}{(t \rhd^e_i a) \rhd^{\overline{e}}_i a} \se \assign{k}{\ione}{t}>
  \]
  as follows.
  \[
    \tmcancel{t}{e}{i}{a}{k} \coloneq
    \ffill*{e}{\overline{e}}{\ell}{
      \begin{array}{l}
        \assign{i}{\izero}{\star} \\
        \assign{i}{\ione}{\hat a(\ell)} \\
        \assign{k}{\izero}{(t \rhd^e_i a) \blacktriangleright^{\overline{e}}_{i,\ell} a} \\
        \assign{k}{\ione}{t \blacktriangleright^e_{i,\ell} a}
      \end{array}
    }{(t \rhd^e_i a)}
  \]
  In the case $e = \izero$, this is the front face of the filler for the open cube pictured below.
  \[
    \dimcube{i}{k}{\ell}
    \hspace{-3em}
    \begin{tikzpicture}[xscale=7,yscale=2.5]
      \tikzmath{\x1=.2; \dx=.5;}
      \tikzmath{\y1=.4; \dy=.5;}
      \tikzmath{\o=1.6;}
      \tikzmath{\x2=\x1+\dx;}
      \tikzmath{\y2=\y1+\dy;}
      \tikzmath{\xp1 = (\x1+\x2)/2;}
      \tikzmath{\yp1 = (\y1+\y2)/2;}

      \draw[->,>=stealth] (-\x2,-\y2) -- ( \x2,-\y2) node [midway,below, fill=none] {$\star$};
      \draw[->,>=stealth] (-\x2, \y2) -- ( \x2, \y2) node [midway,above, fill=none] {$\star$};
      \draw[->,>=stealth] (-\x2,-\y2) -- (-\x2, \y2) node [midway,left, fill=none] {$(t \rhd^\izero_i a) \rhd^{\ione}_i a$};
      \draw[->,>=stealth] ( \x2,-\y2) -- ( \x2, \y2) node [midway,right,fill=none] {$t$};

      \draw[->,>=stealth] (-\x1,-\y1) -- ( \x1,-\y1) node [midway,fill=white,text=black!80] {$\scriptstyle \star$};
      \draw[->,>=stealth] (-\x1, \y1) -- ( \x1, \y1) node [midway,fill=white,text=black!80] {$\scriptstyle \star$};
      \draw[->,>=stealth] (-\x1,-\y1) -- (-\x1, \y1) node [midway,fill=white,text=black!80] {$\scriptstyle t \rhd^\izero_i a$};
      \draw[->,>=stealth] ( \x1,-\y1) -- ( \x1, \y1) node [midway,fill=white,text=black!80] {$\scriptstyle t \rhd^\izero_i a$};

      \draw[->,>=stealth] (-\x1,-\y1) -- (-\x2,-\y2) node [midway,fill=white,text=black!80] {$\scriptstyle \star$};
      \draw[->,>=stealth] ( \x1,-\y1) -- ( \x2,-\y2) node [midway,fill=white,text=black!80] {$\scriptstyle \star$};
      \draw[->,>=stealth] ( \x1, \y1) -- ( \x2, \y2) node [midway,fill=white,text=black!80] {$\scriptstyle \hat a(\ell)$};
      \draw[->,>=stealth] (-\x1, \y1) -- (-\x2, \y2) node [midway,fill=white,text=black!80] {$\scriptstyle \hat a(\ell)$};

      \node at (0,-\yp1) {$\star$};
      \node at (0, \yp1) {$\hat a(\ell)$};
      \node at (-\xp1,0) {$(t \rhd^{\izero}_i a) \blacktriangleright^{\ione}_{i,\ell} a$};
      \node at ( \xp1,0) {$t \blacktriangleright^\izero_{i,\ell} a$};
      \node at (0,0) {$t \rhd_i^\izero a$};
    \end{tikzpicture}
  \]
\end{defi}

Next we construct cells corresponding to the equations in $R$.  For this we will
make use of the following auxiliary construction.

\begin{defi}[Pseudo-$\lor$]
  Given a cell $\isfterm[\Gamma][i]{t}<\assign{i}{\izero}{u} \se \assign{i}{\ione}{v}>$, define the cell
  \[
    \isfterm*[\Gamma][j,k]{\tmcnxor{i.t}{j}{k}}<%
      \begin{array}{lll}
        \assign{j}{\izero}{\face{t}{\inclface{i}{k}}} &\se& \assign{j}{\ione}{v} \\
        \assign{k}{\izero}{\face{t}{\inclface{i}{j}}} &\se& \assign{k}{\ione}{v}
      \end{array}
    >
  \]
  to be
  \[
    \ffill*{\izero}{\ione}{\ell}{
      \begin{array}{l}
        \assign{j}{\izero}{\ffill*{\ione}{k}{m}{\assign{\ell}{\izero}{u} \se \assign{\ell}{\ione}{\face{t}{\inclface{i}{m}}}}{\face{t}{\inclface{i}{\ell}}}} \\
        \assign{k}{\izero}{\ffill*{\ione}{j}{m}{\assign{\ell}{\izero}{u} \se \assign{\ell}{\ione}{\face{t}{\inclface{i}{m}}}}{\face{t}{\inclface{i}{\ell}}}} \\
        \assign{j}{\ione}{\face{t}{\inclface{i}{\ell}}} \\
        \assign{k}{\ione}{\face{t}{\inclface{i}{\ell}}}
      \end{array}}{u}
  \]
  which is the front face of the filler for the open cube pictured below.
  \[
    \dimcube{j}{k}{\ell}
    \hspace{-2em}
    \begin{tikzpicture}[xscale=4,yscale=2.5]
      \tikzmath{\x1=.2; \dx=.5;}
      \tikzmath{\y1=.4; \dy=.5;}
      \tikzmath{\o=1.6;}
      \tikzmath{\x2=\x1+\dx;}
      \tikzmath{\y2=\y1+\dy;}
      \tikzmath{\xp1 = (\x1+\x2)/2;}
      \tikzmath{\yp1 = (\y1+\y2)/2;}

      \draw[->,>=stealth] (-\x2,-\y2) -- ( \x2,-\y2) node [midway,below, fill=none] {$\face{t}{\inclface{i}{k}}$};
      \draw[->,>=stealth] (-\x2, \y2) -- ( \x2, \y2) node [midway,above, fill=none] {$v$};
      \draw[->,>=stealth] (-\x2,-\y2) -- (-\x2, \y2) node [midway,left, fill=none] {$\face{t}{\inclface{i}{j}}$};
      \draw[->,>=stealth] ( \x2,-\y2) -- ( \x2, \y2) node [midway,right,fill=none] {$v$};

      \draw[->,>=stealth] (-\x1,-\y1) -- ( \x1,-\y1) node [midway,fill=white,text=black!80] {$\scriptstyle u$};
      \draw[->,>=stealth] (-\x1, \y1) -- ( \x1, \y1) node [midway,fill=white,text=black!80] {$\scriptstyle u$};
      \draw[->,>=stealth] (-\x1,-\y1) -- (-\x1, \y1) node [midway,fill=white,text=black!80] {$\scriptstyle u$};
      \draw[->,>=stealth] ( \x1,-\y1) -- ( \x1, \y1) node [midway,fill=white,text=black!80] {$\scriptstyle u$};

      \draw[->,>=stealth,dashed] (-\x1,-\y1) -- (-\x2,-\y2);
      \draw[->,>=stealth] ( \x1,-\y1) -- ( \x2,-\y2) node [midway,fill=white,text=black!80] {$\scriptstyle \face{t}{\inclface{i}{\ell}}$};
      \draw[->,>=stealth] ( \x1, \y1) -- ( \x2, \y2) node [midway,fill=white,text=black!80] {$\scriptstyle \face{t}{\inclface{i}{\ell}}$};
      \draw[->,>=stealth] (-\x1, \y1) -- (-\x2, \y2) node [midway,fill=white,text=black!80] {$\scriptstyle \face{t}{\inclface{i}{\ell}}$};

      \node at (0,-\yp1) {$\mathsf{fill}^{\ione\to k}$};
      \node at (0, \yp1) {$\face{t}{\inclface{i}{\ell}}$};
      \node at (-\xp1,0) {$\mathsf{fill}^{\ione\to j}$};
      \node at ( \xp1,0) {$\face{t}{\inclface{i}{\ell}}$};
      \node at (0,0)  {$u$};
    \end{tikzpicture}
  \]
\end{defi}

Note that if we are working relative to the disjunctive, Dedekind, or De Morgan
contortion theory, then there is a much simpler construction of a cell with the
same boundary when $t$ is a contorted cell, namely
$\tmcnxor{i.t}{j}{k} \coloneq \cont{t}{\isubstface{i}{j \lor k}}$ (cf.\
\eqref{or-connection-cell}).

\begin{defi}
  \label{defi:equation-word-rewrite}
  Let $\present{X}{R}$ be a convenient presentation of a group, $abc^{-1} = 1$ be an equation in $R$, and $\isfterm[\presentctxt{X}{R}][i]{t}<\assignbdy{i}{\star}>$ be a cell.
  Define the cell
  \[
    \isfterm[\presentctxt{X}{R}][i,k]{\tmrewrite{t}{a,b,c}{i}{k}}<\assignbdy{i}{\star} \se \assign{k}{\izero}{(t \rhd^{\ione}_i a) \rhd^{\ione}_i b} \se \assign{k}{\ione}{t \rhd^{\ione}_i c}>
  \]
  to be
  \[
    \ffill*{\izero}{\ione}{j}{
      \begin{array}{l}
        \assign{i}{\izero}{\star} \\
        \assign{i}{\ione}{\tmcnxor{i.\hat{b}(i)}{j}{k}} \\
        \assign{k}{\izero}{(t \rhd^{\ione}_i a) \blacktriangleright^{\ione}_{i,j} b} \\
        \assign{k}{\ione}{t \rhd^{\ione}_i c}
      \end{array}
    }{
      \left(\ffill*{\izero}{\ione}{j}{
        \begin{array}{l}
          \assign{i}{\izero}{\star} \\
          \assign{i}{\ione}{s_{a,b,c}(j,k)} \\
          \assign{k}{\izero}{t \blacktriangleright^{\ione}_{i,j} a} \\
          \assign{k}{\ione}{t \blacktriangleright^{\ione}_{i,j} c}
        \end{array}
      }{t}\right)
    }
  \]
  which is the iterated composite pictured below.

  \[
    \dimcube{i}{k}{j}
    \hspace{-2em}
    \begin{tikzpicture}[xscale=3.5,yscale=1.9]
      \tikzmath{\x1=.2; \dx=.5;}
      \tikzmath{\y1=.4; \dy=.5;}
      \tikzmath{\o=1.6;}
      \tikzmath{\x2=\x1+\dx; \x3=\x2+\o*\dx;}
      \tikzmath{\y2=\y1+\dy; \y3=\y2+\o*\dy;}
      \tikzmath{\xp1 = (\x1+\x2)/2; \xp2 = (\x2+\x3)/2;}
      \tikzmath{\yp1 = (\y1+\y2)/2; \yp2 = (\y2+\y3)/2;}
      \draw[->,>=stealth] (-\x3,-\y3) -- (\x3,-\y3) node [midway,below, fill=none] {$\star$};
      \draw[->,>=stealth] (-\x3,\y3) -- (\x3,\y3) node [midway,above, fill=none] {$\star$};
      \draw[->,>=stealth] (-\x3,-\y3) -- (-\x3,\y3) node [midway,left, fill=none] {$(t \rhd^{\ione}_i a) \rhd^{\ione}_i b$};
      \draw[->,>=stealth] (\x3,-\y3) -- (\x3,\y3) node [midway,right,fill=none] {$t \rhd^{\ione}_i c$};
      
      \draw[->,>=stealth] (-\x2,-\y2) -- (\x2,-\y2) node [midway,fill=white,text=black!80] {$\scriptstyle \star$};
      \draw[->,>=stealth] (-\x2,\y2) -- (\x2,\y2) node [midway,fill=white,text=black!80] {$\scriptstyle \hat b(k)$};
      \draw[->,>=stealth] (-\x2,-\y2) -- (-\x2,\y2);
      \draw[->,>=stealth] (\x2,-\y2) -- (\x2,\y2);

      \draw[->,>=stealth] (-\x1,-\y1) -- (\x1,-\y1) node [midway,fill=white,text=black!80] {$\scriptstyle \star$};
      \draw[->,>=stealth] (-\x1,\y1) -- (\x1,\y1) node [midway,fill=white,text=black!80] {$\scriptstyle \star$};
      \draw[->,>=stealth] (-\x1,-\y1) -- (-\x1,\y1) node [midway,fill=white,text=black!80] {$\scriptstyle t$};
      \draw[->,>=stealth] (\x1,-\y1) -- (\x1,\y1) node [midway,fill=white,text=black!80] {$\scriptstyle t$};

      \draw[->,>=stealth] (-\x1,-\y1) -- (-\x2,-\y2) node [midway,fill=white,text=black!80] {$\scriptstyle \star$};
      \draw[->,>=stealth] (\x1,-\y1) -- (\x2,-\y2) node [midway,fill=white,text=black!80] {$\scriptstyle \star$};
      \draw[->,>=stealth] (\x1,\y1) -- (\x2,\y2) node [midway,fill=white,text=black!80] {$\scriptstyle \hat c(j)$};
      \draw[->,>=stealth] (-\x1,\y1) -- (-\x2,\y2) node [midway,fill=white,text=black!80] {$\scriptstyle \hat a(j)$};

      \draw[->,>=stealth] (-\x2,-\y2) -- (-\x3,-\y3) node [midway,fill=white,text=black!80] {$\scriptstyle \star$};
      \draw[->,>=stealth] (\x2,-\y2) -- (\x3,-\y3) node [midway,fill=white,text=black!80] {$\scriptstyle \star$};
      \draw[->,>=stealth] (\x2,\y2) -- (\x3,\y3) node [midway,fill=white,text=black!80] {$\scriptstyle \star$};
      \draw[->,>=stealth] (-\x2,\y2) -- (-\x3,\y3) node [midway,fill=white,text=black!80] {$\scriptstyle \hat b(k)$};

      \node at (0, 0)  {$t$};

      \node at (0, -\yp1)  {$\star$};
      \node at (0, \yp1) {$s_{a,b,c}(j,k)$};
      \node at (-\xp1, 0)  {$t \blacktriangleright^\ione_{i,j} a$};
      \node at (\xp1, 0) {$t \blacktriangleright^\ione_{i,j} c$};

      \node at (0, -\yp2)  {$\star$};
      \node at (0, \yp2) {$\tmcnxor{i.\hat{b}(i)}{j}{k}$};
      \node at (-\xp2, 0)  {$(t \rhd^\ione_i a) \blacktriangleright^\ione_{i,j} b$};
      \node at (\xp2, 0) {$t \rhd^\ione_i c$};
    \end{tikzpicture}
  \]
\end{defi}

Combining the previous constructions, we can obtain a cell from any equation in
the presented group.

\begin{prop}
  \label{prop:equal-words-yield-cell}
  Let $\present{X}{R}$ be a convenient presentation of a group $G$.
  For any pair of words $v,w$ on $X$ such that $v \equiv_G w$, there exists a cell
  \begin{equation}
    \label{eqn:loop-path}
    \isfterm[\presentctxt{X}{R}][i,k]{t_{v,w}}<\assignbdy{i}{\star} \se \assign{k}{\izero}{\tmword{v}{i}} \se \assign{k}{\ione}{\tmword{w}{i}}>
  \end{equation}
\end{prop}
\begin{proof}
  The relation $\equiv_G$ on words on $X$ is generated by the clauses
  \begin{enumerate}[label=(\arabic*)]
  \item \label{word-equivalence-rewrite} $wab \equiv_G wc$ whenever $abc^{-1} = 1$ is an equation in $R$.
  \item \label{word-equivalence-cancel} $waa^{-1} \equiv_G w$ and $wa^{-1}a \equiv_G w$ for a word $w$ and $a \in X$;
  \item \label{word-equivalence-snoc} $va \equiv_G wa$ and $va^{-1} \equiv_G wa^{-1}$ for all words $v,w$ with $v \equiv_G w$ and $a \in X$.
  \item \label{word-equivalence-reflexivity} $w \equiv_G w$ for all words $w$;
  \item \label{word-equivalence-symmetry} $w \equiv_G v$ for all words $v,w$ with $v \equiv_G w$;
  \item \label{word-equivalence-transitivity} $u \equiv_G w$ for all words $u,v,w$ with $u \equiv_G v$ and $v \equiv_G w$.
  \end{enumerate}
  It thus suffices to show that we have operations on cells of the form \eqref{eqn:loop-path} corresponding to each clause.
  Clause \ref{word-equivalence-rewrite} is covered by \autoref{defi:equation-word-rewrite} and \ref{word-equivalence-cancel} is covered by \autoref{defi:equation-word-cancel}.
  Clause \ref{word-equivalence-snoc} corresponds to applying $(-) \rhd^{\ione}_i a$ or $(-) \rhd^{\izero}_i a$.
  Clauses \ref{word-equivalence-reflexivity}, \ref{word-equivalence-symmetry}, and \ref{word-equivalence-transitivity} correspond to reflexivity, symmetry, and transitivity of 2-dimensional paths respectively.
  For the latter two, given $t_{u,v}$ and $t_{v,w}$ as in \eqref{eqn:loop-path}, we have
  \[
    t_{w,v} \coloneq
    \ffill*{\izero}{\ione}{j}{
      \begin{array}{l}
        \assignbdy{i}{\star} \\
        \assign{k}{\izero}{\face{t_{v,w}}{\inclface{k}{j}}} \\
        \assign{k}{\ione}{\tmword{v}{i}}
      \end{array}
    }{\tmword{v}{i}}
  \]
  and
  \[
    t_{u,w} \coloneq
    \ffill*{\izero}{\ione}{j}{
      \begin{array}{l}
        \assignbdy{i}{\star} \\
        \assign{k}{\izero}{\tmword{u}{i}} \\
        \assign{k}{\ione}{\face{t_{v,w}}{\inclface{k}{j}}}
      \end{array}
    }{t_{u,v}}
  \]
  respectively.
\end{proof}

Now we define an inverse mapping, sending 1-cells over $\presentctxt{X}{R}$ to
elements of the presented group and 2-cells to equations between them.  Again,
these definitions make sense for any of the contortion theories we consider; we
rely only of the fact that any dimension term in the unit context $()$ is equal
to either $\izero$ or $\ione$. For each definition, we go by structural
induction on the Kan cell term formers; as mentioned in
\autoref{rem:explicit-substitutions}, we treat the syntax here as coming with
explicit substitutions, which simplifies the reasoning.

\begin{defi}
  Fix a convenient presentation $\present{X}{R}$ of a group $G$.
  For each Kan cell $\isfterm[\presentctxt{X}{R}][\Psi]{t}$, we define a family of elements $\gsem{t}{\psi} \in G$ for each substitution $\isincl{\psi}{(i)}{\Psi}$.
  We go by structural induction on $t$ as follows.
  \begin{itemize}
  \item Define $\gsem{\face{t}{\psi'}}{\psi} \coloneq \gsem{t}{\psi'\psi}$.
  \item Define $\gsem{\star}{\psi} \coloneq 1$.
  \item Define $\gsem{\hat a(\psi')}{\psi} \coloneq g_a(\psi'\psi(\izero))^{-1}g_a(\psi'\psi(\ione))$, where $g_a(e)$ is defined for $e \in \braces{\izero,\ione}$ by $g_a(e) = a^e$, i.e.,
    \[
      \begin{array}{ll}
        g_a(\izero) &\coloneq 1 \\
        g_a(\ione) &\coloneq a
      \end{array}
    \]
    Note that this assignment respects cell equality: we have $\gsem{\hat a(\izero)}{\psi} = \gsem{\hat a(\ione)}{\psi} = \gsem{\star}{\psi}$.
  \item Define $\gsem{s_{a,b,c}(\psi')}{\psi} \coloneq g_{a,c}(\psi'\psi(\izero))^{-1}g_{a,c}(\psi'\psi(\ione))$, where $g_{a,c}(e,e')$ is defined for $e,e' \in \braces{0,1}$ by
    \[
      \begin{array}{ll}
        g_{a,c}(\izero\izero) &\coloneq 1 \\
        g_{a,c}(\izero\ione) &\coloneq 1
      \end{array}
      \qquad
      \begin{array}{ll}
        g_{a,c}(\ione\izero) &\coloneq a \\
        g_{a,c}(\ione\ione) &\coloneq c
      \end{array}
    \]
    Again, we can check that $\gsem{s_{a,b,c}(r,\izero)}{\psi} = \gsem{\hat a(r)}{\psi}$, $\gsem{s_{a,b,c}(r,\ione)}{\psi} = \gsem{\hat c(r)}{\psi}$, $\gsem{s_{a,b,c}(\izero,r)}{\psi} = \gsem{\star}{\psi}$, and $\gsem{s_{a,b,c}(\ione,r)}{\psi} = \gsem{\hat b(r)}{\psi}$ as required by the equational theory.
  \item
    We define $\gsem{\ffill{e}{r}{\ell}{\phi}{u}}{\psi}$ as follows.
    First, for $\isincl{\psi'}{(i)}{(\Psi,\ell)}$, say that \emph{$\phi$ is satisfied at $\psi'$} if we have some $(\assign{s}{e'}{t}) \in \phi$ with $\face{s}{\psi'} = e'$; in this case, we write $\gsem{\phi}{\psi'}$ to mean $\gsem{t}{\ictxtcon{\psi'}{s}{e'}}$.
    Note that if there are multiple applicable clauses in $\phi$, this value is independent of the choice.
    In general, define
    \[
      \gsem*{\phi}{\psi'} \coloneq
      \left\{
        \begin{array}{ll}
          \gsem{\phi}{\psi'}, &\text{if $\phi$ is satisfied at $\psi'$} \\
          1, &\text{otherwise}
        \end{array}
      \right.
    \]
    We now divide into two cases.
    \begin{itemize}
    \item If $\phi$ is satisfied at $\psi$, then set $\gsem{\ffill{e}{r}{\ell}{\phi}{u}}{\psi} \coloneq \gsem{\phi}{(\psi,\face{r}{\psi})}$.
    \item Otherwise, set
      \[
        \gsem{\ffill{e}{r}{\ell}{\phi}{u}}{\psi} \coloneq (\gsem*{\phi}{(\psi(\izero),i)})^{e - \face{r}{\psi(\izero)}} \gsem{u}{\psi} (\gsem*{\phi}{(\psi(\ione),i)})^{\face{r}{\psi(\ione)} - e}
      \]
      Here, for $e' \in \braces{\izero,\ione}$, $\isincl{\face{\psi}{\inclcon{i}{e'}}}{()}{\ictxtcon{\Psi}{i}{e'}}$ is the induced substitution between constrained contexts and thus we have $\isincl{(\face{\psi}{\inclcon{i}{e'}},i)}{(i)}{(\ictxtcon{\Psi}{i}{e'},\ell)}$.
    \end{itemize}
    \noindent 
    Once more, we check that the assignment respects cell equality: we have $\gsem{\ffill{e}{e}{\ell}{\phi}{u}}{\psi} = \gsem{u}{\psi}$ and $\gsem{\ffill{e}{r}{\ell}{\phi}{u}}{\psi} = \gsem{t}{(\psi,r)} = \gsem{\face{t}{\inclface{i}{r}}}{\psi}$ whenever $(\assign{e'}{e'}{t}) \in \phi$.      
  \end{itemize}
\end{defi}

\begin{lem}
  \label{lem:group-interpretation-word}
  Fix a convenient presentation $\present{X}{R}$ of a group $G$.
  For each word $w$ on $X$, we have $\gsem{\tmword{w}{i}}{(i)} = w$ in $G$.
\end{lem}
\begin{proof}
  By calculation using the definition of $\tmword{w}{i}$.
\end{proof}

\begin{lem}
  \label{lem:group-interpretation-constant}
  Fix a convenient presentation $\present{X}{R}$ of a group $G$.  For each Kan
  cell $\isfterm[\presentctxt{X}{R}][\Psi]{t}$, if $\isincl{\psi}{(i)}{\Psi}$ is
  a constant substitution (i.e., either $(\izero)$ or $(\ione))$, then
  $\gsem{t}{\psi} = 1$.
\end{lem}
\begin{proof}
  The cases where $t$ is $\star$, $\hat a (\psi')$, or $s_{a,b,c}(\psi')$ are immediate from the definition, and the case where $t$ is a substitution application is direct by induction hypothesis.
  For $\ffill{e}{r}{\ell}{\phi}{u}$, we have two cases.
  \begin{itemize}
  \item If $\phi$ is satisfied at $\psi$, then $\gsem{\ffill{e}{r}{\ell}{\phi}{u}}{\psi} = \gsem{\phi}{(\psi,\face{r}{\psi})}$. Since $\psi$ is constant, so is $(\psi,\face{r}{\psi})$, and thus the result follows by induction hypothesis.
  \item Otherwise, we have
    \[
      \gsem{\ffill{e}{r}{\ell}{\phi}{u}}{\psi} = (\gsem*{\phi}{(\psi(\izero),i)})^{e - \face{r}{\psi(\izero)}} \gsem{u}{\psi} (\gsem*{\phi}{(\psi(\ione),i)})^{\face{r}{\psi(\ione)} - e}
    \]
    By induction hypothesis, we have $\gsem{u}{\psi} = 1$.
    Moreover, we have $\psi(\izero) = \psi(\ione)$, so
  \end{itemize}
  \[
    \gsem{\ffill{e}{r}{\ell}{\phi}{u}}{\psi} = (\gsem*{\phi}{(\psi(\izero),i)})^{e - \face{r}{\psi(\izero)}} (\gsem*{\phi}{(\psi(\izero),i)})^{\face{r}{\psi(\izero)} - e} = 1 \rlap{.} \qedhere
  \]
\end{proof}
\noindent 
For 2-cells, we intuitively want to show that any cell
\[
  \isfterm[\presentctxt{X}{R}][j,k]{t}<\assign{j}{\izero}{u_0} \se \assign{j}{\ione}{u_1} \se \assign{k}{\izero}{v_0} \se \assign{k}{\ione}{v_1}>
\]
induces an equality between the group elements $\gsem{u_0}{(k)}\gsem{v_1}{(j)}$
and $\gsem{v_0}{(j)}\gsem{u_0}{(k)}$. To make the induction go through, we
prove a more general statement where instead of the boundary of $t$, we consider
any ``quadrilateral'' of 1-cells tracing out a closed loop inside a 2-cell.

\begin{lem}
  \label{lem:group-interpretation-2}
  For a convenient presentation $\present{X}{R}$ of a group $G$, a term
  $\isfterm[\presentctxt{X}{R}][\Psi]{t}$, a substitution
  $\isincl{\psi}{(j,k)}{\Psi}$, and a quadruple of substitutions $\isincl{\delta_{\izero\bullet},\delta_{\ione\bullet},\delta_{\bullet\izero},\delta_{\bullet\ione}}{(i)}{(j,k)}$ such that
  \begin{mathpar}
    \psi\delta_{\izero\bullet}(\izero) = \psi\delta_{\bullet\izero}(\izero) \and
    \psi\delta_{\izero\bullet}(\ione) = \psi\delta_{\bullet\ione}(\izero) \and
    \psi\delta_{\ione\bullet}(\izero) = \psi\delta_{\bullet\izero}(\ione) \and
    \psi\delta_{\ione\bullet}(\ione) = \psi\delta_{\bullet\ione}(\ione)
  \end{mathpar}
  we have $\gsem{t}{\psi\delta_{\bullet\izero}}\gsem{t}{\psi\delta_{\ione\bullet}} = \gsem{t}{\psi\delta_{\izero\bullet}}\gsem{t}{\psi\delta_{\bullet\ione}}$.
\end{lem}
\begin{proof}
  In the situation of the statement, we abbreviate $\delta_{ee'} \coloneq \delta_{e \bullet}(e') = \delta_{\bullet e'}(e)$ for $e,e' \in \braces{\izero,\ione}$.
  We go by structural induction on $t$ as follows.
  \begin{itemize}
  \item For $\face{t}{\psi'}$, we have
    \begin{align*}
      \gsem{\face{t}{\psi'}}{\psi\delta_{\bullet\izero}}\gsem{\face{t}{\psi'}}{\psi\delta_{\ione\bullet}}
      &= \gsem{t}{\psi'\psi\delta_{\bullet\izero}}\gsem{t}{\psi'\psi\delta_{\ione\bullet}} \\
      &= \gsem{t}{\psi'\psi\delta_{\izero\bullet}}\gsem{t}{\psi'\psi\delta_{\bullet\ione}}
       = \gsem{\face{t}{\psi'}}{\psi\delta_{\izero\bullet}}\gsem{\face{t}{\psi'}}{\psi\delta_{\bullet\ione}}
    \end{align*}
    where the middle step is by induction hypothesis.
  \item For $\star$, we have
    $
      \gsem{\star}{\psi\delta_{\bullet\izero}}\gsem{\star}{\psi\delta_{\ione\bullet}} = 1 \cdot 1 = \gsem{\star}{\psi\delta_{\izero\bullet}}\gsem{\star}{\psi\delta_{\bullet\ione}}
    $.
  \item For $\hat a(\psi')$, we have
    \begin{align*}
      \gsem{\hat a(\psi')}{\psi\delta_{\bullet\izero}}\gsem{\hat a(\psi')}{\psi\delta_{\ione\bullet}}
      &= g_a(\psi'\psi\delta_{\izero\izero})^{-1}g_a(\psi'\psi\delta_{\ione\izero})g_a(\psi'\psi\delta_{\ione\izero})^{-1}g_a(\psi'\psi\delta_{\ione\ione}) \\
      &= g_a(\psi'\psi\delta_{\izero\izero})^{-1}g_a(\psi'\psi\delta_{\ione\ione}) \\
      &= g_a(\psi'\psi\delta_{\izero\izero})^{-1}g_a(\psi'\psi\delta_{\izero\ione})g_a(\psi'\psi\delta_{\izero\ione})^{-1}g_a(\psi'\psi\delta_{\ione\ione}) \\
      &= \gsem{\hat a(\psi')}{\psi\delta_{\izero\bullet}}\gsem{\hat a(\psi')}{\psi\delta_{\bullet\ione}} \rlap{.}
    \end{align*}
  \item For $\gsem{s_{a,b,c}(\psi')}{\psi}$, the same argument applies as for $\hat a(\psi')$.
  \item For $\gsem{\ffill{e}{r}{\ell}{\phi}{u}}{\psi}$, we proceed as follows.
    For simplicity we restrict our attention to the case $e = \izero$; a symmetric argument applies for $e = \ione$.

    First, we use our induction hypothesis to prove the following: for all $\isincl{\delta}{(i)}{(j,k)}$, we have
    \[
      \gsem{\ffill{\izero}{r}{\ell}{\phi}{u}}{\psi\delta} = (\gsem*{\phi}{(\psi\delta(\izero),i)})^{-\face{r}{\psi\delta(\izero)}}\gsem{u}{\psi\delta}(\gsem*{\phi}{(\psi\delta(\ione),i)})^{\face{r}{\psi\delta(\ione)}} \rlap{.}
    \]
    If $\phi$ is not satisfied at $\psi\delta$ then this is true by definition.
    Suppose then that $\phi$ \emph{is} satisfied at $\psi\delta$, i.e., there is $(\assign{s}{e'}{t}) \in \phi$ with $\face{s}{\psi\delta} = e'$.
    Define substitutions $\isincl{\delta_{\izero\bullet}',\delta_{\ione\bullet}',\delta_{\bullet\izero}',\delta_{\bullet\ione}'}{(i)}{(i,\ell)}$ by
    \begin{mathpar}
      \delta'_{\izero\bullet} \coloneq (i,\delta) \and
      \delta'_{\ione\bullet} \coloneq {(i,\face{r}{\psi\delta})} \\
      \delta'_{\bullet\izero} \coloneq
      \left\{
        \begin{array}{ll}
          (\izero,\izero) &\text{if $\face{r}{\psi\delta(\izero)} = \izero$} \\
          (\izero,i) &\text{if $\face{r}{\psi\delta(\izero)} = \ione$}
        \end{array}
      \right.
      \and
      \delta'_{\bullet\ione} \coloneq
      \left\{
        \begin{array}{ll}
          (\ione,\izero) &\text{if $\face{r}{\psi\delta(\ione)} = \izero$} \\
          (\ione,i) &\text{if $\face{r}{\psi\delta(\ione)} = \ione$}
        \end{array}
      \right.
    \end{mathpar}
    By induction hypothesis applied to the cell
    $\isfterm[\presentctxt{X}{R}][\ictxtcon{\Psi}{s}{e'},\ell]{t}$, the substitution $\isincl{(\ictxtcon{(\psi\delta)}{s}{e'},\ell)}{(i,\ell)}{(\ictxtcon{\Psi}{s}{e'},\ell)}$, and the four substitutions
    $\delta_{\izero\bullet}',\delta_{\ione\bullet}',\delta_{\bullet\izero}',\delta_{\bullet\ione}'$ just defined, we have
    \begin{equation}
      \label{eqn:sub-square-group-equation}
      \gsem{t}{(\ictxtcon{(\psi\delta)}{s}{e'},\ell)\delta'_{\bullet\izero}}\gsem{t}{(\ictxtcon{(\psi\delta)}{s}{e'},\ell)\delta'_{\ione\bullet}} = \gsem{t}{(\ictxtcon{(\psi\delta)}{s}{e'},\ell)\delta'_{\izero\bullet}}\gsem{t}{(\ictxtcon{(\psi\delta)}{s}{e'},\ell)\delta'_{\bullet\ione}} \rlap{.}
    \end{equation}
    Calculating, we have
    \begin{align*}
      \gsem{t}{(\ictxtcon{(\psi\delta)}{s}{e'},\ell)\delta'_{\izero\bullet}} &= \gsem{t}{(\ictxtcon{(\psi\delta)}{s}{e'},\izero)} = \gsem{\face{t}{\inclface{\ell}{\izero}}}{\ictxtcon{(\psi\delta)}{s}{e'}} = \gsem{u}{\ictxtcon{(\psi\delta)}{s}{e'}} = \gsem{u}{\psi\delta} \\
      \gsem{t}{(\ictxtcon{(\psi\delta)}{s}{e'},\ell)\delta'_{\ione\bullet}} &= \gsem{t}{(\ictxtcon{(\psi\delta)}{s}{e'},\face{r}{\psi\delta})} = \gsem{\phi}{(\psi\delta,\face{r}{\psi\delta})} = \gsem{\ffill{\izero}{r}{\ell}{\phi}{u}}{\psi\delta} \\
      \gsem{t}{(\ictxtcon{(\psi\delta)}{s}{e'},\ell)\delta'_{\bullet\izero}} &= (\gsem{t}{(\ictxtcon{(\psi\delta)}{s}{e'}(\izero),i)})^{\face{r}{\psi\delta(\izero)}} = (\gsem{\phi}{(\psi\delta(\izero),i)})^{\face{r}{\psi\delta(\izero)}} \\
      \gsem{t}{(\ictxtcon{(\psi\delta)}{s}{e'},\ell)\delta'_{\bullet\ione}} &= (\gsem{t}{(\ictxtcon{(\psi\delta)}{s}{e'}(\ione),i)})^{\face{r}{\psi\delta(\ione)}} = (\gsem{\phi}{(\psi\delta(\ione),i)})^{\face{r}{\psi\delta(\ione)}}
    \end{align*}
    where in the last two rows we use case analysis on ${\face{r}{\psi\delta(\izero)}}$ and ${\face{r}{\psi\delta(\ione)}}$ and \autoref{lem:group-interpretation-constant}.
    Rearranging \eqref{eqn:sub-square-group-equation}, we thus have
    \[
      \gsem{\ffill{\izero}{r}{\ell}{\phi}{u}}{\psi\delta} = (\gsem{\phi}{(\psi\delta(\izero),i)})^{-\face{r}{\psi\delta(\izero)}}\gsem{u}{\psi\delta}(\gsem{\phi}{(\psi\delta(\ione),i)})^{\face{r}{\psi\delta(\ione)}}
    \]
    as desired.

    Returning to the main claim, we now have
    \begin{eqnarray*}
      && \gsem{\ffill{\izero}{r}{\ell}{\phi}{u}}{\psi\delta_{\bullet\izero}}\gsem{\ffill{\izero}{r}{\ell}{\phi}{u}}{\psi\delta_{\ione\bullet}} \\
      &=& (\gsem*{\phi}{(\psi\delta_{\izero\izero},i)})^{-\face{r}{\psi\delta_{\izero\izero}}}\gsem{u}{\psi\delta_{\bullet\izero}}\gsem{u}{\psi\delta_{\ione\bullet}}(\gsem*{\phi}{(\psi\delta_{\ione\ione},i)})^{\face{r}{\psi\delta_{\ione\ione}}} \\
      &=& (\gsem*{\phi}{(\psi\delta_{\izero\izero},i)})^{-\face{r}{\psi\delta_{\izero\izero}}}\gsem{u}{\psi\delta_{\izero\bullet}}\gsem{u}{\psi\delta_{\bullet\ione}}(\gsem*{\phi}{(\psi\delta_{\ione\ione},i)})^{\face{r}{\psi\delta_{\ione\ione}}} \\
      &=& \gsem{\ffill{\izero}{r}{\ell}{\phi}{u}}{\psi\delta_{\izero\bullet}}\gsem{\ffill{\izero}{r}{\ell}{\phi}{u}}{\psi\delta_{\bullet\ione}}
    \end{eqnarray*}
    as required. \qedhere
  \end{itemize}
\end{proof}

\begin{cor}
  \label{cor:kan-solution-iff-words-equal}
  Let $\present{X}{R}$ be a convenient presentation of a group $G$.
  For any pair of words $v,w$ on $X$ there exists a cell
  \begin{equation*}
    \isfterm[\presentctxt{X}{R}][i,k]{t}<\assignbdy{i}{\star} \se \assign{k}{\izero}{\tmword{v}{i}} \se \assign{k}{\ione}{\tmword{w}{i}}>
  \end{equation*}
  if and only if $v \equiv_G w$.
\end{cor}
\begin{proof}
  One direction was already proven in \autoref{prop:equal-words-yield-cell}.
  For the converse, suppose we have such a cell $t$. By applying
  \autoref{lem:group-interpretation-2} with the
  $\delta_{\izero\bullet} = (\izero,i)$, $\delta_{\ione\bullet} = (\ione,i)$,
  $\delta_{\bullet\izero} = (i,\izero)$, and
  $\delta_{\bullet\ione} = (i,\ione)$, we get that
  $\gsem{\tmword{v}{i}}{(i)}\gsem{\star}{(i)} =
  \gsem{\star}{(i)}\gsem{\tmword{w}{i}}{(i)}$.  By
  \autoref{lem:group-interpretation-constant} and
  \autoref{lem:group-interpretation-word}, this means that $v \equiv_G w$.
\end{proof}

\begin{thm}
  $\KanCubicalCell$ is undecidable. More specifically, there are contexts
  $\Gamma$ for which there is no algorithmic decision procedure for the problem
  $\isfterm[\Gamma][i,k]{\hole}<\phi>$ uniformly in Kan boundaries
  $\isfbdy[\Gamma][i,k]{\phi}$.
\end{thm}
\begin{proof}
  This follows from \autoref{cor:kan-solution-iff-words-equal} and the fact that
  there are finitely presented groups with undecidable word problem; see, e.g.,
  Collins \cite{collins86} for an example of the latter.
\end{proof}
\vspace{-.5em}

\section{Finding Dedekind and De Morgan contortions}
\label{sec:contortions}

We have seen that \Cartesian and, to a lesser extent, \Disjunctive are problems
that can be feasibly solved by enumeration. In contrast, the search space for
\Dedekind and \DeMorgan quickly explodes, which make a brute-force approach
infeasible even when solving boundary problems in lower dimensions. We hence
need to explore the search space for Dedekind and De Morgan contortions more
carefully. In \S\ref{ssec:ppms}, we will see how an alternative characterisation
based on a Stone-type duality~\cite{johnstone86_stone_spaces}, specifically
Birkhoff's duality between finite distributive lattices and finite posets
\cite{birkhoff37}, gives rise to a lossy but space-saving representation of
collections of Dedekind contortions. We use this representation in
\S\ref{ssec:contortionsolver} to develop an algorithm for solving \Dedekind. We
also have a duality between De Morgan algebras and finite poset maps, which
allows us to adapt our space-saving representation of Dedekind contortions to
also represent collections of De Morgan contortions (\S\ref{ssec:demorgan}).

\vspace{-.5em}
\subsection{Representing Dedekind contortions with potential poset maps}
\label{ssec:ppms}

Recall the example (\ref{or-connection-cell}), where we contorted a path $p$
into a square using a Dedekind contortion $\isfterm*[\isof{p}{i}][j,k]{p(j \lor
  k)}$. We can think of $\join$ as logical disjunction---if either $j$ or $k$ is
$\ione$, the contortion evaluates to $\ione$. Similarly, we can treat the
connection $\meet$ as logical conjunction, which means that we can view any
contortion as a tuple of propositional formulas. In fact a Dedekind contortion
is uniquely determined by its truth table; for example, the contortion above is
determined by the assignment $\brackss{-} \colon \{\izero,\ione\} \times
\{\izero,\ione\} \to \{\izero,\ione\}$ defined by $\brackss{\izero\izero} =
\izero$ and $\brackss{\izero\ione} = \brackss{\ione\izero} =
\brackss{\ione\ione} = \ione$. In general, an $n$-term Dedekind contortion in
$m$ variables gives a truth function $\{\izero,\ione\}^m \to
\{\izero,\ione\}^n$.

Since a Dedekind contortion $\psi$ contains no negations, its truth function is
\emph{monotone}---we cannot make $\psi$ false by setting more variables to true.
Thus the truth function induced by $\psi$ is in fact a map of posets $\pint{m}
\to \pint{n}$, where $\pint{k}$ is the $k$-fold power of the poset $\pint{}
\coloneq \{ \izero < \ione\}$ with its product ordering. Conversely, any map of
posets $\pint{m} \to \pint{n}$ determines a unique $n$-term contortion in $m$
variables.
For example, we can depict the poset map corresponding to $j \join k$ as an
assignment between the posets $\pint{2}$ and $\pint{1}$, which we draw as a
Hasse diagram below.

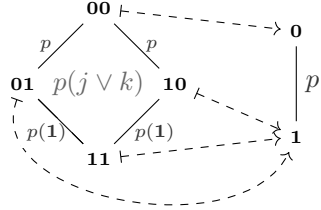
\begin{center}
  \begin{tikzpicture}[scale=2]
    \node (zz) at (0,1) {$\izero\izero$};
    \node (zo) at (-1,0) {$\izero\ione$};
    \node (oz) at (1,0)  {$\ione\izero$};
    \node (oo) at (0,-1) {$\ione\ione$};
    \node (z) at (2.6,.7) {$\izero$};
    \node (o) at (2.6,-.7) {$\ione$};

    \draw[|->,dashed] (zz) -- (z);
    \draw[|->,dashed] (zo) to [out=-110,in=-115] (o);
    \draw[|->,dashed] (oz) -- (o);
    \draw[|->,dashed] (oo) -- (o);

    \node[text=black!60] (label) at (0,0) {$p(j \join k)$};
    \draw (zz) -- (zo) node[midway,left,text=black!80] {$\scriptstyle p$};
    \draw (zz) -- (oz) node[midway,right,text=black!80] {$\scriptstyle p$};
    \draw (zo) -- (oo) node[near end,left,fill=white,inner sep=2pt, outer sep=0pt,text=black!80] {$\scriptstyle p(\ione)$};
    \draw (zo) -- (oo);
    \draw (oz) -- (oo) node[near end,right,text=black!80] {$\scriptstyle p(\ione)$};
    \draw (z) -- (o) node[midway,right,text=black!80] {$p$};
  \end{tikzpicture}
\end{center}

We can read off the boundary of $p(j \join k)$ by looking at the action of the
poset map: the edge from $\izero\izero$ to $\izero\ione$ is sent to the edge
from $\izero$ to $\ione$ in the target, so the $j = \izero$ side of
$p(j \join k)$ is $p(k)$. Between $\izero\ione$ and $\ione\ione$, we stay at
$\ione$, so the boundary at $j = \ione$ is constantly $p(\ione)$.

We will in the following freely switch between regarding Dedekind contortions as
tuples of propositional formulas and as poset maps. We write \formulatopm{\psi}
for the poset map induced by the contortion $\psi$ and \pmtoformula{\sigma} for
the contortion prescribed by the poset map $\sigma$. The poset map perspective
on contortions does not only give geometric intuition for the boundary of a
contorted cell, but also allows for concisely representing a collection of
contortions: by assigning a \emph{set} of values $Y \subseteq \pint{n}$ to each
element of $\pint{m}$, we can at once represent several contortions. The
monotonicity constraint on poset maps entails that only some such assignments
are meaningful; we call these \emph{potential poset maps}.

\begin{defi}\label{def:ppm}
  A \textbf{potential poset map} (\textbf{PPM}) is a map
  $\Sigma \colon \pint{m} \to \pow{\pint{n}}$ s.t.\ $\forall x \leq y$:
  \begin{mathpar}
    \forall u \in \Sigma(y).\ \exists v \in \Sigma(x).\ v \leq u
    \and
    \text{ and }
    \and
    \forall v \in \Sigma(x).\ \exists u \in \Sigma(y).\ v \leq u
  \end{mathpar}

\end{defi}

Readers familiar with bisimulation might intuit PPMs as order-bisimulating,
multivalued relations satisfying a certain back-and-forth condition.

With a PPM, we can represent a collection of contortions with very little data:
representing all $D(m)^n$ poset maps $\pint{m} \to \pint{n}$ with the total PPM
$x \mapsto \pint{n}$ for $x \in \pint{m}$ requires $2^m$ entries of $2^n$
values---the memory requirements are therefore independent of the Dedekind
numbers and grow ``only'' exponentially in $m$ and $n$. This comes with the
trade-off that PPMs are a lossy representation of sets of poset maps. For
example, any PPM containing the two poset maps
$\sigma,\sigma' : \pint{1} \to \pint{2}$ defined by
$\sigma(\izero) = (\izero\izero) , \sigma(\ione) = (\ione\izero)$ and
$\sigma'(\izero) = (\izero\ione) , \sigma'(\ione) = (\ione\ione)$ also contains
the diagonal map sending $\izero \mapsto (\izero\izero)$ and
$\ione \mapsto (\ione\ione)$.

In the following, we unfold a PPM $\Sigma$ into the set of poset maps it
contains with $\Call{UnfoldPPM}{\Sigma}$. We update a PPM $\Sigma$ to restrict
its values at $x$ to some set $vs \subseteq \Sigma(x)$ using
$\Call{UpdatePPM}{\Sigma,x,vs}$, thereby obtaining a new PPM $\Sigma'$ with
$\Sigma'(x) = vs$. Due to space constraints we refer to the source code of the
solver discussed in \S\ref{sec:casestudy} for details.

\subsection{An algorithm for gradually constructing Dedekind contortions}
\label{ssec:contortionsolver}

We now use PPMs in Algorithm~\ref{alg:contsolve} to solve \Dedekind more
efficiently than by brute-force. Given a boundary problem
$\iscterm[\Gamma][\Psi]{\hole}<\phi>$ and a cell $\isof{a}{\Psi'}[\phi']$ in
$\Gamma$, we search for a contortion $\isisubst{\psi}{\Psi}{\Psi'}$ such that
$a(\psi)$ has boundary $\phi$ by gradually restricting a PPM to be
compatible with the faces of $\phi$. If a contortion of $a$ appears as one of
the faces of $\phi$, we can even reduce the search space significantly before
performing any expensive operations.

\begin{algorithm}
  \caption{Constructing a Dedekind contortion}
  \label{alg:contsolve}
  \begin{algorithmic}[1]
    \Require $\iscbdy[\Gamma][\Psi]{\phi}$ and $\isof{a}{\Psi'}[\phi'] \in
    \Gamma$. Let $m \coloneq \card{\Psi}$ and $n \coloneq \card{\Psi'}$.
    \Ensure $\isisubst{\psi}{\Psi}{\Psi'}$ s.t. $\iscterm[\Gamma][\Psi]{a(\psi)}<\phi>$ if
    such a $\psi$ exists, \texttt{Unsolvable} otherwise
    \Procedure{DedekindContort}{$\Gamma , \Psi , \phi , a$}
      \State $\Sigma \coloneq \{ x \mapsto \pint{n} \mid x \in \pint{m} \}$ \label{cont:line:initsigma}
      \For{$(\assign{i}{e}{b(\psi)}) \in \phi$ with $\isisubst{\psi}{\ictxtcon{\Psi}{i}{e}}{\Psi''}$, in descending order of $\card{\Psi''}$} \label{cont:line:faceloop}
        \If{$a = b$}
        \State $\Theta \coloneq \{ x \mapsto \{ \formulatopm{\psi}(x) \} \mid x \in \pintrestr{m}{i}{e}
          \}$ \label{cont:line:peqq}
        \Else
          \State $\Theta \coloneq \{ x \mapsto \emptyset \mid x \in \pintrestr{m}{i}{e} \}$
          \For{$\sigma \in
              \Call{UnfoldPPM}{\restrict{\Sigma}{\pintrestr{m}{i}{e}}}$} \label{cont:line:unfold}
            \If{$a(\pmtoformula{\sigma}) = b(\psi)$}
              \For{$x \in \pintrestr{m}{i}{e}$}
                \State $\Theta(x) \coloneq \Theta(x) \cup \{ \sigma(x) \} $ \label{cont:ln:complete}
              \EndFor
            \EndIf
          \EndFor
        \EndIf
        \For{$x \in \pintrestr{m}{i}{e}$}
          \State \Call{UpdatePPM}{$\Sigma, x, \Theta(x)$} \label{cont:line:update}
        \EndFor
      \EndFor
      \If{$\exists \sigma \in \Call{UnfoldPPM}{\Sigma}$ such that
          $\iscterm[\Gamma][\Psi]{a(\pmtoformula{\sigma})}<\phi>$} \label{cont:line:check}
        \State \Return{$\pmtoformula{\sigma}$}
      \Else
        \State \Return{\texttt{Unsolvable}}
      \EndIf
    \EndProcedure
  \end{algorithmic}
\end{algorithm}

We first initialise $\Sigma$ to the total PPM on line~\ref{cont:line:initsigma}.
We then go through the faces of $\phi$, each of which normalises to the form
$b(\psi)$ for some variable $b$ and contortion $\psi$, and use them to
restrict $\Sigma$. Crucially, we order the boundary faces by descending
dimensionality of the contorted variable on line~\ref{cont:line:faceloop}, as
contortions of higher-dimensional variables constrain the search space more.
Given a face $\assign{i}{e}{b(\psi)}$ of $\phi$, we proceed as follows:
if $b$ is in fact $a$, we can constrain $\Sigma$ to maps that agree with $\psi$
where $i = e$ on line~\ref{cont:line:peqq}. Otherwise, we iterate through the
poset maps $\sigma$ contained in the restriction of $\Sigma$ to
$\pintrestr{m}{i}{e}$ on line~\ref{cont:line:unfold}. For each, we mark its
values for retention only when $a(\pmtoformula{\sigma})$ matches the face
$b(\psi)$. Finally, we propagate our findings to $\Sigma$ on
line~\ref{cont:line:update}.
After restricting $\Sigma$ according to all faces of $\phi$, we unfold $\Sigma$
and brute-force search the results for a valid solution to return. Note that not
all poset maps in $\Sigma$ need be solutions, as a PPM is a lossy representation
of a set of maps. The algorithm is complete: if $a$ can be contorted by some
$\psi$ to solve the goal boundary, then it keeps $\formulatopm{\psi}(x)$ in
$\Sigma(x)$ for all $x \in \pint{m}$ in each iteration of the main loop, whether
in line~\ref{cont:line:peqq} or line~\ref{cont:ln:complete}.

The main computational effort in Algorithm~\ref{alg:contsolve} consists
unfolding all poset maps from a subposet on line~\ref{cont:line:unfold}. For an
unconstrained PPM, we have to check $D(m-1)^n$ poset maps, and as we are doing
this for up to $2m$ faces of $\phi$, we are unfolding $2m \cdot D(m-1)^n$ poset
maps in the worst case. In many boundary problems, the cell to be contorted
appears in the boundary, which means the search space significantly shrinks
before any PPM is unfolded. This allows us to compute many contortions that
would have been impossible to find by na\"ive brute-force.

\begin{exa}[name=Square to cube contortion] \label{exp:contsolve}
  Suppose that we are given the cell context
  $
  \Gamma \coloneqq
      \isof{a},
        \isof{s}{i,j}[\assign{i}{\izero}{a} \se \assign{i}{\ione}{a} \se \assign{j}{\izero}{a} \se \assign{j}{\ione}{a}]
  $ 
  and want to contort the square $s$ to match the following 3-cube boundary,
  which has a contortion of $s$ on one face and squares which are constantly $a$
  otherwise:
  \begin{mathpar}
    \iscterm*[\Gamma][i,j,k]{\hole}<
      \begin{array}{lllll}
        \assign{i}{\izero}{s (j \meet k ,j \join k) } &\se& \assign{j}{\izero}{a} &\se& \assign{k}{\izero}{a} \\
                \assign{i}{\ione}{a} &\se& \assign{j}{\ione}{a} &\se& \assign{k}{\ione}{a}
      \end{array}
      >
  \end{mathpar}

  This is a difficult instance of \Dedekind because most faces of the goal
  are contortions of a 0-cell, which can be obtained in many ways. To
  construct $\isisubst{\psi}{(i,j,k)}{(i,j)}$ such that
  $s(\psi)$ has boundary $\phi$, we search for the equivalent poset map
  $\pint{3} \to \pint{2}$ using \autoref{alg:contsolve}.

  On line~\ref{cont:line:initsigma}, the total PPM $\Sigma : \pint{3} \to
  \pow{\pint{2}}$ is initialised with $x \mapsto \pint{2}$ for all $x \in
  \pint{3}$. We then go through all faces of the goal boundary and use them to
  restrict $\Sigma$, starting with the contortion of $s$ at $i = \izero$. Since
  $s$ is also the cell that we are contorting, the subposet
  $\pintrestr{3}{i}{\izero}$ of the domain of $\Sigma$ is mapped in a unique way
  to the elements of $\pint{2}$. The monotonicity restrictions on PPMs further
  restrict $\Sigma$, which only contains 10 poset maps after this first
  restriction.
  In the next iteration of the outer loop, we only have degenerate $a$ faces
  left in the goal boundary. Going through each face further restricts $\Sigma$,
  as most induced poset maps give rise to a contortion of $s$ which is not the
  constant $a$ square. Afterwards, $\Sigma$ comprises a single poset map:
  $\Sigma(\izero\izero\izero) = \{\izero\izero\}$, $\Sigma(\izero\izero\ione) =
  \Sigma(\izero\ione\izero) = \Sigma(\izero\ione\ione) =
  \Sigma(\ione\izero\izero) = \Sigma(\ione\izero\ione) = \{\izero\ione\}$ and
  $\Sigma(\ione\ione\izero) = \Sigma(\ione\ione\ione) = \{\ione\ione\}$.
  Translating this poset map to a contortion gives rise to a solution for our
  boundary problem: $\iscterm[\Gamma][i,j,k]{s (i \meet j , i \join j \join k)}<\phi>$

  Our algorithm finds this solution quickly since the search space is restricted
  to only 10 possible contortions after looking at the first face of $\phi$.
  This contrasts with brute-force search, where we would have to check $D(3)^2 =
  400$ contortions. The increase in speed gets apparent for a larger goal: a
  6-dimensional analogue of the above proof goal can be found by unfolding less
  than $16 000$ poset maps. A brute-force search would have to find a solution in
  a search space with $D(6)^2 = 7\,828\,354^2 = 61\,283\,126\,349\,316$ contortions.
\end{exa}

\subsection{De Morgan contortions as poset maps}
\label{ssec:demorgan}

The most expressive contortion theory that we consider are the De Morgan
contortions which can be formed with both $\land$, $\lor$ as well as a unary
operator $\inv$ which captures reversal of paths. In fact, the number of De
Morgan contortions grows with the even Dedekind numbers, i.e., there are $D(2m)$
many ways to contort a $1$-cube into an $m$-dimensional cube using a De Morgan
contortion. These combinatorics suggest a connection with Dedekind formulas, and
indeed any De Morgan contortion over $m$ variables corresponds to a monotone
boolean function in $2m$ variables~\cite[Theorem
3.2]{movsisyan14_funct_compl_theor_de_morgan_funct}
\cite{gehrke03_normal_forms_truth_tables_fuzzy_logic}. Intuitively, we can
regard a variable $j$ and its inverse $\inv j$ separately since, e.g., $j \lor
\inv j$ is in normal form and does not reduce to $\ione$. We can hence reuse our
potential poset maps to also represent De Morgan contortions, and thereby obtain
a space-efficient representation for this comprehensive contortion theory. Our
construction is reminiscent of the proof of coNP-hardness of equivalence between
monotone Boolean formulas using a reduction from the tautology problem
\cite{reith03}.

Consider again the cell context from above, but we now contort $p$ into another
1-dimensional path using a De Morgan contortion: $\isfterm*[\isof{p}{i}][j]{p(j
  \lor \inv j)}$. The poset map corresponding to $j \lor \inv j$ has two
variables, one for $j$ and one for $\inv j$, and captures that we take the
disjunction of both literals.%
\begin{center}
  \begin{tikzpicture}[scale=2]
    \node (zz) at (0,1) {$\izero\izero$};
    \node (zo) at (-1,0) {$\izero\ione$};
    \node (oz) at (1,0)  {$\ione\izero$};
    \node (oo) at (0,-1) {$\ione\ione$};
    \node (z) at (2.6,.7) {$\izero$};
    \node (o) at (2.6,-.7) {$\ione$};

    \draw[|->,dashed] (zz) -- (z);
    \draw[|->,dashed] (zo) to [out=-110,in=-115] (o);
    \draw[|->,dashed] (oz) -- (o);
    \draw[|->,dashed] (oo) -- (o);

    \node[text=black!60] (label) at (0,0) {};
    \draw (zz) -- (zo) node[midway,left,text=black!80] {};
    \draw (zz) -- (oz) node[midway,right,text=black!80] {};
    \draw (zo) -- (oo) node[near end,left,fill=white,inner sep=2pt, outer sep=0pt,text=black!80] {};
    \draw (zo) -- (oo);
    \draw (oz) -- (oo) node[near end,right,text=black!80] {};
    \draw (z) -- (o) node[midway,right,text=black!80] {$p$};
  \end{tikzpicture}
\end{center}
\vspace{-2em}

Of interest to us is the antichain consisting of $\izero\ione$ and
$\ione\izero$, which are both sent to $\ione$, which captures that both $j$ and
$\inv j$ are present in the contortion. Note that we cannot read off the
boundary of the contorted term directly from the poset map anymore ($p(j \lor
\inv j)$ is a 1-dimensional path after all), but that we have to focus on the
part of the poset map that corresponds to a ``consistent'' assignment of truth
values to the variables. In this case, these are precisely $\izero\ione$ and
$\ione\izero$, inspecting their values under the poset map allows us to compute
the boundary of the contorted term, i.e., $\isof{p}{j \lor \inv
  j}[\assign{j}{\izero}{p(\ione)} \se \assign{j}{\ione}{p(\ione)}]$. We hence
have another solution to a boundary problem that could have been also solved
simply with $p(\ione)$.

The power of De Morgan contortions comes from the fact that we can directly
reverse paths as discussed in \S\ref{ssec:contorting-cubes}, which gives us a
rich language also for higher contortions. For example, the contortion
$\isfterm*[\isof{p}{i}][j,k]{p(\inv j \lor k)}$ corresponds to a square where
between the bottom left and top right corner we once travel along the inverse
$p$ and $p$, and once constantly stay at $p(\ione)$.
\begin{mathpar}
  \dimsquare{j}{k}
  \quad
  \begin{tikzpicture}[xscale=.8,yscale=.8]
    \draw[->,>=stealth] (0,0) -- (4,0) node [midway,below, fill=none] {$p(\ione)$};
    \draw[->,>=stealth] (0,0) -- (0,4) node [midway,left, fill=none] {$p(\inv j)$};
    \draw[->,>=stealth] (4,0) -- (4,4) node [midway,right, fill=none] {$p(\ione)$};
    \draw[->,>=stealth] (0,4) -- (4,4) node [midway,above, fill=none] {$p(k)$};
    \node at (2,2) {$p(\inv j \lor k)$};
  \end{tikzpicture}
\end{mathpar}

We can intuit the poset map $\sigma : \pint{4} \to \pint{1}$ corresponding to
the contortion $\inv j \lor k$ as follows. An element of the domain $x \in
\pint{4}$ has four indices, where $x_1$ stands for $j$, $x_2$ for $k$, $x_3$ for
$\inv j$ and $x_4$ for $\inv k$. The antichain which determines $\sigma$ is
consequently $\izero\izero\ione\izero$ and $\izero\ione\izero\izero$,
corresponding to $\sim j$ and $k$, respectively.

The total potential poset map $\Sigma : \pint{4} \to \pint{1}$ is again a
space-efficient representation of all $D(4) = 168$ De Morgan contortions of
$p$ into a square, and we can restrict a potential poset map corresponding to De
Morgan contortions similarly to Algorithm~\ref{alg:contsolve} to gradually
construct a contortion involving reversals.

By constructing De Morgan contortions in this way, we can find compact solutions
to lower-dimensional boundary problems for theories which support reversals
(such as \CubicalAgda). However, this approach is not very practical for
higher-dimensional boundary problems as the space of possible contortions grows
too quickly.

\section{Finding Kan fillers}
\label{sec:kan}

We now turn to \KanCubicalCell and develop an algorithm for solving general
boundary problems. Recall that a Kan cell is of the form
$\ffill{e}{r}{i}{\phi}{u}$, where $\phi$ and $u$ constitute an ``open box''
which is filled in direction $e \to r$. Searching for such fillers requires a
different approach depending on whether $r$ is a dimension variable or an
endpoint. In the former case, $\ffill{e}{j}{i}{\phi}{u}$ has the same dimension
as $\phi$ and has $\ffill{e}{\negI{e}}{i}{\phi}{u}$ as its $j = \negI{e}$
face. This means that it is easy to recognise if a boundary problem can be
solved by a filler $e \to j$: we simply have to check if some face of the goal
boundary is an $e \to \negI{e}$ filler. We hence call the filler in direction
$e \to j$ the ``natural filler'' for a goal boundary which has
$\ffill{e}{\negI{e}}{i}{\phi}{u}$ at side $j = \negI{e}$.

In contrast, determining when we have to introduce $e \to \negI{e}$ fillers is
difficult.
We focus our attention on fillers in direction $\izero \to \ione$, since such a
filler can be constructed if and only if we can construct a filler in the converse
direction. Note that a cell $\ffill{\izero}{\ione}{i}{\phi}{u}$ is of one
dimension less than the open box spanned by $\phi$ and $u$---put differently, to
solve a given goal boundary by a $\izero \to \ione$ filler, we need to first
construct a higher-dimensional cube. We hence call fillers in direction
$\izero \to \ione$ ``higher-dimensional'' fillers. Searching for such cells is
difficult because the goal boundary only partially constrains the faces of
$\phi$, while $u$ could be any cell of the correct dimension. In particular, we
could again use higher-dimensional fillers as faces for $\phi$ or $u$, leading
to infinite search spaces that we have to carefully navigate with heuristics.

\begin{rem}
  \label{rem:comp-fill}
  Some presentations of cubical type theories \cite{CoquandHuberMortberg18} use
  separate primitives depending on whether we are filling towards a dimension
  variable or an endpoint, where usually only the former is called ``filling'',
  while the latter is called ``composition'' (\AgdaFunction{hfill} and
  \AgdaFunction{hcomp} in \CubicalAgda, respectively). To simplify our theory,
  we have chosen a more general notion of Kan cells which can be used both for
  filling and composition, but for the purpose of proof search it might be
  helpful of thinking of these two as different operations.
\end{rem}

In our solver we follow the principle that when solving a boundary problem, it's
best to use contorted cells and natural fillers if possible, and only construct
higher-dimensional fillers if necessary. The more powerful our contortion theory
is, the more boundary problems can be solved without Kan filling, but in general
we will often have to construct Kan fillings even when using the most powerful
De Morgan contortion theory. For example, our contortions do not allow us to
concatenate several paths, such that we always have to resort to Kan filling as
soon as a boundary problem cannot be solved using a single cell. Since more
powerful contortion theories mean that the search space for contortions is
larger, it is sometimes expedient to choose a simpler contortion theory for a
given boundary problem. We will hence treat the underlying contortion theory as
a parameter to our solver for Kan fillings, which means the results in this
section are applicable to all contortion theories.

We formulate the problem of finding a higher-dimensional filler which only uses
contorted cells as a constraint satisfaction problem (CSP)~\cite{laurierecsp,
  tsangcsp} in \S\ref{ssec:kancsp}, which allows us to employ finite domain
constraint solvers for this sub-problem of \KanCubicalCell. By carefully calling
this solver, we then give a complete search procedure for \KanCubicalCell in
\S\ref{ssec:solver}.

\subsection{Kan filling as a constraint satisfaction problem}
\label{ssec:kancsp}

When constructing a higher-dimensional cell for a goal boundary, the sides of
the open box that we fill need to match up. This suggests a recipe for
constructing fillers: we formulate the search problem as a CSP. In this section,
we focus on the problem where all the sides of the filler are contortions. Since
there are only finitely many contortions into a given dimension, we can use a
finite domain constraint solver to solve this CSP. Still, the number of
contortions grows very quickly in the case of the Dedekind and De Morgan
contortion theories, making it quickly infeasible to list all contortions.

To rectify this, we may again use PPMs to represent a collection of contortions.
We will write $\Conts{\Psi}{\Psi'}$ for the set of all contortions of an
$n$-dimensional cell into $m$ dimensions, where $n = \card{\Psi'}$ and $m =
\card{\Psi}$. In the case of Dedekind contortions, this set can be represented
by the total PPM ${\Sigma : \pint{m} \to \mathcal{P}(\pint{n})}$, for De Morgan
contortions we could use $\Sigma : \pint{2m} \to \mathcal{P}(\pint{n})$. By
representing a collection of contortions with a PPM, we can quickly construct
our CSP with little memory requirements; a solver such as the one discussed in
\S\ref{sec:casestudy} can then gradually narrow down the PPMs until it arrives
at a solution. For cartesian and disjunctive contortions it is expedient to
simply list all formulas.

Recall that a CSP is given by a set of variables $Var$; an assignment of domains
to $Var$, i.e., a set $D_X$ for each $X \in Var$; and a set of constraints $C
\subseteq D_X \times D_{X'}$ for $X,X' \in Var$. A solution
is a choice
of one element of each domain, i.e., $t_X \in D_X$ for all $X \in Var$, s.t., all constraints are satisfied, i.e., $C(t_X,t_{X'})$ for all $C, X, X'$.

We now state the CSP for filling
boundaries via Kan
fillers that have only contortions as
sides.

\begin{defi} \label{def:compcsp} Given a boundary
  $\isfbdy[\Gamma][\Psi]{\phi}$ and a fresh dimension $k \notin \Psi$, as well
  as a set of indices $Ope \subseteq \{ \csvar{k}{\izero} \} \cup \{ \csvar{i}{e} \mid i \in
  \Psi , e \in \{\izero,\ione\} \}$, the CSP
  $\Call{KanCSP}{\phi , Ope}$ is given as follows:
\begin{itemize}
\item $Var \coloneq \{ X_{\csvar{i}{e}} \mid i \in \Psi, e \in \{\izero, \ione\} ,
  \csvar{i}{e} \notin Ope \} \cup \{ X_{\csvar{k}{\izero}} \text{ if }
  \csvar{k}{\izero} \notin Ope \} $
\item $D_{\csvar{i}{e}} \coloneq \{ (p,\Conts{\Psi}{\Psi'}) \mid \isof{p}{\Psi'}[\phi'] \in \Gamma \}$
\item and constraints for all $\isdim[\Psi]{i,j}$, $e,e' \in \{\izero,\ione\}$:
  \begin{align*}
    &\iscterm[\Gamma][\ictxtcon{\Psi}{i}{e}]{\face{X_{\csvar{i}{e}}}{\inclcon{k}{\ione}}}[\face{\phi}{\inclcon{i}{e}}] \text{ if } (i,e) \text{ specified in } \phi \\
    &\iscterm[\Gamma][\ictxtcon{\ictxtcon{\Psi}{i}{e}}{j}{e'}]{\face{X_{\csvar{i}{e}}}{\inclcon{j}{e'}}}[\face{X_{\csvar{j}{e'}}}{\inclcon{i}{e}}]
  \end{align*}
\end{itemize}
\end{defi}
\noindent 
The CSP contains a variable for any side of the boundary that is not left open,
the domains contain pairs representing all contortions of a cell $p$ into the
needed dimension. The first set of constraints ensures that all sides agree with
the goal boundary, while the second set of constraints makes sure that all sides
have mutually matching boundaries.

If $Ope$ contains only sides which are unspecified in $\phi$, a solution
\Call{KanCSP}{$\phi, Ope$} is a solution to the boundary problem $\phi$:
\[
  \isfterm[\Gamma][\Psi]{\ffill{\izero}{\ione}{k}{ \assign{i}{e}{t_{(i,e)}}
      \text{ for } i \in \Psi, e \in \{\izero, \ione\} , \csvar{i}{e} \notin Ope
    }{t_{(k,\izero)}}}<\phi>
\]

When calling the solver, one has to carefully consider which underlying
contortion theory one should choose. For higher-dimensional problems, the De
Morgan contortions are often too unwieldy since the domain of the poset maps
grows with the even Dedekind numbers.

Take for example the Eckmann-Hilton argument, the cubical version of which we
introduced in \S\ref{sec:intro}. If we were to solve the cube presented in
\autoref{eh-cubical} using De Morgan contortions, we would have to consider of
the order of $D(6)^2 = 7\,828\,354^2$ possible contortions. Since the cube in
the example can be constructed without reversals, it is more expedient to
instead use the Dedekind contortion theory, leading to a different filler than
that depicted in \autoref{eh-inverses} and \autoref{eh-degeneracy}.

\begin{exa}[name=The Eckmann-Hilton cube] \label{exp:eh} Using Dedekind
  contortions, we want to fill the cube from \autoref{eh-cubical}, where we are
  given a cell context $\Gamma$ with a point $\isof{x}$ and two squares $p(i,j)$
  and $q(i,j)$ with boundaries $[\assign{i}{\izero}{x} \se \assign{i}{\ione}{x}
  \se \assign{j}{\izero}{x} \se \assign{j}{\ione}{x}]$, and which are assembled
  into:
  \[\isfbdy[\Gamma][i,j,k]{\boundary*{
        \begin{array}{lll}
      \assign{i}{\izero}{p(j,k)} \se&
      \assign{j}{\izero}{q(i,k)} \se&
      \assign{k}{\izero}{x} \\
      \assign{i}{\ione}{p(j,k)} \se&
      \assign{j}{\ione}{q(i,k)} \se&
      \assign{k}{\ione}{x}
        \end{array}
      }}\]

  Our boundary problem hence corresponds to the cube from \autoref{eh-cubical},
  where the gray 2-cubes of the $i$ sides are filled with $p$, the $j$ sides
  with $q$, and the $k$ sides are constantly $x$ (note also that all corner
  points of the cube are judgmentally equal to $x$):
  \[
    \dimcube{j}{i}{k}
    \hspace{1em}
    \begin{tikzpicture}[scale=1]
      \filldraw[gray] (0,0,0) -- (0,0,2) -- (0,2,2) -- (0,2,0) -- cycle;
      \filldraw[pattern=north west lines] (0,0,0) -- (2,0,0) -- (2,0,2) -- (0,0,2) -- cycle;

      \draw[thick] (0,0,0) -- (0,2,0) -- (2,2,0) -- (2,0,0) -- cycle;
      \draw[thick] (0,0,0) -- (0,0,2) -- (0,2,2) -- (0,2,0) -- cycle;
      \draw[thick] (0,0,0) -- (2,0,0) -- (2,0,2) -- (0,0,2) -- cycle;

      \filldraw[color=gray] (2,0,0) -- (2,0,2) -- (2,2,2) -- (2,2,0) -- cycle;

      \draw[thick,preaction={fill, white, fill opacity=0.8},pattern=north west lines] (0,2,0) -- (0,2,2) -- (2,2,2) -- (2,2,0) -- cycle;
      \draw[thick,fill=white,fill opacity=0.7] (0,0,2) -- (2,0,2) -- (2,2,2) -- (0,2,2) -- cycle;
      \draw[thick,fill=white,fill opacity=0.7] (2,0,0) -- (2,0,2) -- (2,2,2) -- (2,2,0) -- cycle;
    \end{tikzpicture}
  \]

  We try to solve $\Call{KanCSP}{}$ with no open sides. This CSP has 7
  variables corresponding to sides $i$, $j$, $k$ and a backside $l = \izero$.
  If we can construct all 7 cubes in a compatible way, we have solved the
  boundary problem as we can then return a filler in direction $\izero \to
  \ione$. Note in particular that the dimension $\ell$ is only part of the
  filler term, while the cube we are filling is three-dimensional.

  After imposing the first set of constraints, the domains for the $i$ and $j$
  sides are significantly reduced, e.g., $D_{\csvar{i}{\izero}} = \{
  p(\Sigma) \}$ for $\Sigma : \pint{3} \to \mathcal{P}(\pint{2})$ is
  given by:
  \[
  \begin{array}{llll}
    \izero\izero\izero \mapsto \{ \izero\izero \}&
    \izero\izero\ione \mapsto \{ \izero\izero \}&
    \izero\ione\izero \mapsto \{ \izero\izero,\izero\ione\}&
    \izero\ione\ione \mapsto \{ \izero\ione\}\\
    \ione\izero\izero \mapsto \{ \izero\izero,\ione\izero\}&
    \ione\izero\ione \mapsto \{ \ione\izero\}&
    \ione\ione\izero \mapsto \{ \izero\izero,\izero\ione,\ione\izero,\ione\ione\} &
    \ione\ione\ione \mapsto \{ \ione\ione\}
  \end{array}
  \]

  The PPM $\Sigma$ gives rise to 9 contortions of $p$, which contrasts with
  $D(3)^2 = 400$ total contortions of $p$. The domains for
  $D_{\csvar{k}{\izero}}$, $D_{\csvar{k}{\ione}}$, and the back side
  $D_{\csvar{l}{\izero}}$ still contain all contortions of $x$, $p$ and $q$ into
  three dimensions since the $k$ sides of the goal boundary do not give any
  indication which contortion could be used for this side of the filler.

  The second set of constraints ensures that all sides of the Kan
  filler have matching boundaries, after which we find a solution to
  \Call{KanCSP}{} that gives rise to the following
  filler:
  \[\isfterm[\Gamma][i,j,k]{\ffill*{\izero}{\ione}{l}{
      \begin{array}{lll}
        \assign{i}{\izero}{p(j,k \meet l)} & \assign{j}{\izero}{q(i,k)} & \assign{k}{\izero}{x} \\
        \assign{i}{\ione}{p(j,k \meet l)} & \assign{j}{\ione}{q(i,k)} & \assign{k}{\ione}{p(j,l)}
      \end{array}
    }{q(i,k)}}\]
  This filler captures the argument sketched in \autoref{fig:ehcube}, albeit in
  a single step: the $p$ sides are mapped to the $k = \ione$ side such that they
  cancel out as in \autoref{eh-inverses}, while the $q$ sides are constantly mapped
  to the backside of the filler, which is the cube from \autoref{eh-degeneracy}.
\end{exa}

\subsection{A solver for \KanCubicalCell}
\label{ssec:solver}

We now give an algorithm to construct fillers of open cubes which might have
fillers on their faces, and not only contorted terms as in \Call{KanCSP}{}. We
also make use of a procedure $\Call{KanFill}{\Gamma , \Psi , \phi}$ which
produces fillers with the same dimension as $\phi$: we check for any face of
$\phi$ if it gives rise to a natural filler.

The difficult part of \KanCubicalCell is the construction of higher-dimensional
fillers, which might possibly have fillers on their sides. We introduce a
variable $d$ to iteratively deepen the level of such nested fillers, which
effects a sort-of ``breadth-first'' search for nested fillers.

Given a goal boundary $\phi$, we search for solutions either by natural fillers or
by higher-dimensional fillers constructed with \Call{KanCube}{} on
line~\ref{kan:line:fillorcomp}. In \Call{KanCube}{}, we first select a set of
sides that are left open on line~\ref{kan:line:openorder} and then pick a
solution to the corresponding \Call{KanCSP}{} on line~\ref{kan:line:csp}, which
will fill all sides not left open with contorted cells. Finally, we call
\Call{KanSolver}{} recursively on the open sides on line~\ref{kan:line:rec},
where $\boundary{\face{\phi'}{\inclcon{i}{e}}}$ denotes the boundary at
$\inclcon{i}{e}$ induced by the faces already present in $\phi'$.

\begin{algorithm}[H]
      \caption{Finding Kan cells}
      \label{alg:kan}
      \begin{algorithmic}[1]
        \Require $\isfbdy[\Gamma][\Psi]{\phi}$, depth variable $d$
        \Ensure $\isfterm[\Gamma][\Psi]{t}<\phi>$, if
        $\Call{\KanCubicalCell}{\Gamma,\Psi,\phi}$ solvable with $\leq d$
        nested Kan fillers
    \Procedure{KanSolver}{$\Gamma, \Psi, \phi , d$}
      \If{$d = 0$}
        \State \Return{\texttt{Unsolvable}}
      \EndIf
      \State $t \gets$ \Call{KanFill}{$\Gamma,\Psi,\phi$} $\cup$
      \Call{KanCube}{$\Gamma, \Psi, \phi, d$} \label{kan:line:fillorcomp}
    \EndProcedure

    \Procedure{KanCube}{$\Gamma, \Psi, \phi,d$} \label{kan:line:main}
      \State{$Ope \gets \mathcal{P}(\{ \csvar{i}{e} \mid i \in \Psi, e \in
        \{\izero, \ione\} \} \cup \{\csvar{k}{\izero}\}) $}   \label{kan:line:openorder}
        \State $\phi' \gets \mproblem{KanCSP}(\phi,Ope)$ \label{kan:line:csp}
        \For{$\csvar{i}{e} \in Ope$}
        \State $t \gets \Call{KanSolver}{\boundary{\face{\phi'}{\inclcon{i}{e}}}, d-1}$ \label{kan:line:rec}
        \State $\phi' \coloneq \boundary{\phi' \se \assign{i}{e}{t}}$
        \EndFor
        \State \Return{$\isfterm[\Gamma][\Psi]{\ffill{\izero}{\ione}{k}{\phi'
              - \csvar{k}{\izero} }{(\face{\phi'}{\inclcon{k}{\izero}})}}<\phi>$}
    \EndProcedure
  \end{algorithmic}
\end{algorithm}

The choices of solutions and open sides on lines~\ref{kan:line:fillorcomp},
\ref{kan:line:openorder}, \ref{kan:line:csp} and \ref{kan:line:rec} are
non-deterministic, which is implemented using the list monad in the solver
discussed in \S\ref{sec:casestudy}. In practice, the performance of the
algorithm depends heavily on the choices we make at this point. In our
implementation, we first try to solve \Call{KanCSP}{} with $Ope = \emptyset$. If
contortions are not enough to construct all sides, it is useful to first use
natural fillers which are induced by the goal boundary. In addition, it is
expedient to incrementally increase the number of open sides solutions of
\Call{KanCSP}{}, e.g., using the depth-parameter $d$.%

We now devise a complete search procedure for \KanCubicalCell with
Algorithm~\ref{alg:solver}. The \Call{Solver}{} starts by trying to contort some
cell of the context into the goal boundary. If this fails, we perform
iterative deepening on the level of nested Kan cells constructed by
Algorithm~\ref{alg:kan}. Again, the contortion theory is a parameter to our
solver, which means that \Call{Contort}{} will call the solver for Dedekind and
De Morgan contortions introduced in \S\ref{sec:contortions}, or simply look for
a cartesian or disjunctive contortion by brute-force.

\begin{algorithm}[H]
  \caption{A solver for boundary problems}
  \label{alg:solver}
  \begin{algorithmic}[1]
    \Require $\isfbdy[\Gamma][\Psi]{\phi}$
    \Ensure $\isfterm[\Gamma][\Psi]{t}<\phi>$, if $\Call{\KanCubicalCell}{\Gamma,\Psi,\phi}$ is solvable
    \Procedure{Solver}{$\Gamma, \Psi, \phi$}
      \For{$p \in \Gamma$}
        \State $t \gets$ \Call{Contort}{$\Gamma,\Psi,\phi,p$}
        \If{$t \neq$ \texttt{Unsolvable}}
          \State \Return{$t$}
        \EndIf
      \EndFor
      \For{$d \in \{ 1,\ldots \}$}
        \State $t \gets$ \Call{KanSolver}{$\Gamma,\Psi,\phi,d$}
        \If{$t \neq$ \texttt{Unsolvable}}
          \State \Return{$t$}
        \EndIf
      \EndFor
    \EndProcedure
  \end{algorithmic}
\end{algorithm}

\begin{exa}[name=Sq→Comp] \label{exp:sqtocomp}
  To complete the proof of Eckmann-Hilton, we need to fill the cube from
  \autoref{eh-main} using \autoref{eh-cubical}.
  This problem can be solved directly using Dedekind contortions, but finding
  the four-dimensional filler is relatively involved (but can be done using the solver
  presented in \S\ref{sec:casestudy}). Instead, it is easier to consider a
  lower-dimensional---and hence more general---version of the boundary problem.
  Concretely, the cube from \autoref{eh-cubical} is captured with a square

  \begin{figure}[h!!!]
  \centering
  \begin{minipage}{0.68\linewidth}
    \[
      \Gamma \coloneqq \left\{
        \begin{array}{l}
          \isof{x},\\\isof{p}{i}[\assign{i}{\izero}{x} \se \assign{i}{\ione}{x} ],\\
          \isof{q}{i}[\assign{i}{\izero}{x} \se \assign{i}{\ione}{x} ],\\
          \isof*{\alpha}{i,j}[
          \begin{array}{ll}
            \assign{i}{\izero}{p(j)} &\se \assign{j}{\izero}{q(i)} \\
            \assign{i}{\ione}{p(j)} &\se \assign{j}{\ione}{q(i)}
          \end{array}
          ]
        \end{array}
      \right.
    \]
  \end{minipage}
  \begin{minipage}{0.28\linewidth}
    \begin{mathpar}
      \begin{tikzpicture}[scale=.5]
        \draw[->,>=stealth] (0,0) -- (3,0) node [midway,below, fill=none] {$p(j)$};
        \draw[->,>=stealth] (0,0) -- (0,3) node [midway,left, fill=none] {$q(i)$};
        \draw[->,>=stealth] (3,0) -- (3,3) node [midway,right, fill=none] {$q(i)$};
        \draw[->,>=stealth] (0,3) -- (3,3) node [midway,above, fill=none] {$p(j)$};
        \node (filler) at (1.5,1.5) {$\alpha$};
      \end{tikzpicture}
      \dimsquare{i}{j}
    \end{mathpar}
  \end{minipage}
  \end{figure}
  \noindent that we want to turn into a square with both path concatenations on opposite sides:
  \begin{equation}\label{sqtocomp}
    \isfterm[\Gamma][i,j]{\hole}<
        \assign{i}{\izero}{(\pcomp{p}{q})(j)} \se
        \assign{i}{\ione}{(\pcomp{q}{p})(j)} \se
        \assign{j}{\izero}{x} \se
        \assign{j}{\ione}{x}
      >
  \end{equation}

  Incidentally, $\Gamma$ is the list of generators of the HIT capturing the
  \AgdaDatatype{Torus} in \agdaCubical, while boundary (\ref{sqtocomp}) captures
  $T^2$, the definition of the torus in the HoTT book \cite{hott}. A solution to
  this problem thus induces a map from the cubical torus to the
  HoTT book torus.

  We solve (\ref{sqtocomp}) using Algorithm~\ref{alg:solver}. After seeing that we
  cannot solve this goal with a contortion, the algorithm at some point reaches
  depth $d = 3$ and solves \Call{KanCSP}{} with open sides $Ope = \{
  \csvar{i}{\izero}, \csvar{i}{\ione} \}$. A solution to this CSP has the
  constant $x$ square for $j = \izero$, $p(k)$ for $j = \ione$ and $q(j)$ for $k
  = \izero$ as depicted in the left cube below.

  When calling \Call{KanSolver}{} recursively on the two missing sides, we find
  with \Call{KanFill}{} that the $i = \ione$ side can be solved with the natural
  filler for $\pcomp{q}{p}$. To fill side $i = \izero$, we again have to
  construct an open cube. One solution of \Call{KanCSP}{} for this open cube is
  depicted on the right below. The $k = \ione$ side is filled by the natural
  filler for $\pcomp{p}{q}$. The other sides can be filled with contortions,
  where side $j = \ione$ makes use of $\alpha$.

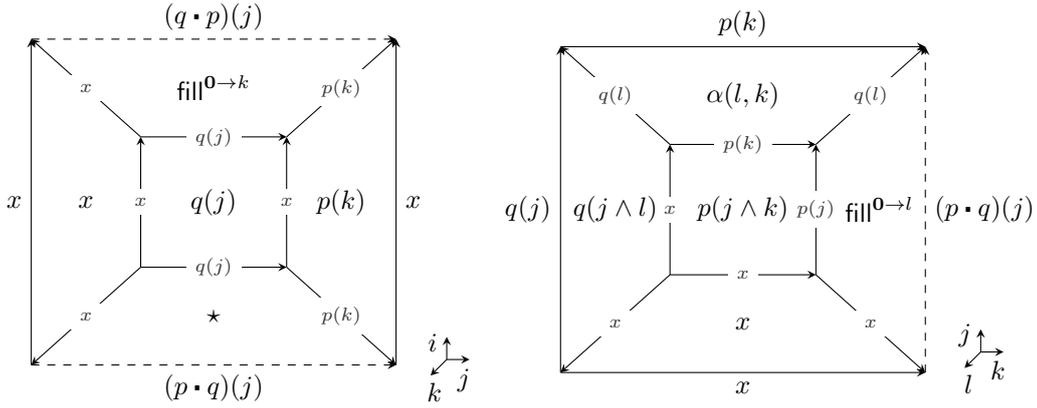
\begin{figure}[h!!!]
  \centering
  \begin{minipage}{0.46\linewidth}
  \begin{tikzpicture}[xscale=1.2,yscale=1.08]
    \draw[->,>=stealth,dashed] (0,0) -- (4,0) node [midway,below, fill=none] {$(\pcomp{p}{q})(j)$};
    \draw[->,>=stealth,dashed] (0,4) -- (4,4) node [midway,above, fill=none] {$(\pcomp{q}{p})(j)$};
    \draw[->,>=stealth] (0,0) -- (0,4) node [midway,left, fill=none] {$x$};
    \draw[->,>=stealth] (4,0) -- (4,4) node [midway,right,fill=none] {$x$};

    \draw[->,>=stealth] (1.2,1.2) -- (2.8,1.2) node [midway,fill=white,text=black!80] {$\scriptstyle q(j)$};
    \draw[->,>=stealth] (1.2,2.8) -- (2.8,2.8) node [midway,fill=white,text=black!80] {$\scriptstyle q(j)$};
    \draw[->,>=stealth] (1.2,1.2) -- (1.2,2.8) node [midway,fill=white,text=black!80] {$\scriptstyle x$};
    \draw[->,>=stealth] (2.8,1.2) -- (2.8,2.8) node [midway,fill=white,text=black!80] {$\scriptstyle x$};

    \draw[->,>=stealth] (1.2,1.2) -- (0,0) node [midway,fill=white,text=black!80] {$\scriptstyle x$};
    \draw[->,>=stealth] (2.8,1.2) -- (4,0) node [midway,fill=white,text=black!80] {$\scriptstyle p(k)$};
    \draw[->,>=stealth] (2.8,2.8) -- (4,4) node [midway,fill=white,text=black!80] {$\scriptstyle p(k)$};
    \draw[->,>=stealth] (1.2,2.8) -- (0,4) node [midway,fill=white,text=black!80] {$\scriptstyle x$};

    \node at (2, .6)  {$?$};
    \node at (2, 3.4) {$\mathsf{fill}^{\izero \to k}$};
    \node at (.6, 2)  {$x$};
    \node at (3.4, 2) {$p(k)$};
    \node at (2, 2)  {$q(j)$};
  \end{tikzpicture}
  \!\!\!\!\!\!\!\!
  \dimcube{i}{j}{k}
\end{minipage}
\begin{minipage}{0.53\linewidth}
  \begin{tikzpicture}[xscale=1.2,yscale=1.08]
    \draw[->,>=stealth] (0,0) -- (4,0) node [midway,below, fill=none] {$x$};
    \draw[->,>=stealth] (0,4) -- (4,4) node [midway,above, fill=none] {$p(k)$};
    \draw[->,>=stealth] (0,0) -- (0,4) node [midway,left, fill=none] {$q(j)$};
    \draw[->,>=stealth,dashed] (4,0) -- (4,4) node [midway,right,fill=none] {$(\pcomp{p}{q})(j)$};

    \draw[->,>=stealth] (1.2,1.2) -- (2.8,1.2) node [midway,fill=white,text=black!80] {$\scriptstyle x$};
    \draw[->,>=stealth] (2.8,1.2) -- (2.8,2.8) node [midway,fill=white,text=black!80] {$\scriptstyle p(j)$};
    \draw[->,>=stealth] (1.2,1.2) -- (1.2,2.8) node [midway,fill=white,text=black!80] {$\scriptstyle x$};
    \draw[->,>=stealth] (1.2,2.8) -- (2.8,2.8) node [midway,fill=white,text=black!80] {$\scriptstyle p(k)$};

    \draw[->,>=stealth] (1.2,1.2) -- (0,0) node [midway,fill=white,text=black!80] {$\scriptstyle x$};
    \draw[->,>=stealth] (2.8,1.2) -- (4,0) node [midway,fill=white,text=black!80] {$\scriptstyle x$};
    \draw[->,>=stealth] (2.8,2.8) -- (4,4) node [midway,fill=white,text=black!80] {$\scriptstyle q(l)$};
    \draw[->,>=stealth] (1.2,2.8) -- (0,4) node [midway,fill=white,text=black!80] {$\scriptstyle q(l)$};

    \node at (2, .6)  {$x$};
    \node at (2, 3.4) {$\alpha(l,k)$};
    \node at (.6, 2)  {$q(j \meet l)$};
    \node at (3.4, 2) {\ \ $\mathsf{fill}^{\izero \to l}$};
    \node at (2, 2)  {$p(j \meet k)$};
  \end{tikzpicture}
  \!\!\!\!\!\!\!\!\!\!\!\!\!\!\!\!\!\!\!\!\!\!\!\!\!\!\!
  \dimcube{j}{k}{l}
\end{minipage}
\end{figure}
\end{exa}

While the Dedekind contortions are quite powerful and were an apt contortion
theory to prove Eckmann-Hilton, it can often be useful to have reversal $\inv$
available, in particular for lower-dimensional proof goals where the even faster
blowup of the number of De Morgan contortions is not so severe.

\begin{exa}[name=Associativity of path concatenation] \label{exp:assoc}
  Given a context $\Gamma$
  \[
    \isof{p}{i}[], \isof{q}{i}[\assign{i}{\izero}{p(\ione)} ],
    \isof{r}{i}[\assign{i}{\izero}{q(\ione)}] \] we want to show that path
  composition as defined in \S\ref{sec:bdryproblems} is associative. This
  amounts to constructing a term with the following boundary.
  \[\isfbdy[\Gamma][i,j]{\boundary{\assign{j}{\izero}{(\pcomp{(\pcomp{p}{q})}{r})(i)} \se
      \assign{j}{\ione}{(\pcomp{p}{(\pcomp{q}{r})})(i)} \se
      \assign{i}{\izero}{p(\izero)} \se \assign{i}{\ione}{r(\ione)}}}\]

  To solve this boundary problem, the solver constructs an open 3-cube and
  finds that almost no sides of it can be filled with a contorted cell: the $j$
  sides have a Kan filler on their boundary and are hence best filled using the
  natural fillers. This means that the back and right side of the box will also
  have a path composition on their boundary. Hence only the left side can be
  filled with a contortion, namely the square which is constantly $p(\izero)$.
  In sum, the CSP solver called at line~\ref{kan:line:csp} returns the
  following box, where the fillers $?_0$ and $?_1$ for sides $i=\ione$
  and $k=\izero$ need to be constructed by calling the solver recursively.

  \[
    \hspace{-3em}
    \begin{tikzpicture}[xscale=7,yscale=2]
      \tikzmath{\x1=.2; \dx=.27;}
      \tikzmath{\y1=.4; \dy=.6;}
      \tikzmath{\o=1.6;}
      \tikzmath{\x2=\x1+\dx;}
      \tikzmath{\y2=\y1+\dy;}
      \tikzmath{\xp1 = (\x1+\x2)/2;}
      \tikzmath{\yp1 = (\y1+\y2)/2;}

      \draw[->,>=stealth] (-\x2,-\y2) -- ( \x2,-\y2) node [midway,below, fill=none] {$(\pcomp{(\pcomp{p}{q})}{r})(i)$};
      \draw[->,>=stealth] (-\x2, \y2) -- ( \x2, \y2) node [midway,above, fill=none] {$(\pcomp{p}{(\pcomp{q}{r})})(i)$};
      \draw[->,>=stealth] (-\x2,-\y2) -- (-\x2, \y2) node [midway,left, fill=none] {$p(\izero)$};
      \draw[->,>=stealth] ( \x2,-\y2) -- ( \x2, \y2) node [midway,right,fill=none] {$r(\ione)$};

      \draw[->,>=stealth] (-\x1,-\y1) -- ( \x1,-\y1) node [midway,fill=white,text=black!80] {$(\pcomp{p}{q})(i)$};
      \draw[->,>=stealth] (-\x1, \y1) -- ( \x1, \y1) node [midway,fill=white,text=black!80] {$p(i)$};
      \draw[->,>=stealth] (-\x1,-\y1) -- (-\x1, \y1) node [midway,fill=white,text=black!80] {$p(\izero)$};
      \draw[->,>=stealth,dashed] ( \x1,-\y1) -- ( \x1, \y1);

      \draw[->,>=stealth] (-\x1,-\y1) -- (-\x2,-\y2) node [midway,fill=white,text=black!80] {$p(\izero)$};
      \draw[->,>=stealth] ( \x1,-\y1) -- ( \x2,-\y2) node [midway,fill=white,text=black!80] {$r(k)$};
      \draw[->,>=stealth] ( \x1, \y1) -- ( \x2, \y2) node [midway,fill=white,text=black!80] {$(\pcomp{q}{r})(k)$};
      \draw[->,>=stealth] (-\x1, \y1) -- (-\x2, \y2) node [midway,fill=white,text=black!80] {$p(\izero)$};

      \node at (0,-\yp1) {$\mathsf{fill}^{\izero \to k}$};
      \node at (0, \yp1) {$\mathsf{fill}^{\izero \to k}$};
      \node at (-\xp1,0) {$p(\izero)$};
      \node at ( \xp1,0) {$?_1$};
      \node at (0,0) {$?_0$};
    \end{tikzpicture}
    \dimcube{j}{i}{k}
  \]

  For the back side at $k = \izero$, the boundary problem $?_0$ has an open
  right side as we do not know what the common boundary of $?_0$ and $?_1$
  should be. The solver hence constructs a constraint satisfaction problem with
  the right-hand side $i = \ione$ free. The goal boundary of this line is
  $[\assign{j}{\izero}{q(\ione)} \se \assign{j}{\ione}{p(\ione) = q(\izero)} ]$,
  which suggests that a De Morgan contortion involving reversal $\inv$ will come
  in handy. Indeed, when calling the solver for $?_0$, it again constructs an
  open 3-cube, uses the natural filler for $\pcomp{p}{q}$ for the $j = \izero$
  square, and is able to fill all other sides with contortions as follows.

    \[
    \hspace{-3em}
    \begin{tikzpicture}[xscale=7,yscale=2]
      \tikzmath{\x1=.2; \dx=.37;}
      \tikzmath{\y1=.4; \dy=.6;}
      \tikzmath{\o=1.6;}
      \tikzmath{\x2=\x1+\dx;}
      \tikzmath{\y2=\y1+\dy;}
      \tikzmath{\xp1 = (\x1+\x2)/2;}
      \tikzmath{\yp1 = (\y1+\y2)/2;}

      \draw[->,>=stealth] (-\x2,-\y2) -- ( \x2,-\y2) node [midway,below, fill=none] {$(\pcomp{p}{q})(i)$};
      \draw[->,>=stealth] (-\x2, \y2) -- ( \x2, \y2) node [midway,above, fill=none] {$p(i)$};
      \draw[->,>=stealth] (-\x2,-\y2) -- (-\x2, \y2) node [midway,left, fill=none] {$p(\izero)$};
      \draw[->,>=stealth] ( \x2,-\y2) -- ( \x2, \y2) node [midway,right,fill=none] {$q(\inv j)$};

      \draw[->,>=stealth] (-\x1,-\y1) -- ( \x1,-\y1) node [midway,fill=white,text=black!80] {$p(i)$};
      \draw[->,>=stealth] (-\x1, \y1) -- ( \x1, \y1) node [midway,fill=white,text=black!80] {$p(i)$};
      \draw[->,>=stealth] (-\x1,-\y1) -- (-\x1, \y1) node [midway,fill=white,text=black!80] {$p(\izero)$};
      \draw[->,>=stealth] ( \x1,-\y1) -- ( \x1, \y1)  node [midway,fill=white,text=black!80] {$p(\ione)$};

      \draw[->,>=stealth] (-\x1,-\y1) -- (-\x2,-\y2) node [midway,fill=white,text=black!80] {$p(\izero)$};
      \draw[->,>=stealth] ( \x1,-\y1) -- ( \x2,-\y2) node [midway,fill=white,text=black!80] {$q(k)$};
      \draw[->,>=stealth] ( \x1, \y1) -- ( \x2, \y2) node [midway,fill=white,text=black!80] {$p(\ione)$};
      \draw[->,>=stealth] (-\x1, \y1) -- (-\x2, \y2) node [midway,fill=white,text=black!80] {$p(\izero)$};

      \node at (0,-\yp1) {$\mathsf{fill}^{\izero \to k}$};
      \node at (0, \yp1) {$p(i)$};
      \node at (-\xp1,0) {$p(\izero)$};
      \node at ( \xp1,0) {$q(\inv j \land k)$};
      \node at (0,0) {$p(i)$};
    \end{tikzpicture}
    \dimcube{j}{i}{k}
  \]

  The contortion $\inv j \land k$ was picked out among $D(4) = 168$ possible
  contortions of $q$, which can be represented with a PPM $\pint{4} \to
  \mathcal{P}(\pint{})$ containing only 16 entries.

  For the remaining side $?_1$ we are left to construct a square
  with the boundary
    \[\isfbdy[\Gamma][j,k]{\boundary{\assign{j}{\izero}{r(k)} \se
      \assign{j}{\ione}{(\pcomp{q}{r})(k)} \se
      \assign{k}{\izero}{q(\inv j)} \se \assign{k}{\ione}{r(\ione)}}}\]
  which can be done in much the same spirit as $?_0$ was solved.

  In summary, having the powerful De Morgan theory at hand makes the proof of
  associativity of path concatenation relatively straightforward, while already
  in the case of Dedekind contortions we would have had to come up with
  additional fillers wherever $\inv$ was used, let alone the cartesian theory
  where the solver has to come up with quite involved nested fillers.
\end{exa}

\section{A practical solver for Cubical Agda boundary problems}
\label{sec:casestudy}

We have implemented the solver in \Haskell,\footnote{We have implemented a
  solver which is parametric over all contortion theories
  (\url{https://github.com/maxdore/cubetac}) as well as a solver specialised to
  the Dedekind contortions which comes with an interface to \CubicalAgda which
  was used to generate the code in this paper
  (\url{https://github.com/maxdore/dedekind}).} providing the first experimental
solver for boundary problems coming from \CubicalAgda. The implementation of
\Call{KanCSP}{} is based on a monadic solver for finite domain constraint
satisfaction problems \cite{overton15_const_progr_haskel}. The user inputs
problems in a \texttt{.cube} file which contains a cell context and boundary
problems over that context. If the solver finds a solution, it is printed in
\CubicalAgda syntax so that it can be copied and pasted into proof goals. Proper
integration into \CubicalAgda that allows the solver to be called as a tactic
from \Agda is left to future work.

We have curated a small benchmarking suite of boundary problems, many of which are
from the \agdaCubical library. The problems are common proof obligations,
such as associativity of path concatenation,
rearrangements of sides of cubes, etc. On a standard laptop, all problems are quickly solved (often in $<50$ms). This means that the solver is fast enough to fit
seamlessly into a formalisation workflow and can be used as a tactic
for solving routine proof goals. It can also solve some more complex goals such as \autoref{exp:eh}.

In \CubicalAgda, the constant path at $x$ of type \AgdaBound{x}~\AgdaFunction{≡}~\AgdaBound{x} is expressed with λ-abstraction as
\AgdaSymbol{λ}~\AgdaBound{i}~\AgdaSymbol{→}~\AgdaBound{x}.
We can use the \AgdaFunction{PathP} type to describe higher-dimensional
boundaries, e.g., \AgdaPostulate{PathP}\AgdaSpace{}%
\AgdaSymbol{(λ}~\AgdaBound{j}~\AgdaSymbol{→}~\AgdaBound{x}~\AgdaOperator{\AgdaFunction{≡}}~\AgdaBound{x}\AgdaSymbol{)}
\AgdaSymbol{(λ}~\AgdaBound{i}~\AgdaSymbol{→}~\AgdaBound{x}\AgdaSymbol{)}~\AgdaSymbol{(λ}~\AgdaBound{i}~\AgdaSymbol{→}~\AgdaBound{x}\AgdaSymbol{)}
is the boundary of a square with reflexive paths on its sides. Given two such
squares $p$ and $q$, \textsf{\nameref{exp:eh}} is derived in $\sim$150ms:
\agdaeckmannhiltoncube{}%

The \CubicalAgda primitive \AgdaFunction{hcomp}
captures Kan fillers in direction $\izero \to \ione$.
The solution to the boundary problem discussed in
the \textsf{\nameref{exp:sqtocomp}} example is found in $\sim$15ms,
its translation into \CubicalAgda looks as follows (manually compressed to not use too much space in the paper; the actual pretty-printed output is more readable):%
\agdasqtocomp{}%
The function \AgdaFunction{hfill} $\phi$ $t$ $i$ is used in
\agdaCubical to define fillers in direction $\izero \to i$. The term $t$ has
to be embedded into the cube structure using \AgdaFunction{inS}, which is
inserted automatically by the \CubicalAgda syntax pretty-printer of the solver.

Using these two automatically constructed proofs, we can readily establish by
hand the classical formulation of the Eckmann-Hilton argument in terms of
path concatenations:%
\agdaeckmannhilton{}

The boundary problem posed by \AgdaFunction{EckmannHilton} can also be passed
directly to our solver as a single problem instance (without requiring a manual
decomposition into \AgdaFunction{Sq→Comp} and \AgdaFunction{EckmannHilton-Cube}),
and our search strategy should in principle yield a solution to this boundary
problem. However, our solver is not yet able to find such a solution within
100s. We have also curated some further boundary problems which cannot be solved
at the moment, these include a 7-dimensional analogue of the
\textsf{\nameref{exp:contsolve}} example and the syllepsis
\cite{sojakova22_syllep_homot_type_theor}, which establishes a higher coherence
property of the Eckmann-Hilton proof.

In summary, while there is room to make the solver more performant, it
can quickly prove technical lemmas for us that would be tedious to prove by
hand, taking significant proof burden from a user of \CubicalAgda. Furthermore,
some deeper results of synthetic homotopy theory, like the Eckmann-Hilton
argument, can also be proved if the statement is phrased carefully.

\section{Future and related work}

There are many ways in which our work can be extended: the performance of the
solver can be improved by exploring other heuristics and refinements of the
algorithms; the solver should be properly integrated into theorem provers such
as \CubicalAgda and \redtt. The solver could be extended to problems involving
multiple types and functions and to use cubical type theory's \emph{transport}
primitive.

Early work on proof automation in HoTT is Brunerie's work on computer-generated
proofs for the monoidal structure of smash products \cite{brunerie18} which used
path-induction and metaprogramming in \Agda.
\cite{grzybowski23} generates visual presentations of \CubicalAgda proof terms.
The problem of deciding equality in
the cofibration logic of cubical type theories has been studied by
\cite{rose25_compl_cubic_cofib_logic_i}. Among other things, they also establish
complexity-related results, in particular, that the entailment problems of the
cofibration languages of \cite{FootballHockeyLeague19} and
\cite{cohen18_cubic_type_theor} are coNP-complete. Another line of related work
is on higher-dimensional algebraic rewriting, in particular, on
$\infty$-categories \cite{finster22}, operads
\cite{thanh19_sequen_calcul_opetop}, polygraphs~\cite{polygraphs} and
associative $n$-categories~\cite{dorn18_assoc}. For the latter, the tool
\systemname{homotopy.io}~\cite{reutter21_high} gives a graphical user interface
for constructing cells based on a higher-dimensional generalisation of string
diagrams. Recently, there has been work on automatically constructing coherences
for globular theories which are ``weak'' in the sense of having, e.g., unitality
and associativity of path concatenation hold not definitionally, but only up to
some computational witness (similar to cubical type theory).
\cite{benjamin25_natur} have devised a ``naturality construction'' which gives
rise to several such coherences, for instance, they derive associativity of path
concatenation in their setting, akin to our \autoref{exp:assoc}. Relatedly,
\cite{benjamin25_higher_eckman_hilton_argum_type_theor} devise a general recipe
for proving (higher versions of) the Eckmann-Hilton argument, which we
established for 2-loops in \autoref{exp:eh}. Our approach differs to theirs
however in that we start with a (possibly unsolvable) boundary problem and try
to construct a coherence for it, while
\cite{benjamin25_natur,benjamin25_higher_eckman_hilton_argum_type_theor}
construct a class of coherences for some given construction.

More work is necessary to understand the precise relationship between geometric
accounts of higher-dimensional paths, due to Kan and at the heart of cubical
type theories, and theories based on algebraic rewriting, such as associative
$n$-categories. It would be highly beneficial if methods and ideas could more
easily be exchanged between both approaches, e.g., to devise a semi-automated
proof search which allows the user to interact with higher-dimensional cells as
pioneered by \systemname{homotopy.io}, but also resolves some problems using an
automatic search similar to the present system. Besides, given that the search
space for higher-dimensional paths is generally vast, it could be expedient to
employ recent advances in machine learning for proof search.

\section*{Acknowledgement}
\noindent We are grateful to Axel Ljungström for discussions about the solver
and to him and Tom Jack for cubical versions of Eckmann-Hilton and syllepsis for
us to test it with; and to both the FSCD and LMCS reviewers for their careful reading
and constructive feedback.

\bibliographystyle{alphaurl}
\bibliography{bibliography}

\end{document}